\documentclass[12pt]{article}
\usepackage{ws-rv-van}             
\usepackage{graphicx,color}
\usepackage{amssymb}
\usepackage{amsmath}
\usepackage{amsfonts}
\usepackage{array,hhline,dcolumn} 
\usepackage{epstopdf}
\makeindex

\title{Neutrino Experiments}

\begin{document}

\def\ie{{\sl i.e.}}
\def\kov{$\check{\rm C}$erenkov }
\def\hz{\,{\rm Hz}}
\def\khz{\,{\rm kHz}}
\def\mhz{\,{\rm MHz}}
\def\kev{\,{\rm keV}}
\def\mev{\,{\rm MeV}}
\def\gev{\,{\rm GeV}}
\def\cmsq{\,{\rm cm}^2}
\def\nm{\,{\rm nm}}
\def\ms{\,{\rm m}}
\def\msec{\,{\rm ms}}
\def\nsec{\,{\rm ns}}
\def\dmsq{\Delta m^{2}~}
\def\musec{\mu s}
\def\numu{\nu_{\mu}}
\def\numubar{\bar\nu_{\mu}}
\def\nue{\nu_e}
\def\nuebar{\bar\nu_e}
\def\nutau{\nu_{\tau}}
\def\piplusdecay{\pi^+ \rightarrow \mu^+ \numu}
\def\piminusdecay{\pi^- \rightarrow \mu^- \numubar}
\def\pimue{\pi \rightarrow \mu \rightarrow e}
\def\muplusdecay{\mu^+ \rightarrow e^+ \nue \numubar}
\def\muminusdecay{\mu^- \rightarrow e^- \nuebar \numu}
\def\sinsqtheta{\sin^2 2 \theta~}
\def\mutoe{\numu \rightarrow \nue}
\def\mutotau{\numu \rightarrow \nutau}
\def\pizero{\pi^0~}
\def\reaction{\nuebar + p \to e^+ + n\ \ \ ,}
\def\evsq{{eV}^2~}
\def\flux{\,\nu/{\rm cm}^2/p}
\def\barflux{\,\bar \nu/{\rm cm}^2/p}
\def\ga{\gamma}
\def\gtwid{\mathrel{\raise.3ex\hbox{$>$\kern-.75em\lower1ex\hbox{$\sim$}}}}
\def\ltwid{\mathrel{\raise.3ex\hbox{$<$\kern-.75em\lower1ex\hbox{$\sim$}}}}
\def\mbne{MiniBooNE\ }
\def\BN{BooNE\ }
\def\Gp{GeV/c\ }
\def\gev{GeV\ }
\def\Bo{Booster\ }
\def\MI{Main Injector\ }

\maketitle

\begin{center}

{\bf J.M. Conrad}\\
{\it Columbia University, New York, NY 10027}

\end{center}

\abstract{ This article is a summary of four introductory lectures on
  ``Neutrino Experiments,'' given at the 2006 TASI summer school.  The
  purposes were to sketch out the present questions in neutrino physics
  and to discuss the experimental challenges in addressing them. This
  article concentrates on specific, illustrative examples rather than
  providing a complete overview of the field of neutrino
  physics. These lectures were meant to lay the ground-work for the
  talks which followed on specific, selected topics in neutrino
  physics.}

\newpage

This article is a summary of four introductory 
lectures on ``Neutrino Experiments,'' 
given at the 2006 TASI summer school.   The purpose was to sketch out 
the present questions in neutrino physics, and discuss
the experiments that can address them.  The ideas were then explored 
in depth by later lecturers.  

This article begins with an overview of neutrinos in the Standard
Model and what we know about these particles today.  This is followed
by a discussion of the direction of the field, divided into the three
themes identified in the {\it APS Study on the Future of Neutrino
  Physics.}\cite{APSstudy} This APS study represented the culmination
of a year-long effort by the neutrino community to come to a consensus
on future directions.  The report is recommended reading for students,
along with the accompanying working group white papers, especially the
Theory Group Whitepaper.\cite{APSwhitepaper}

While these lectures used the APS Neutrino Study themes as the core,
the emphasis here is different from the APS report. The point of a
summer school is to teach specific ideas rather than provide a
perfectly balanced overview of the field. The result is that, with
apologies, some experiments were necessarily left out of the
discussion.  Students are referred to the Neutrino Oscillation
Industry Website\cite{oscind} for a complete list of all neutrino
experiments, by category.

\section{Neutrinos As We Knew Them}

Neutrinos are different from the other fermions.  Even before the 
recent evidence of neutrino mass, neutrinos were peculiar members 
of the Standard Model.  They are the only fermions 
\begin{itemize}
\item to carry no  electric charge.   
\item for which we have no evidence of a 
right-handed partner.  
\item that are defined as massless.
\end{itemize}
These ideas are connected by the fact that, unlike
other spin 1/2 particles, neutrinos can only
interact through the weak interaction.

Even though the Standard Model picture is now demonstrably wrong,
this theoretical framework provides a good place to start the discussion.
This section begins by expanding on the Standard Model
picture of the neutrino sketched above.   It then discusses 
how neutrinos interact.  This is followed by an overview of 
neutrino sources and detectors.

\subsection{Neutrinos in the Standard Model}
\label{sub:SM}

\begin{figure} [t]
\centering
\scalebox{.5}{\includegraphics[clip=T, bb=0 300 600 600]{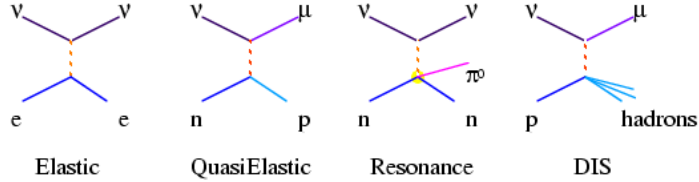}}
\caption{Examples of the four types of neutrino interactions which
  appear throughout this discussion and are defined in
  sec.~\ref{sub:interacts}.  The first two diagrams show an NC and CC
  interaction, respectively.}
\label{interacttypes}
\end{figure}

Neutrinos are the only Standard Model fermions to interact strictly
via the weak interaction. This proceeds through two types of boson
exchange.  Exchange of the $Z^0$ is called the neutral current (NC)
interaction. Exchange of the $W^\pm$ is called the charged
current (CC) interaction. When a $W$ is emitted, charge conservation
at the vertex requires that a charged lepton exits the interaction.
We know the family of an incoming neutrino by the charged partner
which exits the CC interaction. For example, a scattered electron tags
a $\nu_e$ interaction, a $\mu$ tags a $\nu_\mu$ interaction, and a
$\tau$ tags a $\nu_\tau$ interaction.  The neutrino always emits the
$W^+$ and the antineutrino always emits the $W^-$ in the CC
interaction.  In order to conserve charge at the lower vertex, the CC
interaction is flavor-changing for target quarks.  For example, in a
neutrino interaction, if a neutron, $n$, absorbs a $W^+$, a proton,
$p$, will exit the interaction.  The $W$ has converted a $d$ quark to a
$u$ quark. The first two diagrams shown on Fig.~\ref{interacttypes}
illustrate a NC and a CC interaction, respectively.

In 1989, measurements of the $Z^0$ width at LEP\cite{LEP3} and
SLD\cite{SLD3} determined that there are only three families of
light-mass weakly-interacting neutrinos, although we will explore this
question in more depth in section 3 of these lectures.  These are the
$\nu_e$, the $\nu_\mu$, and the $\nu_\tau$.  The interactions of the
$\nu_e$ and $\nu_\mu$ have been shown to be consistent with the
Standard Model weak interaction. Until recently, there has only been
indirect evidence for the $\nu_\tau$ through the decay of the $\tau$
meson. In July 2000, however, the DoNuT Experiment (E872) at Fermilab
presented direct evidence for $\nu_\tau$ interactions\cite{DoNuT}.

Within the Standard Model, neutrinos are massless.  This assumption is
consistent with direct experimental observation.  It is also an
outcome of the feature of ``handedness'' associated with neutrinos.
To understand handedness, it is simplest to begin by discussing
``helicity,'' since for massless particles helicity and handedness are
identical.

For a spin 1/2 Dirac particle, helicity is the projection of a particle's 
spin (${\bf \Sigma}$) along its direction of
motion ${\bf {\hat p}}$, with operator ${\bf \Sigma \cdot {\hat p} }$.  
Helicity has
two possible states: spin aligned opposite the direction of motion
(negative, or ``left helicity'') and spin aligned along the direction
of motion (positive or ``right helicity'').   
If a particle is massive, then the sign of the helicity of the
particle will be frame dependent.
When one boosts to a frame where one is moving 
faster than the particle, the sign of the momentum will change but the
spin will not, and therefore the helicity
will flip.    For massless particles, which must travel at the speed of light, 
one cannot boost to a frame where helicity changes sign.

Handedness (or chirality) is the Lorentz invariant ({\it i.e.},
frame-independent) analogue of helicity for both massless and massive
Dirac particles.  There are two states: ``left handed'' (LH) and
``right handed'' (RH).  For the case of massless particles, including
Standard Model neutrinos, helicity and handedness are identical.  A
massless fermion is either purely LH or RH, and, in principle, can
appear in either state.  Massive particles have both RH and LH
components.  A helicity eigenstate for a massive particle is a
combination of handedness states.  It is only in the high energy
limit, where particles are effectively massless, that handedness and
helicity coincide for massive fermions.  Nevertheless, people tend to
use the terms ``helicity'' and ``handedness'' interchangeably.  Unlike
the electromagnetic and strong interactions, the weak interaction
involving neutrinos has a definite preferred handedness.  

In 1956, it was shown that neutrinos are LH and outgoing antineutrinos
are RH \cite{RH}.  This effect is called ``parity violation.''  If
neutrinos respected parity, then an equal number of LH and RH
neutrinos should have been produced in the 1956 experiment.  The fact
that all neutrinos are LH and all antineutrinos are RH means that,
unlike all of the other fermions in the Standard Model, parity appears
to be maximally violated for this particle.  This is clearly very
strange.

We need a method to enforce parity violation within the weak 
interaction theory.   To this end, consider a fermion wavefunction,
$\psi$, broken up into its LH and RH components:
\begin{equation}
\psi = \psi_L + \psi_R.
\label{firsteqinlrdiscussion}
\end{equation}
We can introduce a projection operator which selects out each component:
\begin{equation}
\gamma^5\psi_{L,R}=\mp\psi_{L,R}.
\end{equation}
To force the correct handedness in calculations involving the weak
interaction, we can require a 
factor of $(1-\gamma^5)/2$ at every weak vertex involving a neutrino.  
As a result of this factor, which corresponds to the LH projection operator,
we often say the charged weak interaction ($W$ exchange) is ``left handed.'' 

Note that by approaching the problem this way, 
RH neutrinos (and LH antineutrinos) 
could in principle exist but be undetected because they do not interact.
They will not interact via the electromagnetic interactions
because they are neutral, or via the strong interaction because they are
leptons.  RH Dirac neutrinos 
do not couple to the Standard Model $W$, because this interaction  
is ``left handed,'' as discussed above.  
Because they are non-interacting, they 
are called ``sterile neutrinos.''    By definition, the Standard Model
has no RH neutrino.

With no RH partner, the neutrino can have no Dirac mass term in the Lagrangian.
To see this, note that the free-particle Lagrangian
for a massive, spin 1/2 particle is
\begin{equation}
{\cal L} = i \overline{\psi}\gamma_\mu \partial^\mu \psi - {m \overline{\psi}\psi}, 
\end{equation}
However, $\overline{\psi}\psi$  can be rewritten using
\begin{eqnarray}
\psi_{L,R} & = & 1/2 (1 \mp \gamma^5)\psi, \\
\bar \psi_{L,R} & = & 1/2 \bar \psi (1 \pm \gamma^5),
\end{eqnarray}  
giving 
\begin{equation}
\bar \psi \psi = \bar \psi \bigg [  {{1+\gamma^5}\over{2}} + 
{{1-\gamma^5}\over{2}}\bigg ] \bigg [  {{1+\gamma^5}\over{2}} + 
{{1-\gamma^5}\over{2}}\bigg ] \psi
=\bar \psi_L \psi_R + \bar \psi_R
\psi_L.
\label{lasteqinlrdiscussion}
\end{equation}
In other words, an $m \bar \psi \psi$ (``mass'') term in a Lagrangian
mixes RH and LH states of the fermion.  If the fermions have only one
handedness (like $\nu$s), then the Dirac mass term will automatically
vanish.  In the Standard Model, there is no Dirac mass term for
neutrinos.

\subsection{Neutrino Interactions}
\label{sub:interacts}

\begin{figure} [t]
\vspace{5mm}
\centering
\scalebox{0.3}{\includegraphics[angle=90.]{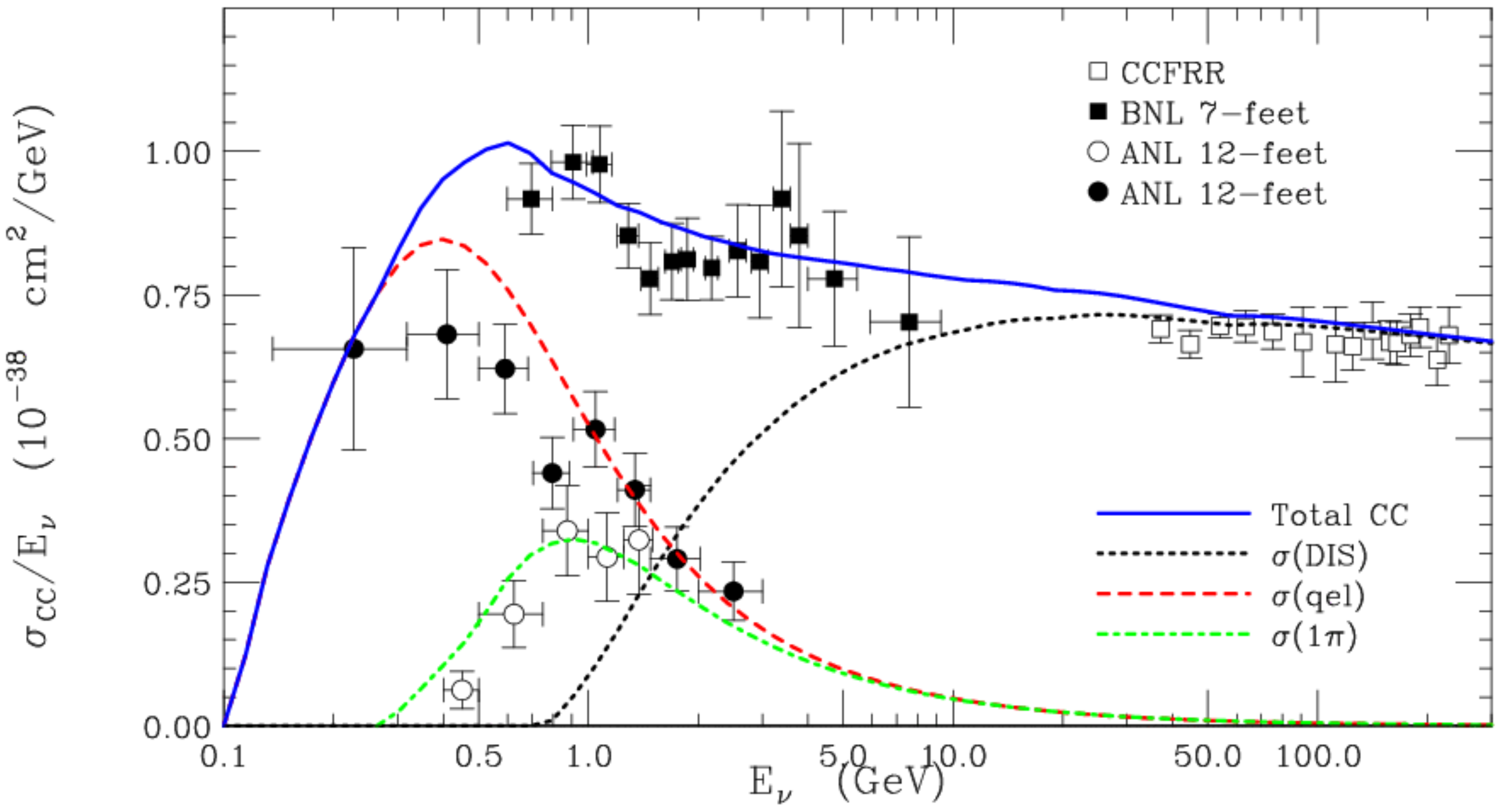}}
\caption{Current status of $\nu_\mu$ CC cross section measurements in
  the 1 to 100 GeV range.  This plot shows $\sigma/E$, thus removing
  the linear energy dependence at high energies.  Note the low energy
  cut-off due to the muon mass suppression. Components of the total
  cross section are indicated by the curves.\cite{Lipari}}
\label{fig:xsec_summary}
\end{figure}

Neutrino interactions in the Standard Model come in four basic types.
Fig.~\ref{interacttypes} shows examples of the four interactions.  In
{\it Elastic} scattering, ``what goes is what comes out,'' just like
two billiard balls colliding. An example is a NC interaction where the
target is does not go into an excited state or break up, {\it e.g.,}
$\nu_e + n \rightarrow \nu_e + n$.  A more complicated example is
electron-neutrino scattering from electrons, where the $W$ exchange
yields a final state which is indistinguishable from the $Z$ exchange
on an event-by-event basis, so this is categorized as an elastic
scatter.  {\it Quasi-elastic} scattering is, generally, the CC analogue
to elastic scattering.  Exchange of the $W$ causes the incoming lepton
and the target to change flavors, but the target does not go into an
excited state or break apart.  An example is $\nu_\mu + n \rightarrow
\mu + p$.  {\it Single pion} production may be caused by either NC or
CC interactions.  In resonant single pion production, the target
becomes a $\Delta$ which decays to emit a pion.  In coherent
scattering, there is little momentum exchange with the nucleon and a
single pion is produced diffractively in the forward direction.  The
case of NC single $\pi^0$ production is particularly important,
because this forms a background in many neutrino oscillation searches.
Finally, {\it DIS}, or Deep Inelastic Scattering, is the case where
there is large 4-momentum exchange, breaking the nucleon apart.  One
can have NC or CC deep inelastic scattering.

Fig.~\ref{fig:xsec_summary} summarizes the low energy behavior of
$\sigma/E$ for CC events (solid line), as predicted by the NUANCE
neutrino event generator \cite{nuance}.  The quasi-elastic, single
pion and deep inelastic contributions are indicated by the broken
curves.  The data indicate the state of the art for neutrino cross
section measurements.  One can see that if precision neutrino studies
are to be pursued in the MeV to few GeV range, that more accurate
measurements are essential.  The MiniBooNE \cite{MiniBooNE}, SciBooNE
\cite{SciBooNE}, and MINERvA \cite{MINERvA} experiments are expected
to improve the situation in the near future.

Above a few GeV, the total neutrino cross section rises linearly with
energy.  The total cross section is the sum of many partial cross
sections: quasi-elastic + single pion + two pions + three pions +
etc. As the energy increases, each of these cross sections
sequentially ``turns on'' and then becomes constant with energy.  Thus
the sum, which is the total cross section, increases continuously and
linearly with $E$.

Nevertheless, even at high energies, this interaction is called
``weak'' for good reason.  The total cross section for most neutrino
scattering experiments is small.  For 100 GeV $\nu_\mu$ interactions
with electrons, the cross section is $\sim 10^{-40}$ cm$^2$.  For 100
GeV $\nu_\mu$ interactions with nucleons, the cross section is $\sim
10^{-36}$ cm$^2$.  This is many orders of magnitude less than the
strong interaction.  For example, for $pp$ scattering, the cross
section is $\sim 10^{-25}$ cm$^2$.  The result is that a 100 GeV
neutrino will have a mean free path in iron of $3 \times 10^9$ meters.
Thus most neutrinos which hit the Earth travel through without
interacting.  It is only at ultra-high energies that the Earth becomes
opaque to neutrinos, as discussed in sec.~\ref{subsub:ultra}.

\begin{figure}[t]
\vspace{-1.in}
\hspace{2.in}
\scalebox{0.4}{\includegraphics{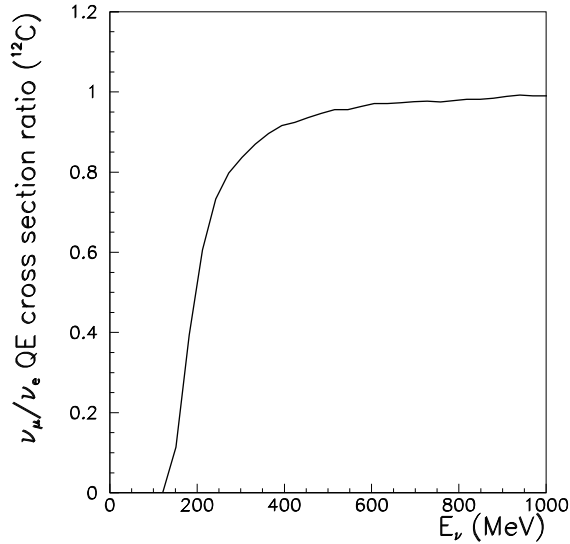}}
\caption{The ratio of the $\nu_e$ to $\nu_\mu$ CC cross sections as
a function of neutrino energy, showing the suppression due
to the lepton mass. }
\label{fig:nuenumu}
\end{figure}

In principle, the interactions of the $\nu_e$, $\nu_\mu$, and
$\nu_\tau$ should be identical (``universal'').  In practice, the mass
differences of the outgoing leptons lead to considerable differences
in the behavior of the cross sections.  In the CC interaction, you
must have enough CM energy to actually produce the outgoing charged
lepton.  Just above mass threshold, there is very little phase space
for producing the lepton, and so production will be highly suppressed.
The cross section increases in a non-linear manner until well above
threshold.  Consider, for example, a comparison of the $\nu_e$ and
$\nu_\mu$ CC quasielastic cross section on carbon, shown in
Fig,~\ref{fig:nuenumu}.  At very low energy the CC $\nu_\mu$ cross
section is zero, while the $\nu_e$ cross section is non-zero, because
the 105 MeV muon cannot be produced.  The ratio approaches one at
about 1 GeV.  A similar effect occurs for the $\nu_\tau$ CC
interaction cross sections.  The mass of the $\tau$ is 1.8 GeV,
resulting in a cross section which is zero below 3.5 GeV and
suppressed relative to the total $\nu_\mu$ CC scattering cross section
for $\nu_\tau$ beam energies beyond 100 GeV.  At 100 GeV, which
corresponds to a center-of-mass energy of $\sqrt{2ME}\approx 14$ GeV,
there is still a 25\% reduction in the total CC $\nu_\tau$ interaction
rate compared to $\nu_\mu$ due to leptonic mass suppression.

For low energy neutrino sources, the CC interaction may also be
suppressed due to conversion of the nucleon at the lower vertex.  For
example, the CC interaction commonly called ``inverse beta decay''
(IBD), $\bar \nu_e p \rightarrow e^+ n$, which is crucial to reactor
neutrino experiments, has a threshold of 1.084 MeV, driven by the mass
difference between the proton and the neutron plus the mass of the
positron.  In the case of bound nuclei, the energy transferred in a CC
interaction must overcome the binding energy difference between the
incoming and outgoing nucleus as well as the mass suppression due to
the charged lepton.  This leads to nuclear-dependent thresholds for
the CC interaction.  For example:
\begin{eqnarray*}
{\rm ^{35}Cl (75.8\%) \rightarrow ^{35}Ar}: &~~~ 5.967~{\rm MeV};\\
{\rm ^{37}Cl (24.2\%) \rightarrow ^{37}Ar}: &~~~ 0.813~{\rm MeV};\\ 
{\rm ^{69}Ga (60.1\%) \rightarrow ^{69}Ge}: &~~~ 2.227~{\rm MeV};\\
{\rm ^{71}Ga (39.9\%) \rightarrow ^{71}Ge}: &~~~ 0.232~{\rm MeV}.  
\end{eqnarray*}
are the thresholds for isotopes which have
been used as targets in past solar neutrino ($\nu_e$) detectors.
\cite{chlorine, sage, gallex}

In discussing neutrino scattering at higher energies, several
kinematic quantities are used to describe events.  The squared center
of mass energy is represented by the Mandelstam variable, $s$. The
energy transferred by the boson is $\nu$, and $y = \nu/E_\nu$ is the
fractional energy transfer, or ``inelasticity.''  The distribution of
events as a function of $y$ depends on the helicity.  For neutrino
scattering from quarks, the $y$-dependence is flat, but for
antineutrinos, the differential cross section is peaked at low $y$.
The variable $Q^2$ is the negative squared four-momentum transfer.
Deep inelastic scattering begins to occur at $Q^2 \sim 1$ GeV$^2$. If
$x$ is the fractional momentum carried by a struck quark in a deep
inelastic scatter, then $x=Q^2/2M\nu$, where $M$ is the target mass.
Elastic and quasieleastic scattering occur at $x=1$, hence $Q^2 = 2 M
\nu \approx sxy$, valid for large $s$.

\subsection{Sources of Neutrinos}
\label{sub:sources}

With such a
small interaction probability, it is clear that 
intense neutrino sources are needed 
to have high statistics
in a neutrino experiment.   
The primary sources of neutrino for interactions observed on Earth are
the Sun, cosmic-ray interactions, reactors, and accelerator beams.

\begin{figure}[t]
\centering
\scalebox{0.4}{\includegraphics{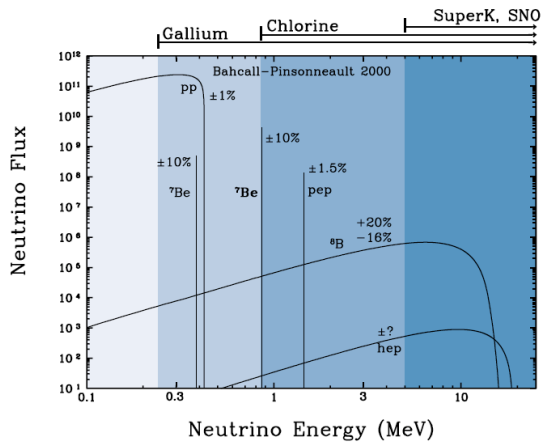}}
\caption{The flux predicted by the Standard Solar Model.\cite{http://www.sns.ias.edu/tildejnb}  The sensitivity 
of past solar neutrino detectors varies due to CC threshold in the target
material.\cite{chlorine, sage, gallex}   The thresholds for various experiments is shown at the
top of the plot.}
\label{fig:soflux}
\end{figure}

\begin{table}[t]
\centering
{
\begin{tabular}{c|c}
\hline
Common Terminology & Reaction \\
\hline
``$pp$ neutrinos'' & $p+p \rightarrow ^{2}{\rm H}+ e^- +\nu_e$ \\
``$pep$ neutrinos'' & $p+e^-+p \rightarrow ^{2}{\rm H}+ \nu_e$ \\
``$^7{\rm Be}$ neutrinos'' & $^7{\rm Be} + e^- \rightarrow ^7{\rm Li} + \nu_e$ \\
``$^8{\rm B}$ neutrinos'' & $^8{\rm B} \rightarrow ^8{\rm Be}^* + e^+ + \nu_e$ \\
``$hep$ neutrinos'' &~~ $^3{\rm He} + p \rightarrow ^4{\rm He} + e^+ +\nu_e$~~ \\
\hline
\end{tabular}}
\caption{
Reactions from the Sun producing neutrinos.
}\label{tab:cheatsheet}
\end{table}

At present, there are two intense sources in the few MeV range that
allow for low energy neutrino interaction studies.  First, the
interactions in the Sun produce a pure $\nu_e$ flux, as listed in
tab.~\ref{tab:cheatsheet}.  The energy distribution of neutrinos
produced by these reactions is shown in Fig.~\ref{fig:soflux}.  The
sensitivity of various solar neutrino experiments, due to the CC
threshold, is shown at the top of the figure.  There is no observable
antineutrino content. The best limit on the solar neutrino $\bar
\nu_e/\nu_e$ ratio for $E_\nu>8.3$ MeV is $2.8\times 10^{-4}$ at 90\%
CL \cite{hep-ex/0310047}.  The second source is from reactors.  In
contrast to the Sun, reactors produce a nearly pure $\bar \nu_e$ flux.
The energy peaks from $\sim 3$ to 7 MeV.  Neutrinos from $\beta$ decay
of accelerated isotopes could, in principle, represent a third intense
source of neutrinos in the MeV range (or higher), once the technical
issues involved in designing such an accelerator are overcome.  Such
a``beta beam'' would produce a very pure $\nu_e$ or $\bar \nu_e$ beam,
depending on the accelerated isotope \cite{nufactworkgroup}.

At present, higher energy experiments use neutrinos produced at
accelerators and in the atmosphere.  In both cases, neutrinos are
dominantly produced via meson decays.  In the atmospheric case, cosmic
rays hit atmospheric nuclei producing a shower of mesons which may
decay to neutrinos along their path through the atmosphere to Earth.
In a conventional neutrino beam, protons impinge on a target, usually
beryllium or carbon, producing secondary mesons.  In many experiments,
the charged mesons are focussed (bent) toward the direction of the
experiment with a magnetic device called a horn.  These devices are
sign-selecting -- they will focus one charge-sign and defocus the
other -- and so produce beams which are dominantly neutrinos or
antineutrinos depending on the sign-selection.  The beamline will have
a long secondary meson decay region, which may be air or vaccuum.
This is followed by a beam dump and an extended region of dirt or
shielding to remove all particles except neutrinos. There are 
excellent reviews of methods of making accelerator-produced neutrino 
beams\cite{kopp}.  Tab.~\ref{tab:accelcheatsheet} summarizes the
common sources of neutrino production in the atmosphere and
conventional accelerator based beams.

\begin{table}[t]
\centering
{
\begin{tabular}{c|c}
\hline
2-body pion decay & $\pi^+ \rightarrow \mu^+ \nu_\mu$, $\pi^- \rightarrow \mu^- \bar \nu_\mu$ \\
2-body kaon decay & $K^+ \rightarrow \mu^+ \nu_\mu$, $K^- \rightarrow \mu^- \bar \nu_\mu$ \\
muon decay & $\mu^+ \rightarrow e^+ \bar \nu_\mu \nu_e$, $\mu^- \rightarrow e^- \nu_\mu \bar \nu_e$ \\
$K_{e3}$ decay & $K^+ \rightarrow \pi^0 e^+ \nu_e$, $K^- \rightarrow \pi^0 e^- \bar \nu_e$, $K^0 \rightarrow \pi^- e^+ \nu_e$, $K^0 \rightarrow \pi^+ e^- \bar \nu_e$ \\
\hline
\end{tabular}}
\caption{
Common sources of neutrinos in atmospheric and accelerator experiments.
}\label{tab:accelcheatsheet}
\end{table}

Many atmospheric and accelerator-based neutrino experiments are
designed to study 100 MeV to 10 GeV neutrinos.  The atmospheric
neutrino flux drops as a power-law with energy, and the 1 to 10 GeV
range dominates the event rate.  Accelerator beams can be tuned to a
specific energy range and, using present facilities, can extend to as
high as 500 GeV.  From the viewpoint of sheer statistics, one should
use the highest energy neutrino beam which is practical for the
physics to be addressed, since the cross section rises linearly with
energy.  However, lower neutrino energy beams, from $\sim 1$ to 10
GeV, are typically used for oscillation experiments.  In these
experiments, having a cleanly identified lepton in a low multiplicity
event trumps sheer rate, and so $\sim 1$ GeV beams are selected to
assure that CCQE and single pion events dominate the interactions. 

Both atmospheric and accelerator based neutrino sources are dominantly
$\nu_\mu$-flavor.  The main source of these neutrinos is pion decay.
To understand why pions preferentially decay to produce $\nu_\mu$
rather than $\nu_e$, consider the case of pion decay to a lepton and
an antineutrino: $\pi^- \rightarrow \ell^- \bar \nu_\ell$.  The pion
has spin zero and so the spins of the outgoing leptons from the decay
must be opposite from angular momentum conservation.  In the center of
mass of the pion, this implies that both the antineutrino and the
charged lepton have spin projected along the direction of motion
(``right'' or ``positive'' helicity).  However, this is a weak decay,
where the $W$ only couples to the RH antineutrino and the LH component
of the charged particle.  The amplitude for the LH component to have
right-helicity is proportional to $m/E$.  Thus it is very small for an
electron compared to the muon, producing a significant suppression for
decays to electrons.  Calculating the expected branching ratios:
\begin{eqnarray}
R_{theory}& = {{\Gamma (\pi^\pm \rightarrow e^\pm \nu_e)}\over
{\Gamma (\pi^\pm \rightarrow \mu^\pm \nu_\mu)}}\\
 & =
{\bigl( {{m_e}\over{m_\mu}} \bigr)}^2 
{\bigl( {{{{m_\pi}^2-{m_e}^2}} \over {{{m_\pi}^2-{m_\mu}^2}}} \bigr)}^2 \\
& =1.23\times 10^{-4}; 
\end{eqnarray}
This compares well to the data:\cite{PDG} $R_{exp}=(1.230 \pm
0.004)\times10^{-4}$.   

The above discussion assumed the neutrino was massless.  If the
neutrino is massive, then it too can be produced with wrong helicity
with an amplitude proportional to $m_\nu/E$ and thus a probability
proportional to $(m_\nu/E)^2$.  As discussed in sec.~\ref{sub:directmass},
below, neutrino mass is limited to be very small ($\sim$ eV) and thus the
rate of wrong-helicity neutrino production is too low a level for any
chance of observation in the near future.

\begin{figure}[tp]
\centering
\scalebox{0.4}{\includegraphics{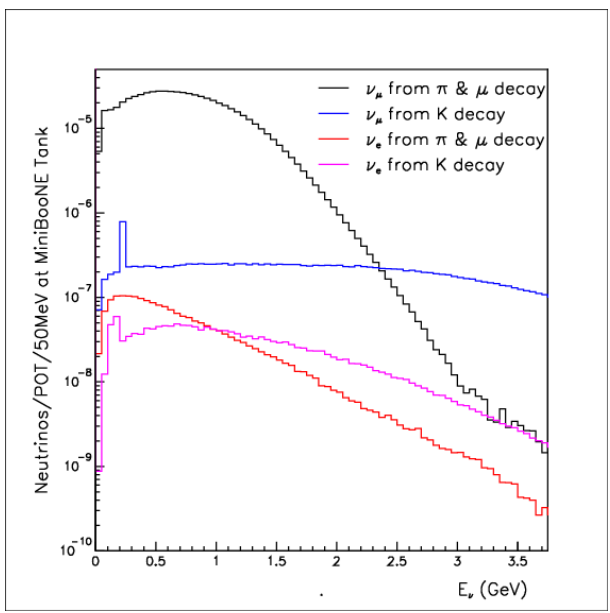}}
\caption{The contributions from pion and kaon production to the total predicted
$\nu_\mu$ flux in the MiniBooNE experiment.  The spikes at low energy 
in the K-produced fluxes are due to decays of stopped kaons in the beam dump.\cite{MiniBooNE}}
\label{fig:minipik}
\end{figure}

Depending on the energy, there may also be significant neutrino production
from kaon decays.   The charged kaon preferentially decays to the $\nu_\mu$ 
for the same reason as the charged pion.  However, for equal 
energy mesons, the kinematic limit for a neutrino from $K^+$ decay is 
much higher than for $\pi^+$ decay: $E_{\nu}^{max, K} = 0.98 E_K$ compared to 
$E_{\nu}^{max, \pi} = 0.43 E_\pi$.   Thus the neutrinos from kaon decays 
can be isolated by studying the highest energy component of a beam.
Fig.~\ref{fig:minipik} shows the contributions of pion and kaon 
decays to the $\nu_\mu$ flux in the MiniBooNE experiment, which uses 
an 8 GeV primary proton beam.

\begin{figure}[tp]
\centering
\scalebox{0.5}{\includegraphics[clip=T, bb= 0 0 600 400]{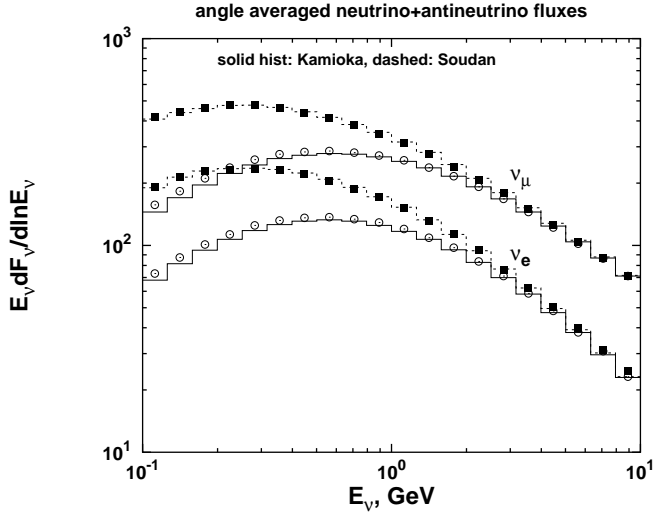}}
\caption{The variation in the atmospheric neutrino flavor content as 
a function of energy for two locations, Japan (solid line, open circles) 
and Minnesota (dashed line, closed squares).   The points are from a 
full 3-dimensional monte carlo of the flux, while the histograms
are from a simpler model.\cite{hep-ex/0403630}}
\label{fig:atmosnuenumu}
\end{figure}

Electron neutrino flavors are produced in these beams through
$K\rightarrow \pi \nu_e e$ (called ``Ke3'') decay and through the
decay of the muons which were produced in the pion decay.  These are
three-body decays which avoid substantial helicity suppression.
Helicity does, however, affect the energy spectrum of the outgoing
decay products.  In an accelerator-based experiment, the level of
electron-flavor content can be regulated, at some level, by the choice
of primary beam energy and the length of the decay region.  A low
primary beam energy will suppress kaon production because of the
relatively high mass of this meson (494 MeV).  A short decay pipe will
suppress $\nu_e$ from $\mu$ decay, which tends to occur downstream,
because it is produced in a multi-step decay chain ($\pi \rightarrow
\mu \rightarrow \nu_e$).  Both of these methods of suppressing $\nu_e$
production also lead to a reduction in the $\nu_\mu$ production rate,
so an experimenter must balance competing goals in the beam design.
In the case of atmospheric neutrinos, the ratio is roughly 2:1 for
$\nu_\mu$:$\nu_e$, though the fraction $\nu_e$'s changes with energy
(see Fig.~\ref{fig:atmosnuenumu}).  The atmospheric flux depends on
the location of the detector because charged particle are bent by the
Earth's magnetic field.  The variation between fluxes at the Kamioka
mine in Japan and Soudan mine in Minnesota are shown in
Fig.~\ref{fig:atmosnuenumu}.

\begin{figure}[t]
\centering
{\includegraphics[clip=T, bb= 0 0 300 300]{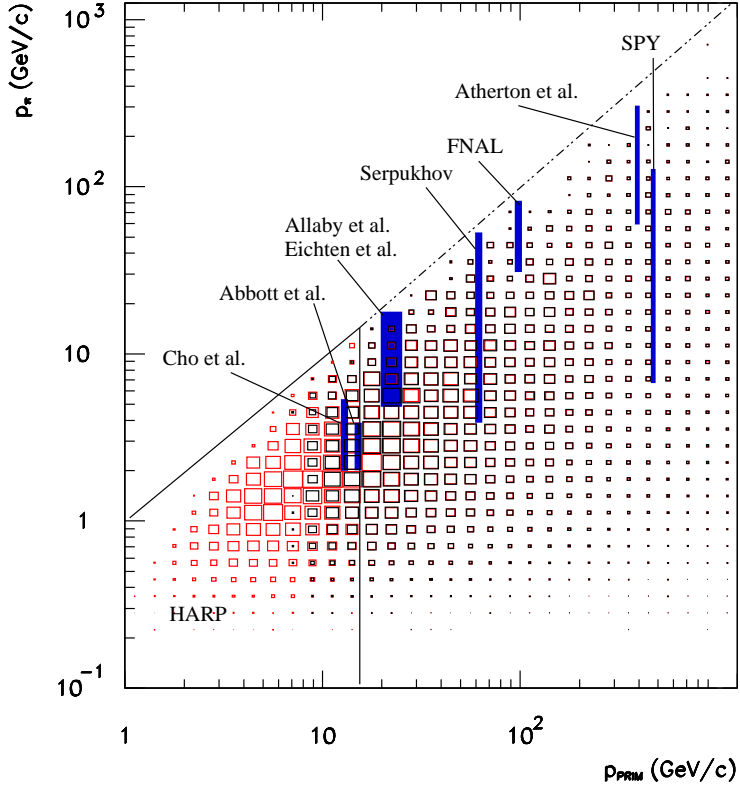}}
\caption{The kinematic range covered by recent experiments measuring
secondary pion and kaon production.\cite{hep-ex/0611266}}
\label{fig:secondary}
\end{figure}

As we move to a precision era in neutrino physics, precise 
``first-principles'' predictions of the flux are becoming 
very important.    For conventional accelerator-based neutrino
beams and for the atmospheric flux, this requires well-measured
cross-sections for production of secondary pions and kaons.  This
has motivated a range of secondary production experiments.  The
kinematic coverage is shown on Fig.~\ref{fig:secondary}.    

The future of high intensity $\nu_\mu$ and $\nu_e$ beams is likely to
lie in beams produced from muon decay.  Because of the potential for
very high intensity, these beams are called ``Neutrino
Factories.''  The concept is very attractive because it produces beams
which are very pure $\nu_\mu$ and $\bar \nu_e$ from $\mu^-$ and vice
versa from $\mu^+$.  Each flavor has no ``wrong sign''
(antineutrino-in-neutrino-beam or neutrino-in-antineutrino beam)
background.  However, neutrino factory designs \cite{nufactworkgroup}
necessarily produce high energy neutrinos, since the muons must be
accelerated to high energies in order to live long enough to be
captured and circulated in an accelerator.  The Neutrino Factory is
seen as a promising first machine for testing ideas for a muon
collider \cite{nufactworkgroup}, and thus has attracted interest beyond the
neutrino community.

A beam enriched in $\nu_\tau$ can be produced by impinging very high
energy protons on a target to produce $D_s$-mesons which are sufficiently 
massive to decay to $\tau~\nu_\tau$.  The $\tau$ lepton is very
massive, at 1.8 GeV, compared to the muon, at 106 MeV, and thus
helicity considerations for the $D_s$ decay strongly favor the $\tau~
\nu_\tau$ mode compared to $\mu~\nu_\mu$, by a ratio of about 10:1.
The $\tau$ then subsequently decays, also producing $\nu_\tau$s.

Unfortunately, because of the short lifetime, it is not possible to
separate $D_s$ mesons from the other mesons prolifically produced by
the primary interaction.  As a result, the beam is dominated by the
$\nu_\mu$s produced by decays of other mesons.  To reduce the
production of $\nu_\mu$, experiments use a ``beam dump'' design where
protons hit a very thick target where pions can be absorbed before
decaying.  The only enriched-$\nu_\tau$ beam created to date was
developed by DoNuT \cite{DoNuT}.  They used an 800 GeV proton on a
beam dump, to produce a ratio of $\nu_e$:$\nu_\mu$:$\nu_\tau$ of about
6:9:1.

\subsection{Typical Neutrino Detectors}
\label{sub:detectors}

Because neutrinos interact so weakly, the options for detectors are
limited to designs which can be constructed on a massive scale.  There
are several general styles in use today: unsegmented scintillator
detectors,  unsegmented Cerenkov detectors, segmented
scintillator-and-iron calorimeters, and segmented scinitillator
trackers.  The most promising future technology is the noble-element
based detector, which is effectively an electronic bubble chamber.
Liquid argon detectors are likely to be the first large-scale working
example of such technology.  There are a few variations on these five
themes, which are considered in later sections in the context of the
measurement.

Unsegmented scintillator detectors are typically used for low energy
antineutrino experiments.  Recent examples include Chooz \cite{chooz},
KamLAND \cite{KamLAND} and LSND \cite{LSND}.  These consist of large
tanks of liquid scintillator surrounded by phototubes.  Usually the
scintillator is oil based, hence the target material is CH$_2$ and its
associated electrons.  Often the tubes are in an pure oil buffer.
This reduce backgrounds from radiation emitted from the glass which
would excite scintillator.  The free protons in the oil provide a
target for the interaction, $\bar \nu_e p \rightarrow e^+ n$, which is
the key for reactor experiments.  The reaction threshold for this
interaction is 1.806 MeV due to the mass differences between the
proton and neutron and the mass of the positron.  The scintillation
light from the $e^+$, as well as light from the Compton scattering of
the 0.511 MeV annihilation photons provide an initial (``prompt'')
signal.  This is followed by $n$ capture to produce deuterium and a
2.2 MeV.  This sequence -- positron followed by neutron capture --
provides a clean signal for the interaction.  Doping the liquid
scintillator with gadolinium substantially increases the neutron capture cross
section as well as the visible energy produced in the form of 
gammas upon neutron capture.

Unsegmented scintillator detectors are now being introduced for low
energy solar neutrino measurements at Borexino \cite{Borexino},
KamLAND \cite{kamsolar} and SNO+ \cite{sno+}.  These provide
energy information on an event-by-event basis, unlike most past solar neutrino experiments, 
such as Homestake \cite{chlorine}, SAGE
\cite{sage} and GallEx \cite{gallex},  which intergrated
over time and energy.  However, these are very
difficult experiments to perform because a neutron is not produced and so the
scattering does not produce a two-fold coincidence, but only a prompt
flash of light.

\begin{figure}[tp]
\centering
\scalebox{0.3}{\includegraphics[clip=T]{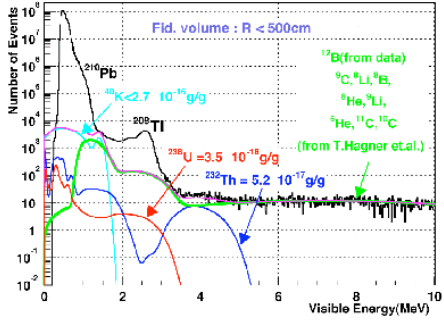}}~~~~~~~~
\caption{Energy distribution and sources of singles events in KamLAND as a function of visible 
  energy.\cite{karsten}.}
\label{fig:UTh}
\end{figure}

Environmental backgrounds are by far the most important issue in low
energy experiments.  These fall into two categories: naturally
occurring radioactivity and muon-induced backgrounds.  To get a sense
for what is expected, Fig.~\ref{fig:UTh} shows the visible energy
distribution of singles events from the KamLAND experiment with the
sources of environmental background identified.  The naturally
occurring radioactive contaminants mainly populate the low energy
range of Fig.~\ref{fig:UTh}, with isotopes from the U and Th chain
extending to the highest energies.  These isotopes must be kept under
control by maintaining very high standards of cleanliness.  The second
source of environmental background, the $\beta$-decays of isotopes
produced by cosmic ray muons.  These dominate the background for
$E_{vible}>4$ MeV (see Fig.~\ref{fig:UTh}).   These can only be eliminated
by shielding the detector from cosmic rays.    As a result, we 
must build deep underground laboratories with many thousands
of meters-water-equivalent (``mwe'') of rock shielding.

In these scintillator detectors, the CC interaction with the carbon in
the oil (which produces either nitrogen or boron depending on whether
the scatterer is a neutrino or antineutrino) has a significantly
higher energy threshold than scattering from free protons.  $\nu_e + C
\rightarrow e^- + N$ has a threshold energy of 13.369 MeV, which
arises from the carbon-nitrogen mass difference (plus the mass of the
electron).  In the case of both reactor and solar neutrinos, the flux
cuts off below this energy threshold.

Existing unsegmented Cerenkov detectors include MiniBooNE
\cite{MBdet}, Super K \cite{SuperK}, and AMANDA \cite{AMANDA}.  These
detectors make use of a target which is a large volume of a clear
medium (undoped oil, water and ice, respectively) surrounded by or
interspersed with phototubes.  Undoped oil has the advantages of a
larger refractive index, leading to larger Cerenkov opening angle, and
of not requiring a purification system to remove living organisms.
Water is the only affordable medium once a detector is larger than a
few ktons.  For ultra-high energy neutrino experiments, a vast natural
target is needed.  Sea water \cite{Antares} and ice \cite{AMANDA} have
been used.  Ice is, to date, more successful because it does not
suffer from backgrounds from bioluminescence.

\begin{figure}[tp]
\centering
\scalebox{0.15}{\includegraphics{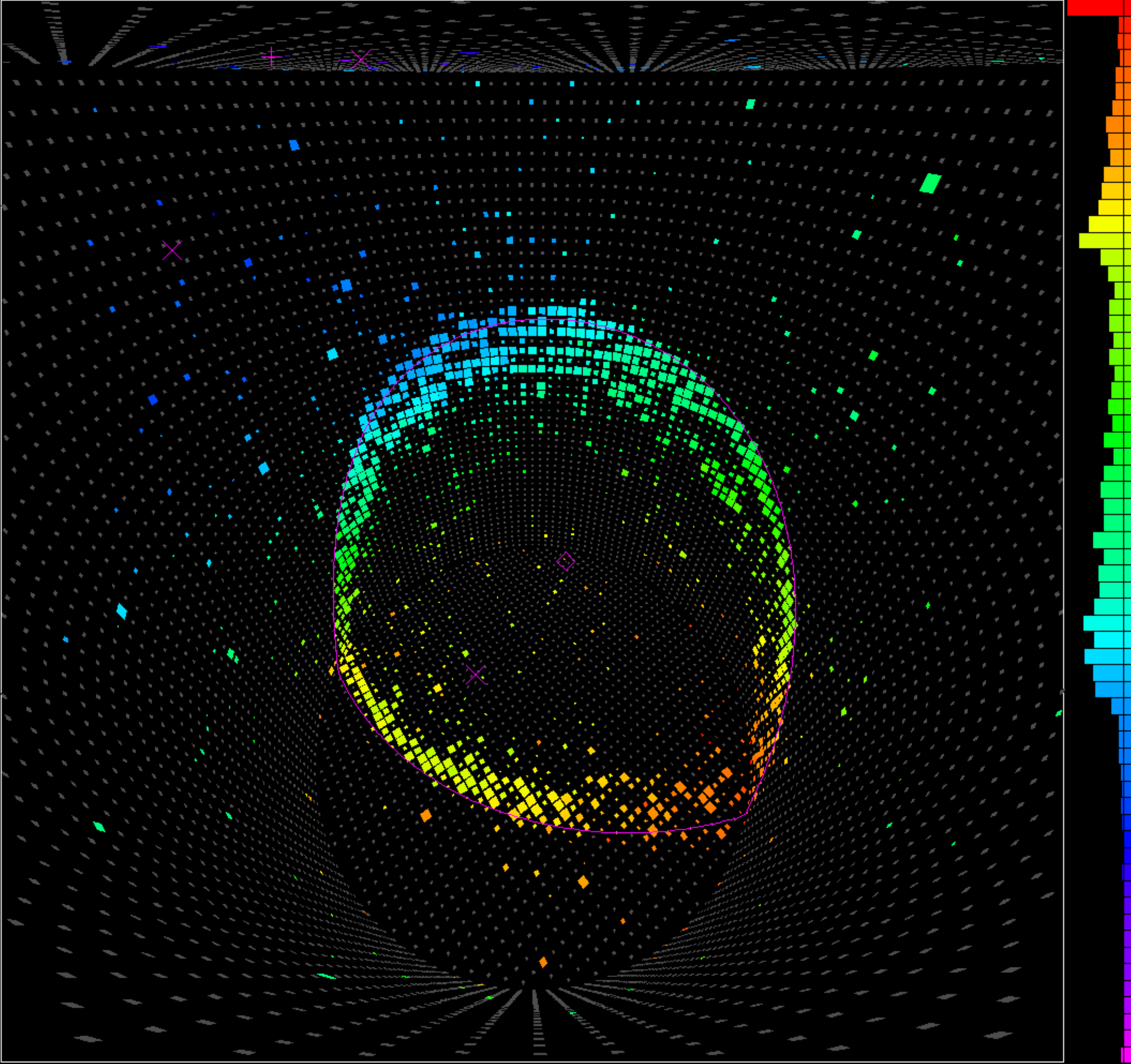}}~
\scalebox{0.15}{\includegraphics{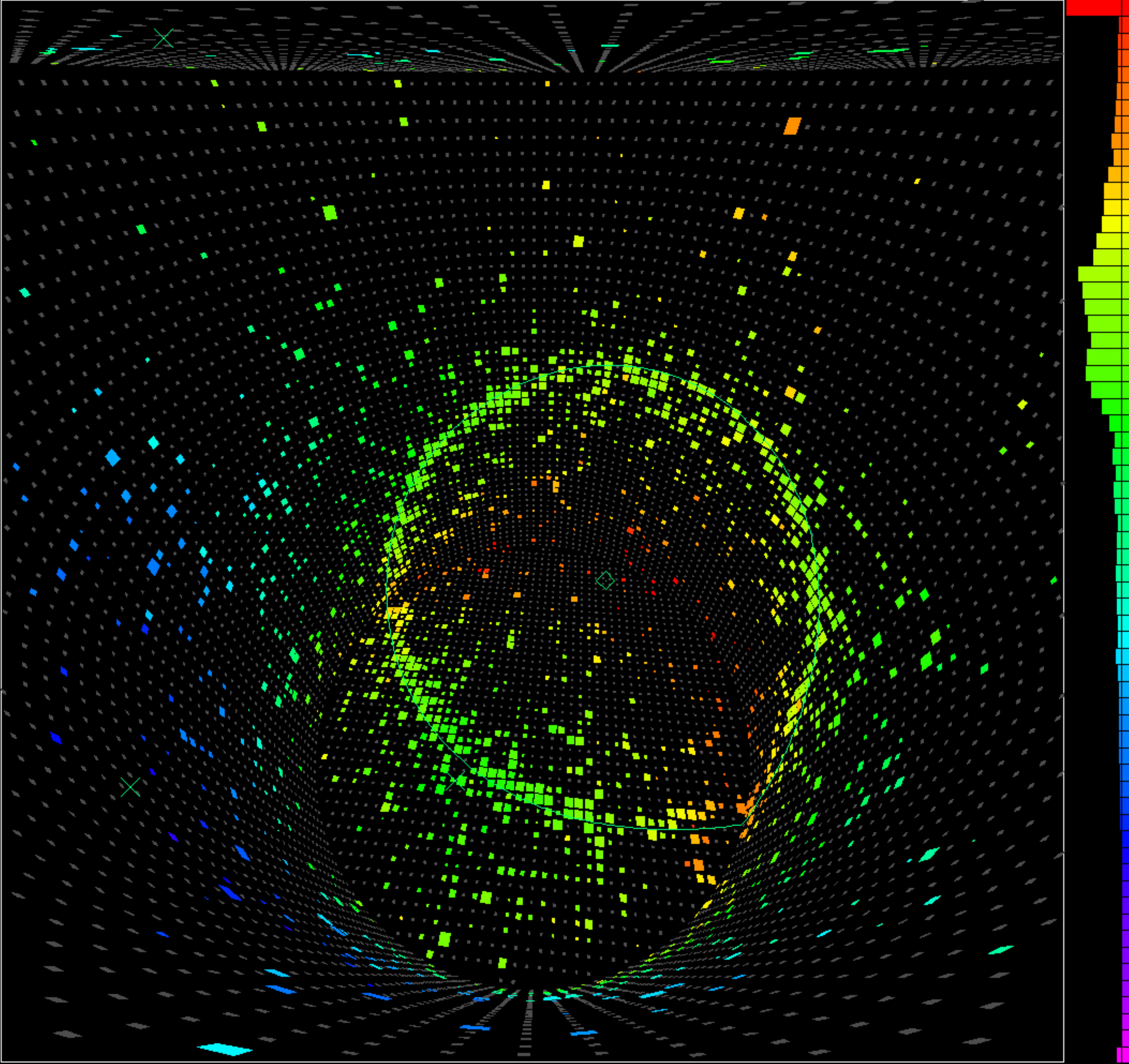}}
\caption{An example of a muon ring (left) and electron ring (right) in the 
Super K Cerenkov detector.\cite{SuperK}}
\label{fig:skring}
\end{figure}

In most cases of these detectors, the tubes surround the medium and
the projected image of the Cerenkov ring is used for particle
identification.  To understand how this works, first consider the case
of a perfect, short track.  This will project a ring with a sharp
inner and outer edge onto the phototubes.  Next consider an electron
produced in a $\nu_e$ CC quasielastic interaction.  Because the
electron is low mass, it will multiple scatter and easily
bremsstrahlung, smearing the light projected on the tubes and
producing a ``fuzzy'' ring.  A muon produced by a CC quasileastic
$\nu_\mu$ interaction is heavier and thus will produce a sharper outer
edge to the ring.  For the same visible energy, the track will also
extend farther, filling the interior of the ring, and perhaps exit
the tank.  Fig.~\ref{fig:skring} compares an electron and muon ring
observed in the Super K detector.  If the muon stops within the tank
and subsequently decays, the resulting ``michel electron'' provides an
added tag for particle identification.  In the case of the $\mu^-$,
18\% will capture in water, and thus have no michel electron tag,
while only 8\% will capture in oil.

Scintillator and iron calorimeters provide affordable detection for
$\nu_\mu$ interactions in the range of $\sim$1 GeV and higher.  Recent
examples include the MINOS \cite{MINOS} and NuTeV \cite{NuTeV}
experiments.  In these detectors, the iron provides the target, while
the scintillator provides information on energy deposition per unit
length.  This allows separation between the hadronic shower, which
occurs in both NC and CC events, and the minimum ionizing track of an
outgoing muon, which occurs in CC events.  Transverse information can
be obtained if segmented scintillator strips are used, or if drift
chambers are interspersed.  The light from scintillator strips is
transported to tubes by mirrored wave-length-shifting fibers.
Transverse information improves separation of electromagnetic and
hadronic showers.  The iron can be magnetized to allow separation of
neutrino and antineutrino events based on the charge of the outgoing
lepton.

In all three of the above detector designs, it is difficult to reconstruct
multi-particle events.  Tracking is not an option for an unsegmented
scintillator detector.   Cerenkov detectors can typically resolve 
two tracks per event.   Segmented calorimeters reduce multiple 
hadrons to a shower, obscuring any track-by-track information other than
from muons.   

\begin{figure}[t]
\centering
\scalebox{0.4}{\includegraphics{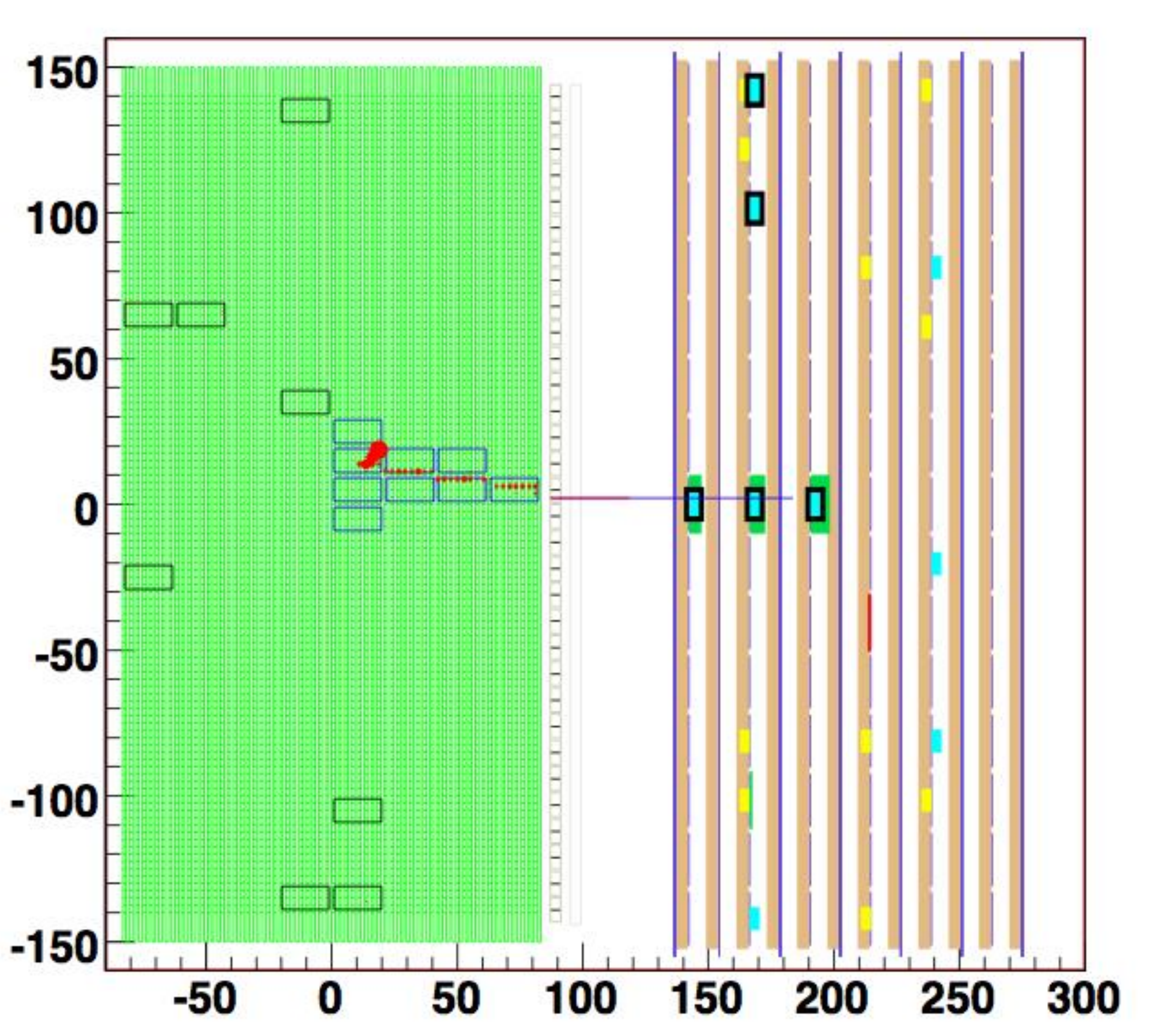}}
\caption{A CCQE($\nu_\mu + n \rightarrow \mu + p$) event observed
  in the SciBooNE detector.  The long, minimum-ionizing red track is identified as the muon,
  the short, heavily-ionizing red track is identifed as the proton.\cite{SciBooNE}}
\label{fig:sibarccqe}
\end{figure}

To address the problem of track reconstruction in low energy ($\lesssim 1$
GeV), low multiplicity events, there has been a move toward
all-scintillator tracking detectors.  This began with the SciBar
detector in K2K \cite{scibar}.  This detector used scintillator strips,
as in MINOS, but
without interspersing iron. As a result, low
energy (few MeV) tracks were clearly observable and
quasielastic and single pion events could be fully reconstructed.
SciBar has since been incorporated into the SciBooNE experiment at
Fermilab \cite{SciBooNE}.  
The CCQE event in SciBooNE, shown in Fig.~\ref{fig:sibarccqe}, makes clear the 
benefits of fine segmentation.  The position of the vertex
and the short track from the proton are well-resolved in the 
SciBar detector (green region).
The technology has been taken further by
the MINERvA experiment, which has attained 2 mm resolution with their
prototype 
\cite{MINERvAstrip}.  Scibar and MINERvA are relatively small (few
ton) detectors.  The first very large scale application of this
technology will be NOvA, which is a future 15 kton detector
\cite{NOvA}.  This detector will use PVC tubes filled with liquid
scintillator, which is more cost-effective than extruded scintillator
strips for very large detectors.  Their design also loops the
wave-length shifting fiber, so that there are, effectively, two
perfectly mirrored fibers are in each cell.  This elegant solution
increases the collected light by a factor of four, which is necessary
for $\sim 15$ m strips.

The most promising new technology for high resolution track
reconstruction in neutrino physics is the liquid argon TPC.  A TPC, or
time projection chamber, uses drift chambers to track in the $x$ and
$y$ views and drift time to determine the $z$ view.  Liquid argon
(LAr), which provides the massive target for the neutrino interaction,
also scintillates, providing the start for the drift-time measurement.
A key point for future neutrino experiments is the high efficiency for
identifying electron showers (expected to be 80-90\%) with a rejection
factor of 70 for NC $\pi^0$ events.  In particular, these detectors
can differentiate between converted photons and electrons through the
$dE/dx$ in the first few centimeters of the track.  Typical energy
resolution for an electromagnetic shower is $3\%/\sqrt{E}$. 

There is a great deal of activity on development of LAr detectors.
Data have been taken successfully on a 50 liter LArTPC prototype in
the NOMAD neutrino beam at CERN, resulting in reconstruction of
$\sim$100 CC quasielastic events\cite{L50}.  Also, recently, a 600 ton
Icarus module has been commissioned at Gran Sasso \cite{T600}.  A 0.8
ton LAr test detector will begin taking data at Fermilab in
January, 2008 \cite{T962}.  As discussed in sec.~\ref{sub:unexpected}, the
microBooNE experiment is a proposed 100 ton detector which would take
data in 2010 \cite{Bonnieletter}. In principle, these detectors can be
scaled up to tens of ktons, as is discussed in the ``Ash River Proposal''
\cite{Ashletter}.

\section{Neutrinos As We Know Them Now}

The recent discovery of neutrino oscillations requires that we
reconsider the Standard Model Lagrangian of sec.~\ref{sub:SM}.  It
must now incorporate, preferably in a motivated fashion, both neutrino
mass and neutrino mixing.  This represents both a challenge and an
opportunity for the theory, which I will discuss in the following
section.  This section concentrates on the experimental discovery.  It
is interesting to note that while neither neutrino mass nor mixing
were ``needed'' in the Standard Model theory, both are required for
the discovery of neutrino oscillations.  The probability for neutrino
oscillations will be zero unless {\it both} effects are present.

The outcome of the observation of neutrino oscillations is typically
summarized by the statement that ``neutrinos have mass.''  To be
clear: we still have no direct measurement of neutrino mass.  At this
point, we have clear evidence of mass differences between neutrinos
from the observation of neutrino oscillations. A mass difference
between two neutrinos necessarily implies that at least one of the
neutrinos has non-zero mass.  All experimental evidence indicates that
the actual values of the neutrino masses are tiny in comparison to the
masses of the charged fermions. At the end of this section, attempts
at direct measurement of neutrino mass are described.

\subsection{Neutrino Oscillations}
\label{sub:osc}

Recent results on neutrino oscillations provide indisputable evidence
that there is a spectrum of masses for neutrinos.  In this section, I
describe the formalism for neutrino oscillations, and then review
the experimental results which have now been confirmed at the
5$\sigma$ level.  This is covered briefly because these results are
well known and covered extensively elsewhere\cite{APSwhitepaper}.

\subsubsection{The Basic Formalism}

\begin{figure}[t]
\centering
\scalebox{0.3}{\includegraphics{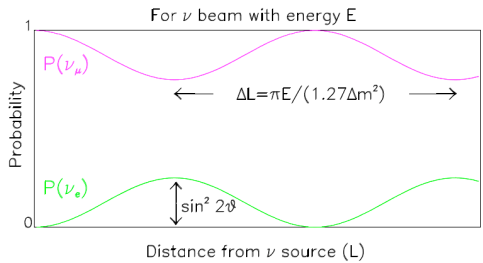}}
\caption{Example of neutrino oscillations as a function of distance from the
source, $L$.   The wavelength depends upon the experimental parameters
$L$ and $E$ (neutrino energy) and the fundamental parameter $\Delta
m^2$.  The amplitude of the oscillation is constrained by the mixing
term, $\sin^2 2\theta$.}
\label{fig:length}
\end{figure}

Neutrino oscillations requires that 
neutrinos have mass, that the difference between the masses
be small, and that the mass eigenstates 
be different from the weak interaction eigenstates.
In this case, the weak eigenstates
can be written as mixtures of the mass eigenstates. For example,
in a simple 2-neutrino model:

\[
\begin{array}{l}
\nu _e=\cos \theta \;\nu _1+\sin \theta \;\nu _2 \\ 
\nu _\mu =-\sin \theta \;\nu _1+\cos \theta \;\nu _2
\end{array}
\]
where $\theta$ is the ``mixing angle.''
In this case, 
a pure flavor (weak) eigenstate born through a weak decay can
oscillate into another flavor as the state propagates in space. This
oscillation is due to the fact that each of the mass eigenstate components
propagates with different frequencies if the masses are different, $\Delta
m^2=\left| m_2^2-m_1^2\right|>0$.  In such a two-component model, 
the oscillation probability for 
$\nu_\mu \rightarrow \nu_e$ oscillations is then given by:
\begin{equation}
{\rm Prob}\left( \nu _\mu \rightarrow \nu _e\right) = \sin ^22\theta \;\sin
^2\left( \frac{1.27\;\Delta m^2\left( {\rm eV}^2\right) \,L\left({\rm km}%
\right) }{E \left({\rm GeV}\right) }\right), 
  \label{prob}
\end{equation}
where
$L$ is the distance from the source, and $E$ is the neutrino energy.
As shown in Fig.~\ref{fig:length}, the oscillation wavelength will
depend upon $L$, $E$, and $\Delta m^2$.  The amplitude will depend
upon $\sin^2 2\theta$.

Neutrino oscillations only occur if the two mass states involved have
sufficiently small $\Delta m^2$ that the neutrino flavor is produced
in a superposition of two mass states.  If the mass splitting is
sufficiently large, a given neutrino flavor would be produced in one
or the other of the two mass eigenstates and interference ({\it i.e.},
oscillations) would not occur.

Most neutrino oscillation analyses consider only two-generation mixing
scenarios, but the more general case includes oscillations among all
three neutrino species.  This can be expressed as:
\[
\left( 
\begin{array}{l}
\nu _e \\ 
\nu _\mu \\ 
\nu _\tau
\end{array}
\right) =\left( 
\begin{array}{lll}
U_{e1} & U_{e2} & U_{e3} \\ 
U_{\mu 1} & U_{\mu 2} & U_{\mu 3} \\ 
U_{\tau 1} & U_{\tau 2} & U_{\tau 3}
\end{array}
\right) \left( 
\begin{array}{l}
\nu _1 \\ 
\nu _2 \\ 
\nu _3
\end{array}
\right). 
\]
This formalism is analogous to the quark sector, where strong 
and weak eigenstates are not identical and the resultant mixing is described 
conventionally by a unitary mixing matrix.    The oscillation
probability is then:
\begin{eqnarray}
{\rm Prob}\left( \nu _\alpha \rightarrow \nu _\beta \right)&
\hspace{-2.in}=\delta_{\alpha \beta }- \nonumber \\
& 4\sum\limits_{j>\,i}U_{\alpha \,i}U_{\beta \,i}^*U^*_{\alpha
\,\,j}U_{\beta \,\,j}\sin ^2\left( \frac{1.27\;\Delta m_{i\,j}^2 \,L }{E }%
\right),  \label{3-gen osc}
\end{eqnarray}
where $\Delta m_{i\,j}^2=m_j^2-m_i^2$, $\alpha$ and $\beta$ are
flavor-state indices $(e, \mu, \tau)$ and $i$ and $j$ are mass-state
indices $(1,2,3)$.

For three neutrino mass states, 
there are three different $\Delta m^2$ parameters, although only two
are independent since the two small $\Delta m^2$ parameters must
sum to the largest.  The neutrino mass states, $\nu_1$, $\nu_2$ and
$\nu_3$ are defined such that the difference between $\nu_1$ and
$\nu_2$ always represents the smallest splitting.  However, the mass
of $\nu_3$ relative to $\nu_1$ and $\nu_2$ is arbitrary and so the sign
of the $\Delta m^2$ parameters which include the third mass state may
be positive or negative.  That is, if $\nu_3 > \nu_1, \nu_2$, then
$\Delta m^2_{23}$ will be positive, but if $\nu_1, \nu_2 > \nu_3$,
then $\Delta m^2_{23}$ will be negative.  The former is called a
``normal mass hierarchy'' and the latter is the ``inverted mass
hierarchy.'' At this point, the sign is irrelevant because $\Delta
m^2$ appears in a term which is squared.  However, in sec.~\ref{sub:paradigm},
this point will become important.

The mixing matrix above can be described in terms of three mixing angles,
$\theta_{12}$,
$\theta_{13}$ and $\theta_{23}$:
\begin{equation}
U =
\begin{pmatrix}
c_{12}c_{13} & s_{12}c_{13} & s_{13}  \cr
-s_{12}c_{23}-c_{12}s_{23}s_{13}  &
c_{12}c_{23}-s_{12}s_{23}s_{13} 
& s_{23}c_{13} \cr
s_{12}s_{23}-c_{12}c_{23}s_{13}  &
-c_{12}s_{23}-s_{12}c_{23}s_{13} & c_{23}c_{13}
\end{pmatrix},
\label{mns}
\end{equation}
where $c_{ij} \equiv \cos\theta_{ij}$ and $s_{ij} \equiv
\sin\theta_{ij}$, with $i$and $j$ referring to the mass states.  In
fits to the oscillation parameters, people variously quote the results
in terms of the matrix element of $U$, $\sin$-squared of the given
angle, $\sin$-squared of twice the angle and a variety of other forms,
all of which are related.  Using the $13$ case as an example, the
quoted parameters are related by:
\begin{equation}
U_{e3}^2 \approx sin^2\theta_{13} \approx {1\over 4} sin^22\theta_{13}.
\end{equation}

Thus, in total, there are five free parameters in the 
simplest three-neutrino oscillation model, which can be
taken to be $\Delta m^2_{12}$, $\Delta m^2_{23}$, $\theta_{12}$,
$\theta_{13}$ and $\theta_{23}$.  

Although in general there will be mixing among all three flavors
of neutrinos, two-generation mixing is often assumed for simplicity.
If the mass scales are quite different ($m_3 >> m_2  >> m_1$, for
example), then the 
oscillation phenomena tend to decouple and the two-generation 
mixing model is a good approximation in limited regions.
In this case, each transition can be described 
by a two-generation mixing equation.
However, it is possible that experimental results
interpreted within the two-generation 
mixing formalism may indicate very different $\dmsq$ 
scales with quite different apparent strengths for the same oscillation.
This is because, as is evident from equation \ref{3-gen osc},
multiple terms involving different mixing strengths and $\Delta m^2$ 
values contribute to the transition probability for $\nu_\alpha \rightarrow
\nu_\beta$.

\subsubsection{Matter Effects}
\label{subsub:matter}

The probability for neutrino oscillations is modified in the presence
of matter.  This is true in any material, however the idea was first
explored for neutrino oscillations in the Sun, by Mikheyev, Smirnov
and Wolfenstein.  Therefore, matter effects are often called ``MSW''
effects \cite{MSW}.  In general, matter effects arise in
neutrino-electron scattering.  The electron neutrino flavor experiences
both CC and NC elastic forward-scattering with electrons.  However,
the $\nu_\mu$ and $\nu_\tau$ experience only NC
forward-scattering, because creation of the $\mu$ or $\tau$ is
kinematically forbidden or suppressed ({\it e.g.} $\nu_\mu + e^- \rightarrow
\nu_e + \mu^-$).  This difference produces the matter effect.

For neutrinos propagating through a constant density of electrons, if
$V_e$ is the elastic forward scattering potential for the $\nu_e$
component, and $V_{other}$ is the potential for the other neutrino
flavors, then the additional scattering potential is 
\begin{equation}
V=V_e-V_{other}
= \sqrt{2}G_F n_e, 
\end{equation}
where $G_F$ is the Fermi constant and $n_e$ is the electron density.  This
potential modifies the Hamiltonian, so that, if $H_0$ is the vacuum
Hamiltonian, then in matter the Hamiltonian is $H_0 + V$.  This means
that the eigenstates are modified from those of a vacuum, $\nu_1$ and
$\nu_2$, to become $\nu_{1m}$ and $\nu_{2m}$.  Effectively, the
neutrino mass spectrum is not the same as in vacuum.  The solutions to
the Hamiltonian are also modified.  From this, one can see that the
presence of electrons may substantially change the oscillatory
behavior of neutrinos.

The simplest outcome is that matter induces a shift in the mass state,
which is a combination of flavor eigenstates, propagates through the
material.  This leads to a change in the oscillation probability:
\begin{equation}
{\rm Prob}\left( \nu _e \rightarrow \nu _\mu\right) =\left(\sin
^22\theta/W^2 \right) \;
\sin^2\left(1.27 W \Delta m^2 L/E \right) 
  \label{probMSW}
\end{equation}
where $W^2 = \sin^2 2\theta + (\sqrt{2}G_F n_e (2E/\Delta m^2) - \cos
2\theta)^2$ (Note that in a vacuum, where $n_e=0$, this reduces to
equation~\ref{prob}.)  From this, one can see that if a neutrino,
passing through matter, encounters an optimal density of electrons, a
``resonance,'' or large enhancement of the oscillation probability,
can occur.  The Sun has a wide range of electron densities and thus is a
prime candidate for causing matter effects.  Also, neutrinos traveling
through the Earth's core, which has a high electron density, might
experience matter effects.  This will produce a
``day-night effect,'' or siderial variation, for neutrinos from the Sun.

For situations like the Sun, with very high electron densities which
vary with the position of the neutrino (and hence the time which
the neutrino has lived), the situation is complex.  If the electron
density is high and the density variation occurs slowly, or
adiabatically, then transition (not oscillation!) between flavors in
the mass state can occur as the neutrino propagates.  Thus it is
possible for neutrinos to be produced in the core of the Sun in a
given mass and flavor state, and slowly evolve in flavor content until
the neutrino exits the Sun, still in the same mass state.  In other
words, in the Sun, a $\nu_e$ produced in a mass eigenstate
$\nu_{2m}(r)$, which depends on the local electron density at radius $r$,
propagates as a $\nu_{2m}$ until it reaches the $r=R_{solar}$, where
$\nu_{2m}(R_{solar})=\nu_2$.  This peculiar effect is called the Large
Mixing Angle MSW solution.

\subsubsection{Designing an Oscillation Experiment}
\label{subsub:design}

From equation \ref{prob}, one can see that 
three important issues confront the designer
of the ideal
neutrino experiment.   
First, if one is searching for oscillations 
in the very small $\Delta m^2$ region, then large $L/E$ must be 
chosen in order to enhance the $\sin^2 (1.27 \Delta m^2 L/E)$ term.
However if $L/E$ is too large in comparison to $\Delta m^2$, then  
oscillations occur rapidly.  Because experiments have finite
resolution on $L$ and $E$, and a spread in beam energies, the 
$\sin^2 (1.27 \Delta m^2 L/E)$ averages to $1/2$ when $\Delta m^2 \gg
L/E$ and one loses
sensitivity to $\Delta m^2$.   Finally, because the probability is
directly proportional to $\sin^2 2\theta$, if the mixing angle is
small, then high statistics are required to observe an oscillation
signal.

There are two types of oscillation searches: ``disappearance'' and
``appearance.'' To be simplistic, consider a pure source of neutrinos
of type $x$. In a disappearance experiment, one looks for a deficit in
the expected flux of $\nu_x$.  This requires accurate knowledge of the
flux, which is often difficult to predict from first principles.
Therefore, most modern disappearance experiments employ a near-far
detector design.  The near detector measures the flux prior to
oscillation (the design goal is to effectively locate it at $L=0$ in
Fig.~\ref{fig:length}).  This is then used to predict the unoscillated
event rate in the far detector.  A deficit compared to prediction
indicates disappearance.  Appearance experiments search for
$\nu_\alpha \rightarrow \nu_\beta$ by directly observing interactions
of neutrinos of type $\beta$. The case for oscillations is most
persuasive if the deficit or excess has the ($L/E$) dependence
predicted by the neutrino oscillation formula (equation \ref{prob}).

\begin{figure}[t]
\centering
\scalebox{.5}{\includegraphics[clip=.true]{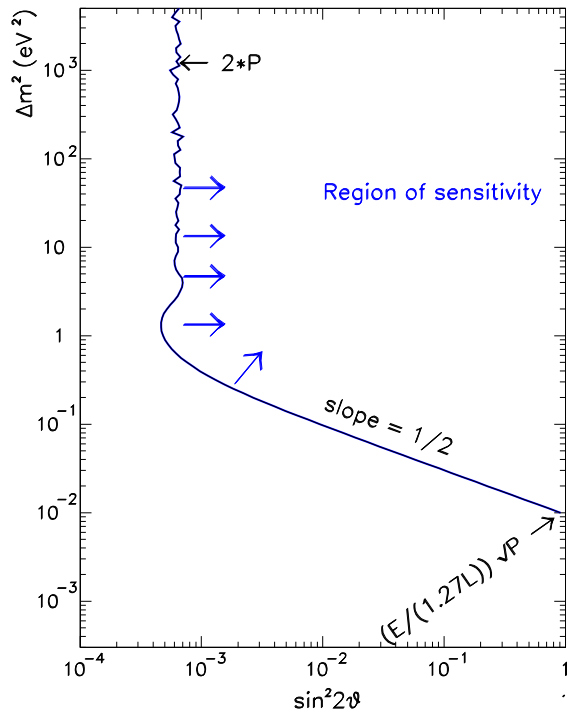}}
\vspace{-0.5in}
\caption{An illustration of the sensitivity of an imaginary
  oscillation experiment.  The region of sensitivity for an experiment
  depends on the oscillation probability, $P$, where one can set a
  limit at some confidence level.  Most experiments use 90\% CL.  The
  boundaries depend on $P$, $L$ and $E$.}
\label{fig:sensitivity}
\end{figure}

The ``sensitivity'' of an experiment is defined as the average
expected limit if the experiment were performed many times with no
true signal (only background).  Let us consider the sensitivity for a
hypothetical perfect (no-systematic error) disappearance neutrino
oscillation experiment with $N$ events.  A typical choice of
confidence level is 90\%, so in this case, the limiting probability,
assuming there is no signal, is
\begin{equation}
P=\sigma \sqrt{N}/N.
\label{setprob}
\end{equation}   
There are two possible choices of $\sigma$ associated with a 90\% CL
sensitivity, depending on the underlying philosophy. If one assumes
there is no signal in the data, then one quotes the sensitivity based
on 90\% of a single-sided Gaussian, which is $\sigma=1.28$.  If the
philosophy is that there is a signal which is too small to measure,
then one quotes the sensitivity using $\sigma=1.64$, which is
appropriate for a double-sided Gaussian.  Historically, $\sigma=1.28$
was used in most publications.  Physicists engage in arguments as to
which is most correct, but what is most important from a practical
point of view is for the reader to understand what was used.  The
reader can always scale between 1.28 and 1.64 depending on personal
opinion.

There is only one measurement, $P$, and there are two unknowns,
$\Delta m^2$ and $\sin^2 2\theta$; so this translates to a region of 
sensitivity
within $\Delta m^2$ -- $\sin^2 2\theta$ space.  This is typically
indicated by a solid line, with the allowed region on the right on a
plot (see illustration in Fig~\ref{fig:sensitivity}).  For the perfect
(no-systematic error) experiment, the high $\Delta m^2$ limit on
$\sin^2 2\theta$ is driven by the statistics.  On the other hand, the
$L$ and $E$ of the experiment drive the low $\Delta m^2$ limit, which
depends on the fourth root of the statistics.  If our perfect
experiment had seen a signal, the indications of neutrino oscillations
would appear as ``allowed regions,'' or shaded areas on plots of
$\Delta m^2$ {\it vs.} $\sin^2 2\theta$.

This rule of thumb -- that statistics drives the $\sin^2 2
\theta$-reach and $L/E$ drives the $\Delta m^2$ reach -- becomes more
complicated when systematics are considered.  The imperfections of a
real experiment affect the limits which can be set.  Systematic
uncertainties in the efficiencies and backgrounds reduce the
sensitivity of a given experiment.  Background sources introduce
multiple flavors of neutrinos in the beam.  Misidentification of the
interacting neutrino flavor in the detector can mimic oscillation
signatures. In addition, systematic uncertainties in the relative
acceptance versus distance and energy need to be understood and
included in the analysis of the data.

For a real experiment, with both statistical and systematic errors,
finding the sensitivity and final limit or allowed region requires a
fit to the data.  The data are compared to the expectation for
oscillation across the range of oscillation parameters, and the set of
parameters where the agreement is good to 90\% CL are chosen.
Historically, there are three main approaches which have been used in
fits. The first method is the ``single sided raster scan."  In this
case one chooses a $\Delta m^2$ value and scans through the $sin^2
2\theta$-space to find the 90\% CL limit.  The second method, the
``global scan," explores $\Delta m^2$- and $sin^2 2\theta$-space
simultaneously.  Thus there are two parameters to fit and two degrees
of freedom.  The third method is the frequentist, or
``Feldman-Cousins" approach \cite{FeldmanCousins}, in which one simulates
``fake-experiments'' for each $\Delta m^2$ and $sin^2 2\theta$ point, and
determines the limit where, in 90\% of the cases, no signal is
observed.  Each method has pros and cons and the choice is something
of a matter of taste.  As with the question of a single- or
double-sided gaussian, what is important is to compare sensitivities,
limits, and signals from like methods.

It is possible for an experiment which does not observe a signal to
set a limit which is better than the sensitivity.  This occurs if the
experiment observed a downward fluctuation in the background.  In this
case, a limit is hard to interpret.  The latest standard practice is
to show the sensitivity and the limit on plots, and the readers can
draw their own interpretation \cite{FeldmanCousins}.

\subsubsection{Experimental Evidence for Oscillations}
\label{subsub:evidence}

Two separate allowed regions in $\Delta m^2$-and-$\sin^2 2 \theta$
-space for neutrino oscillations have been observed at the $> 5\sigma$
level.  These are called the ``Atmospheric $\Delta m^2$'' and ``Solar
$\Delta m^2$'' regions.  The
names are historical, as will be seen below. Many reviews have been
written on these results (see, for example, \cite{APSwhitepaper}, \cite{rev2},
and \cite{rev3}) and so here the results are briefly outlined.

The highest $\Delta m^2$ signal was first observed using 
neutrinos produced in the upper atmosphere. 
These atmospheric neutrinos are produced through collisions of cosmic
rays with the atmosphere.  The neutrinos are detected through their
charged--current interactions in detectors on the Earth's surface.

\begin{figure}[t]
\centering
\scalebox{.5}{\includegraphics[clip=.true]{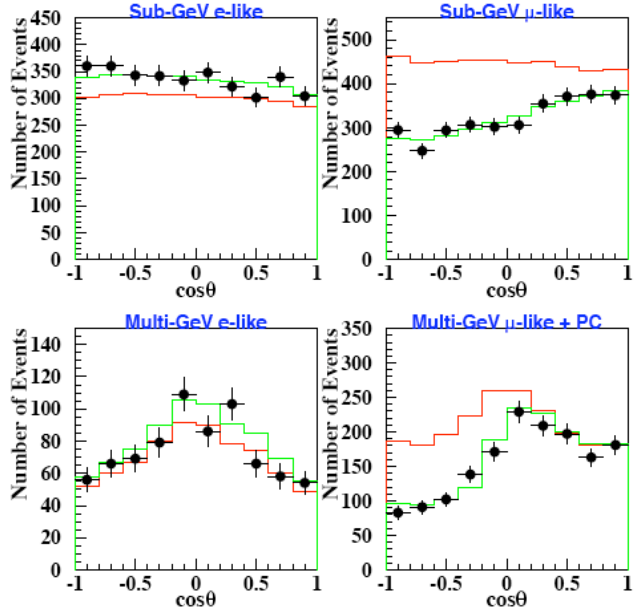}}
\caption{Event rates observed in Super K 
as a function of zenith angle for two energy ranges.
Candidate $\nu_e$ events are on the left, $\nu_\mu$ are on the right.
The red line indicates the predicted rate.  The green line
is the best fit including oscillations.\cite{E.Kearn_nu_2004_talk}}
\label{fig:zenith}
\end{figure}

The first evidence for atmospheric neutrino oscillations came from the
Kamioka \cite{Kamioka} and IMB \cite{IMB} experiments.   This was followed
by the convincing case presented by the Super K experiment
\cite{SuperKatmos}.  These were single detector experiments observing
atmospheric neutrino interactions as a function of zenith angle
(see Fig.~\ref{fig:zenith}).
Several striking features were observed. The first was that the
$\nu_\mu$ flavor neutrinos showed clear evidence of disappearance
while the $\nu_e$ flavor CC scatters were in good agreement with
prediction.  The second striking observation was that the apparent
mixing was nearly maximal.  In other words, the experiments were
seeing a 50\% reduction of the $\nu_\mu$ event rate compared to
expectation.

Complications in the analysis arise from the difficulty in
understanding production of atmospheric neutrinos (affecting the
understanding of $E$) and in the accurate reconstruction of events as
a function of zenith angle (affecting knowledge of $L$).  The $\Delta
m^2$ extracted from the Kamoiokande data is an order of magnitude
higher than that extracted from the Super K data, indicating a clear
systematic effect.  Thus, it was absolutely crucial for
accelerator-based ``long-baseline'' neutrino experiments to confirm
this result.  In these experiments, the $L$ is well defined by the
distance from source to detector, and the $E$ is well understood from
a near detector measurement.  

The challenge for long-baseline experiments is that the $L/E$ required
to access the atmospheric signal is on the order of 1000 km/GeV.  If
the beam is relatively low energy, so that the easy-to-reconstruct
CCQE interaction dominates the events, then $L$ is on the order of
1000 km.  This leads to two major technical challenges.  First,
because the Earth is a sphere, if the source and detector are to be
located on (or near) the surface, the beam must be directed downward,
into the Earth.  Engineering a beamline at a steep angle requires
overcoming substantial hurdles in tunneling.  Second, the beam spreads
as it travels outward from the source, resulting in low intensity at
the detector.  Therefore, very high rates are needed.  However, these
challenges have now been overcome at three accelerator complexes: KEK,
FNAL and CERN, and a new long-baseline beam from the JPARC facility
will be available soon.  Making use of these lines, initial
confirmation of the atmospheric neutrino deficit came from the
KEK-to-Kamiokande (K2K) long baseline experiment \cite{K2K}.  This has
since been followed up by the MINOS experiment to high precision
\cite{MINOSpaper, Nikkitalk}.

In the atmospheric data, the $\nu_e$ CC signal is in agreement with
expectation, and in the long-baseline experiments, no $\nu_e$ excess
has been observed.  Therefore, one cannot interpret this oscillation
signal as $\nu_\mu \rightarrow \nu_e$.  This leaves only $\nu_\mu
\rightarrow \nu_\tau$ as an explanation for the deficit in a
three-neutrino model.  Observation of $\nu_\tau$ CC interactions is
experimentally difficult in these experiments for a number of reasons.
First, the $L/E$ of the signal is such that for lengths available
to present experiments, the energy of the beam must be low ($\lesssim
10$ GeV).  As discussed in sec.~\ref{sub:interacts}, because the
$\tau$ mass is 1.8 GeV, there is substantial mass suppression for $\tau$
production at low energies, so the CC event rate is low.
Second, the $\tau$ decays quickly, leaving behind a complicated event
structure which can be easily confused with $\nu_\mu$ and $\nu_e$ low
multiplicity interactions in calorimeter or Cerenkov detectors.  The
difficulty of identifying $\nu_\tau$ events even in a specialized
emulsion-based detector with a high energy neutrino beam, was made
clear by the DoNuT experiment \cite{DoNuT}, which provided the first, and
so far only, direct observation of CC $\nu_\tau$ interactions.  Thus,
while some studies claim observation of $\nu_\tau$ CC interactions in
SuperK \cite{SuperKnutau}, these results are not very convincing to
this author.  Fortunately, a specialized experiment called OPERA
\cite{OPERA}, which is an emulsion-based long-baseline detector, is
presently taking data.  The average energy of the CNGS beam used by
this experiment is 17 GeV, sufficiently high to produce $\nu_\tau$ CC
events.  This experiment is expected to observe $\sim 15$ events in 5
years of running if the atmospheric neutrino deficit is due to
$\nu_\mu \rightarrow \nu_\tau$ with $\Delta m^2=2.5\times 10^{-3}$
eV$^2$ \cite{OPERAExpect}.

\begin{figure}[t]
\centering
\scalebox{.4}{\includegraphics[clip=.true]{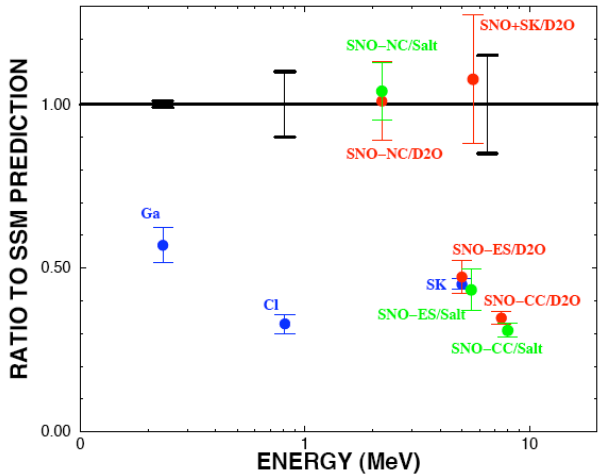}}
\caption{Ratio of observed event rates in solar neutrino
experiments compared to the Standard Solar Model.   Experiments
are plotted at the average energy of the detected signal,
which varies due to detection threshold.  Black error bars
indicate Standard Solar Model error.\cite{hep-ex/0412016}}
\label{fig:solarratios}
\end{figure}

The lower $\Delta m^2$ signal is called the ``Solar Neutrino
Deficit,'' as it was first observed as a low rate of observed
$\nu_e$'s from the Sun.  The first observation of this effect was a
$\nue$ deficit observed using a Cl target \cite{chlorine} by Ray Davis
and collaborators at Homestake, using $\nu_e + {\rm Cl} \rightarrow e + {\rm Ar}$.
Only about 1/3 of the total expected neutrino event rate was observed.
By 1999, four additional experiments had confirmed these observations.
The GALLEX \cite{gallex} and SAGE \cite{sage} experiments confirmed a
deficit for CC electron neutrino interactions in a Ga target producing
Ge.  The Super Kamiokande experiment observed a deficit for $\nu_e + e
\rightarrow \nu_e + e$ reactions in water \cite{SKsolar}.
The deficit is shown on Fig.~\ref{fig:solarratios}, indicated by the blue
points.   This plot shows the ratio to the Standard Solar Model 
prediction, which is indicated by the solid line at unity.  

A few aspects of the initial solar neutrino deficit studies should be
noted.  First, the three types of experiments, chlorine-based,
gallium-based, and water-based, measured different levels of deficit.
Given that each type of nucleus has a different low energy threshold
for observation of CC events, as previously discussed, one can
interpret the varying levels of deficit as an energy dependent effect.
Second, all of the above experiments rely upon the CC interaction.
The energy of neutrinos from the Sun is so low, that should $\nu_\mu$
or $\nu_\tau$ be produced through oscillations, the CC interaction
could not occur.  This is because of the relatively high mass of the
$\mu$ (106 MeV) and the $\tau$ (1.8 GeV).  Thus all of these
experiments can observe that $\nu_e$s disappeared, but they cannot
observe if the neutrinos reappear as one of the other flavors.  This
makes a decisive statement that the effect is due to neutrino
oscillations problematic.

For some time, people argued the apparent deficit was due to an
incomplete picture of solar processes.  The two important theoretical
issues related to the solar neutrino fluxes were the fusion cross
sections and the temperature of the solar interior.  A comprehensive
analysis of the available information on nuclear fusion cross sections
important to solar processes has been compiled \cite{Adelberger} and
shows that the important cross sections are well-known.  Results in
helioseismology provided an important further test of the ``Standard
Solar Model'' \cite{bahcallseis}.  The Sun is a resonant cavity, with
oscillation frequencies dependent upon $P/\rho$, the ratio of
pressure to density.  Helioseismological data confirmed the SSM
prediction of $U$ to better than 0.1\% \cite{helioseis}.  With the
results of these studies, most physicists were convinced that the
Standard Solar Model was substantially correct.
The error bars on the black line at unity in Fig.~\ref{fig:solarratios}
shows the side of the estimated systematic error on the Standard Solar Model.

Interpreting the results as neutrino oscillations resulted in a
complicated picture.  The vacuum oscillation probability, calculated
using equation \ref{prob}, results in allowed regions of $\Delta m^2$
which are very low ($\Delta m^2 \sim 10^{-10} {\rm eV}^2$).  This is
because the energy of the neutrinos is only a few MeV, and the Sun to
the Earth pathlength is very long ($\sim 10^{11}$m) .  On the other
hand, the Sun has high electron content and density, so matter
effects (sec.~\ref{subsub:matter}) could interfere with the picture, allowing
higher true values of $\Delta m^2$.  The MSW effect yielded two
solutions in fits to the data.  One was at mixing angles of $\sim
10^{-3}$.  Until very recently, this was regarded as the most likely
solution based on analogy with mixing in the quark sector.  The other
solution gave a very large, although not maximal, mixing angle.

\begin{figure}[t]

\centering
\scalebox{0.4}{\includegraphics{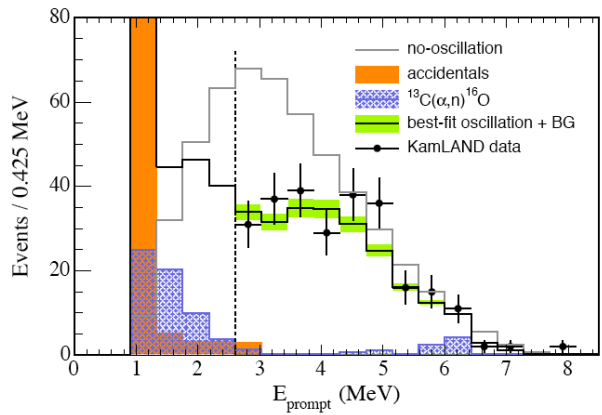}}
\caption{Events in KamLAND as a function of energy.   The grey line
indicates the expectation for no oscillation.\cite{KamLANDresult}}
\label{kamresult}
\end{figure}

Two dramatic results of the early 2000's demonstrated that the solar
neutrino deficit was due to oscillations with the MSW effect and with
large mixing angle.  The first result was from the SNO experiment
\cite{SNO}.  SNO used a D$_2$O target which allowed for measurement of
both CC $\nu_e$ interactions as well as $\nu + d \rightarrow \nu + n +
p$.  In the former measurement, SNO sees a deficit consistent with the
other measurements within an oscillation interpretation, and which
yields a $\nu_e$ flux of $(1.76\pm0.05 {\rm (stat)} \pm0.09 {\rm
  (sys)})\times10^6/{\rm cm}^2{\rm s}$ \cite{SNOtotal}. The later measurement is an NC
interaction, and thus is flavor-blind. It yields a total NC flux of
$(5.09 ^{0.44}_{0.43}  {\rm (stat)} ^{+0.46}_{-0.43} {\rm (sys)} ) \times 10^6/{\rm
  cm}^2 s$ \cite{SNOtotal} which can be compared with the theoretical
prediction of $(5.69\pm0.91)\times 10^6/{\rm cm}^2{\rm s}$ \cite{TheoryFlux}.  In other
words, SNO observed the expected total event rate, within errors.
This implied that the $\nu_e$s are oscillating to neutrinos which
participate in the NC interaction, $\nu_\mu$s and/or $\nu_\tau$s,
with the total $\nu_\mu + \nu_\tau$ flux equal to   
$(3.41 \pm 0.45 {\rm (stat)} ^{+0.48}_{-0.45} {\rm (sys)} ) \times 10^6/{\rm
  cm}^2 s$ \cite{SNOtotal}.  
The results of two runs of the SNO experiment
are shown by the red and green points on Fig.~\ref{fig:solarratios}.
The
second result was from the KamLAND experiment.  This was a
reactor-based experiment located in Japan.  Using many reactors which
were hundreds of kilometers away, the KamLAND experiment was able to reach
$L/E \sim 10^{-6}$ m/MeV.  This covered the MSW allowed-$\Delta m^2$
solution.  The statistics were on the order of hundreds of events, but
this was enough to probe the large mixing-angle MSW solution.  KamLAND
expected 365 events and observed 258 events, and thus had clear
evidence for oscillations with large mixing, $\tan^2 \theta =0.40
^{+0.010}_{-0.07}$ and relatively high $\Delta m^2$, of
$7.9^{+0.6}_{-0.5} \times 10^{-5}$ eV$^2$ \cite{KamLANDresult}.
The energy distribution of the events observed in KamLAND is shown
in Fig.~\ref{kamresult}.

Based on the atmospheric and solar studies, there are two squared mass
differences: $\Delta m^2_{solar}$ and $\Delta m^2_{atmos}$.  The
smaller is identified with the mass splitting between $\nu_1$ and
$\nu_2$: $\Delta m^2_{12}=\Delta m^2_{solar}$.  The atmospheric
deficit measures a combination of $\Delta m^2_{23}$ and $\Delta
m^2_{13}$.  However, since $\Delta m^2_{13}=\Delta m^2_{12}+\Delta
m^2_{23}$ and $\Delta m^2_{12}$ is small, $\Delta m^2_{13} \approx
\Delta m^2_{23} \approx \Delta m^2_{atmos}$.

A recent global analysis of the data \cite{Schwetzglob} from the above
experiments yields a consistent picture for three neutrino
oscillations with five free parameters.  The mass differences are:
$\Delta m^2_{12} = (7.9 \pm 0.3) \times 10^{-5} {\rm eV}^2$ and
$|\Delta m^2_{13}| = (2.5^{+0.20}_{-0.25}) \times 10^{-5} {\rm eV}^2$,
where the absolute value is indicated in the second case because the
sign ({\it i.e.} the mass hierarchy) is unknown.  The two
well-measured mixing angles are determined to be: $\sin^2 \theta_{12}
= 0.30 ^{+0.02}_{-0.03}$ and $\sin^2
\theta_{23}=0.50^{+0.08}_{-0.07}$.  One mixing angle, $\theta_{13}$ is
yet to be measured, but a limit of $\sin^2 \theta_{13} < 0.025$ can be
placed based on global fits.

Based on the measurements, the mixing
matrix of eq.~\ref{mns}, translates roughly into:
\begin{equation}
\hspace*{-0.5cm}
U =
\left( \begin{array}{ccc}
  0.8  & 0.5  & ? \\
  0.4  & 0.6  & 0.7 \\
  0.4  & 0.6  & 0.7 \\
\end{array} \right).
\label{mnsnumbers}
\end{equation}
This matrix, with its large off-diagonal components, looks very
different from the quark-sector mixing matrix where the off-diagonal
elements are all relatively small.  In this matrix, the ``odd element
out'' is $U_{e3}$ which is clearly substantially smaller than the
others.  At this point, there is no consensus on what this matrix may
be telling us about the larger theory, but there is a sense that the
value of $\theta_{13}$ is an important clue.  Theories which attempt
to explain this matrix tend to fall into two classes -- those where
$\theta_{13}$ is just below the present limit and those with very
small values.  As an illustration of this point,
Tab.~\ref{tab:PredictionsT13} shows order of magnitude predictions
for a variety of theories.  Thus a measurement of $\sin^2
2\theta_{13}$ which is greater than about 1\%, or a limit at this
level, can point the way to the larger theory.

\begin{table}[t]
\centering
{
\begin{tabular}{l|c|r}
\hline
Model(s) & Refs.\ & approximate $\sin^2 2\theta_{13}$\\
\hline
\hline
Minimal SO(10) & \cite{Goh:2003hf} & 0.13 \\
Orbifold SO(10) & \cite{Asaka:2003iy} & 0.04 \\
SO(10) + Flavor symmetry
 & \cite{Babu:1998wi} & $1.2 \cdot 10^{-6}$ \\
 & \cite{Albright:2001uh} & $7.8 \cdot 10^{-4}$ \\
 & \cite{Blazek:1999hz,Ross:2002fb,Raby:2003ay} &
   0.01 .. 0.04 \\
 & \cite{Kitano:2000xk,Maekawa:2001uk,Chen:2002pa} &
   0.09 .. 0.18 \\
SO(10) + Texture
 & \cite{Bando:2003ei} & $4 \cdot 10^{-4}$ .. 0.01 \\
 & \cite{Buchmuller:2001dc} & 0.04 \\
$\mathrm{SU}(2)_\mathrm{L} \times \mathrm{SU}(2)_\mathrm{R} \times \mathrm{SU}(4)_c$ &
 \cite{Frampton:2004vw} & 0.09 \\
\hline
Flavor symmetries
 & \cite{Grimus:2001ex,Grimus:2003kq,Grimus:2004rj} & 0 \\
 & \cite{Chen:2004rr,Aizawa:2004qf,Mohapatra:2004mf} &
   $\lesssim 0.004$ \\
 & \cite{Antusch:2004xd,Antusch:2004re,Rodejohann:2004qh} & $10^{-4}$ .. 0.02 \\
 & \cite{Babu:2002dz,Ohlsson:2002rb,King:2003rf,Shafi:2004jy,Mohapatra:2004mf} & 0.04 .. 0.15 \\
\hline
Textures
 & \cite{Bando:2003wb} & $4 \cdot 10^{-4}$ .. 0.01 \\
 & \cite{Honda:2003pg,Lebed:2003sj,Ibarra:2003xp,Harrison:2004he} & 0.03 .. 0.15 \\
\hline
$3 \times 2$ see-saw & \cite{Frampton:2002qc} & 0.04 \\
\hline
Anarchy & \cite{deGouvea:2003xe} & $>0.04$ \\
\hline
Renormalization group enhancement & \cite{Mohapatra:2003tw} & 0.03 .. 0.04 \\
\hline
M-Theory model & \cite{Arnowitt:2003kc} & $10^{-4}$ \\
\hline
\end{tabular}}
\caption{
  Selected predictions for $\sin^2 2\theta_{13}$.\cite{hep-ex/0509019}.
}\label{tab:PredictionsT13}
\end{table}

The best method for measuring  $\theta_{13}$ is 
from reactor experiments
which constrain this oscillation by searching for $\bar \nu_e$
disappearance. The oscillation probability is given by:
\begin{equation}
  P_{reactor}  \simeq \quad \sin^2 2 \theta_{13} \,
  \sin^2 \Delta + \alpha^2 \, \Delta^2 \, \cos^4 \theta_{13}
  \, \sin^2 2 \theta_{12} , \label{OscProb}
\end{equation}
with 
\begin{eqnarray}
\alpha &\equiv& \Delta m_{21}^2 / \Delta m_{23}^2\\ 
\Delta &\equiv& \Delta m^2_{31}L / (4E_\nu).  
\label{alfdelt}
\end{eqnarray}
Events are detected through the
inverse beta decay (IBD) interaction.  The CHOOZ experiment
\cite{chooz}, with a baseline of 1.1 km and typical neutrino event
energies between 3 and 5 MeV $\langle E \rangle = 3.5$ MeV) has set
the best reactor-based limit to date, of $\sin^2 2\theta_{13}<0.27$ at
$\Delta m^2 =2.5\times10^{-3}$ eV$^2$.  This limit can be improved
with a global fit, as quoted above.

Significant improvement is expected from the upcoming round of reactor
experiments results due to introducing a near-far detector design.  The near
detector measures the unoscillated event rate, and the far detector is
used to search for a deficit as a function of energy. The Double Chooz
experiment, beginning in 2009, is expected to reach $\sin^2 2
\theta_{13} \sim 0.03$. \cite{DoubleChooz}.  This will be followed by
the Daya Bay experiment which will reach $\sim$0.01.
\cite{DayaBay}.

\subsection{Direct Measurements of Neutrino Mass}
\label{sub:directmass}

For neutrinos, there are no mass measurements, only mass limits.
Observations of neutrino oscillations are sensitive to the mass
differences between neutrinos, not the actual mass of the
neutrino. Therefore, they do not fall into the category of a ``direct
measurement''.  One can, however, use these oscillation results to
estimate the required sensitivity for a direct mass measurement.  The
upper limit comes from assuming that one of the neutrino masses is
exactly zero.  Given that the largest $\Delta m^2$ is $\sim 3
\times 10^{-3}$ eV$^2$, this implies there is a neutrino with mass
$\sqrt{\Delta m^2} \sim 0.05$ eV.  The mass of the neutrino can be
directly measured from decay kinematics and from time of flight from 
supernovae.  Neither method has reached the 0.05 eV range yet, although the next
generation of decay-based experiments comes close.

We know from neutrino oscillations that there is a very poor correspondence 
between neutrino flavors and neutrino masses, {\it {i.e.,}} the mixings are 
large.   However, it is easiest to conduct the discussion of these limits
in terms of specific flavors.   Thus, what is actually being studied 
is an average mass associated with each flavor.  For 
example, for the $\nu_e$ mass measured from $\beta$ decay,
which will be expanded upon below, what is actually probed is: 
\begin{equation}
m_\beta=\sqrt{\Sigma_i |U_{ei}|^2 m_i^2}.
\label{betamass}
\end{equation}

The simplest method for measuring neutrino mass is applied to the $\nu_\mu$.
The mass is obtained from the 2-body decay-at-rest 
kinematics of $\pi \rightarrow
\mu \nu_\mu$.  
One begins in the center of mass with the 4-vector
relationship: $p_\pi = p_\mu + p_\nu$.   Squaring and solving for
neutrino mass gives: 
$m_\nu^2 = m_\pi^2 + m_\mu^2 - \sqrt{4 m_\pi^2 (|{\bf p_\mu}|^2 + m_\mu^2)}$.
From this, one can see that this technique requires accurate
measurement of the muon momentum, ${\bf p_\mu}$, as well as the masses
of the muon, $m_\mu$ and the pion, $m_\pi$.   In fact, the uncertainty 
on the mass of the pion is what dominates the $\nu_\mu$ mass
measurement.   As a result, a limit is set at $m_{\nu_\mu}<170$ keV \cite{PSI, Jeckel}.

The mass for the $\nu_\tau$ is obtained from the kinematics of $\tau$
decays.  The $\tau$ typically decays to many hadrons.   However, the
four vectors for each of the hadrons can be summed.  Then the decay
can be treated as a two-body problem with the neutrino as one 4-vector
and the sum of the hadrons as the other vector.   At this point, the
same method described for the $\nu_\mu$ can be applied.   Measurements
are again error-limited, so a limit on the mass is placed.
The best limit, which is $m_{\nu_\tau}<18.2$ MeV, comes from fits to  
$\tau^- \rightarrow 2 \pi^- \pi^+
\nu_\tau$
and $\tau^- \rightarrow 3 \pi^- 2\pi^+ (\pi^0) \nu_\tau$ decays
observed by 
the ALEPH experiment \cite{Alephnutau}.

\begin{table}

\centering
\begin{tabular}{c|c|c|c}
\hline
Experiment  &   measured $m^2$ (eV$^2$) &  limit (eV), 95\% C.L. 
&      Year\\ \hline
      Mainz \cite{newmainz}
&    -0.6$\pm$ 2.2$\pm$ 2.1 
& 2.2 
&      2004 \\
     Troitsk \cite{Troitsk}
&  -1.0 $\pm$ 3.0$\pm$ 2.1 
& 2.5 
&      2000 \\
      Mainz\cite{Mainz} 
&    -3.7 $\pm$ 5.3 $\pm$ 2.1
& 2.8 
&      2000 \\
    LLNL \cite{Liver} 
&    - 130 $\pm$ 20$\pm$ 15 
& 7.0 
&      1995 \\
      CIAE \cite{china}
&     - 31 $\pm$ 75$\pm$ 48 
& 12.4 
&      1995 \\ 
      Zurich \cite{zurich}
&     -24 $\pm$ 48$\pm$ 61 
& 11.7 
&      1992 \\
    Tokyo INS \cite{tokyo}
&     - 65 $\pm$ 85$\pm$ 65 
& 13.1 
&      1991 \\
    Los Alamos \cite{Hamish}
&    - 147 $\pm$ 68$\pm$ 41 
& 9.3 
&      1991 \\ \hline
\end{tabular}}
\caption{Overview of $\nu_e$ squared mass measurements.}
{
\label{tab:nuemass}

\end{table}

The experimental situation for the $\nu_e$ mass measurement is more
complicated. The endpoint of the electron energy spectrum from tritium
$\beta$ decay is used to determine the mass. Just as in the case of
the $\nu_\mu$ and $\nu_\tau$, the experiments measure a value of $m^2$.
The problem is that the measurements have been systematically
negative.  A review of measurements, as a function of time, is given
in Tab.~\ref{tab:nuemass}.  Recent measurements at
Troitsk \cite{Troitsk} 
and Mainz\cite{Mainz} are negative, but in
agreement with zero.   Following the Particle Data Group prescription
for setting limit in the case of an unphysical results, $m^2=0$ is assumed,
with the quoted errors.
Based on these results, one can extract a limit
of approximately $<2$ eV for the mass of the $\nu_e$.

The next big step in the measurement of neutrino mass from decay
kinematics will come from the Katrin Experiment \cite{Katrin}.  Katrin
will use tritium beta decay to measure the mass of the neutrino to 0.2
eV.  This does not reach the range of 0.05 eV, which our simplistic
argument presented at the top of this section indicated.  However,
that argument assumed that the lightest neutrino had zero mass.  A
small offset from zero easily boosts the spectrum into the range
observable by Katrin.  On the other hand, Katrin is sensitive to only
electron flavor.  Thus, its sensitivity depends up on the amount of
mixing of $\nu_e$ within the heaviest neutrino.  

Another method for measuring neutrino mass from simple
kinematics is to use time of flight for neutrinos from supernovae.
Neutrinos carry away $\sim 99\%$ of the energy from a supernova.
The mass limit is obtained from the spread in the propagation times of
the neutrinos.  The propagation time for a single neutrino is given by
\begin{equation}
t_{obs}-t_{emit} = t_0 \Big  ( 1 + {{m^2}\over{2E^2}} \Big ) 
\end{equation}
where $t_0$ is the time required for light to reach Earth from the
supernova.   Because the neutrinos escape from a
supernova before the photons, we do not know $t_{emit}$.   But we can
obtain the time difference between 2 events:
\begin{equation}
 \Delta t_{obs} - \Delta t_{emit} \approx 
{{t_0 m^2}\over{2}} \Big( {{1}\over{E^2_1}} - {{1}\over{E^2_2}} \Big ).
\end{equation}
using the assumption
that all neutrinos are emitted at the same time, one can obtain a 
mass limit of  {$\sim 30$ eV} from the 
$\sim 20$ events observed from SN1987a at 2 sites.\cite{imbsn,kamsn}

This is actually an oversimplified argument. The models for neutrino
emission are actually quite complicated. The pulse of neutrinos has a
prompt peak followed by a broader secondary peak with a long tail
distributed over an interval which can be 4 s or more. The prompt peak
is from ``neutronization'' and is mainly $\nu_e$, while all three
neutrino flavors populate the secondary peak. However, the rate of
$\nu_e$ escape is slower compared to $\nu_\mu$ and $\nu_\tau$ produced
at the same time, because the $\nu_e$s can experience CC
interactions, while the kinematic suppression from the charged lepton
mass prevents this for the other flavors. However, when all of the
aspects of the modeling are put together, the bottom line remains the
same: it will be possible to set stringent mass limits if we observe
neutrinos from nearby supernovae.   

Some argue that cosmology provides a ``direct measurement.''
Cosmological fits have sensitivity to neutrino masses, but the results
are dependent on the cosmological parameters \cite{Hannestad} and the
model for relic neutrino production. There are
many examples of models with low relic
neutrino densities which would significantly change the present
interpretation of the cosmological data.\cite{cosmoalternatives}  In the opinion of the
author, until these issues are settled, cosmological measurements
cannot convincingly compete with kinematic decays and supernova
measurements, despite aggressive claims.

\section{Neutrinos We Would Like to Meet}

Now that we know that neutrinos have mass, and thus are outside of 
expectations, the obvious question is: ``what other Beyond Standard
Model properties do they possess?''  {\it The APS Study on the Future of
Neutrino Physics} focussed on this question.  The plan for attack was
divided into three fronts: (1) Neutrinos and the New Paradigm, (2)
Neutrinos and the Unexpected and 3) Neutrinos and the Cosmos.  The
remainder of this paper follows this structure.

The consequences of the discovery of neutrino mass leads to a rich
array of ideas.  It is was beyond the scope of these lectures to
cover the entire spectrum.  So, in each of the three areas, two topics
are chosen for extensive discussion.  The reader is referred to the
study\cite{APSstudy} and the accompanying theory white
paper\cite{APSwhitepaper} for further ideas.

\subsection{Neutrinos and the New Paradigm}
\label{sub:paradigm}

The first step in creating a ``New Standard Model'' is to 
incorporate neutrino mass.  
The simplest method is to introduce a Dirac mass, by analogy
with the electron.   This allows us to introduce a small neutrino mass,
simply by arguing  that the coupling to the Higgs is remarkably 
small.    However, the unlikely smallness of the coupling has
pushed theorists to look
for other approaches. Among the oldest of these ideas is that
neutrinos may be ``Majorana particles,'' {\it i.e.}, they are their
own antiparticle.    This leads to a new type of mass term in the 
Lagrangian.   Through the 
``see-saw'' mechanism, which fits well with Grand Unified Theories,
this can also give a motivation for the apparently small value
of the neutrino masses.

A direct consequence of the Majorana See-Saw Model is a heavy 
neutrino, with mass near the GUT scale.   Because the heavy neutrino
gets its mass through the Majorana rather than Dirac term of the 
Lagrangian, this neutrino was massive during the earliest periods
of the universe, before the electroweak phase transition.    The decays 
of such a heavy lepton could be CP violating.   This would provide
a mechanism for producing the observed matter-antimatter imbalance 
seen today.

The tidiness of the the above theoretical ideas has caused this
paradigm to emerge as the consensus favorite for the ``New Standard Model.''
However, there is absolutely no experimental evidence for this 
theory at this time.   We have no evidence for the Majorana nature
of neutrinos.  Nor do we have any evidence for CP violation in 
the neutrino system.    The great challenge of the next few years, 
then, is to find any sign at all that this theory is correct.  

This section reviews how one introduces mass into the Lagrangian.
The search for evidence of the Majorana nature of neutrinos 
though neutrinoless double beta decay is considered.   Then,
the prospects for finding evidence for CP violation is considered.

\subsubsection{How Neutrinos Might Get Their Mass}

The simplest assumption is that the neutrino mass should appear in the
Lagrangian in the same way as for the charged fermions -- via a Dirac
mass term.  In general, the Dirac mass term in the Lagrangian will be
of the form
\begin{equation}
 m (\bar\psi_L {\psi_R} + \bar\psi_R {\psi_L}) . 
\end{equation}
From the arguments presented
in eqs.~\ref{firsteqinlrdiscussion} through
\ref{lasteqinlrdiscussion}, we saw that the scalar ``mass'' term mixes
the RH and LH states of the fermion.  If the fermion has only one
chirality, then the Dirac mass term will automatically vanish.  For this
reason, a standard Dirac mass term for the neutrino will require the RH
neutrino and LH antineutrino states.

To motivate the mass term, the most straightforward approach is to use
the Higgs mechanism, as was done for the electron in the Standard
Model.  In the case of the electron, when we introduce a spin-0 Higgs
doublet ,($h^0,h^+$), into the Lagrangian, we find terms like: 
\begin{equation}
g_e
 \bar\psi_{e_R} (\psi_{\nu_L} (h^+)^\dagger +   \psi_{e_L} (h^0)^\dagger) + h.c.,
\end{equation} 
where $g_e$ is the coupling constant and ``h.c.'' is the Hermetian
conjugate.  The piece of this term proportional to 
$ \bar\psi_{e_R}  \psi_{e_L} (h^0)^\dagger$, combined with its Hermetian 
conjugate,
can be identified with the Dirac mass term, $m_e \bar \psi_e \psi_e$.
We set $\langle h^0 \rangle = v/\sqrt{2}$, so that we obtain $g
\langle h^0 \rangle \bar \psi_e \psi_e$ and $m_e = g_e v / \sqrt{2}$.
This is the Standard Model method for conveniently converting the {\it
  ad hoc} electron mass, $m_e$, into an {\it ad hoc} coupling to the
Higgs, $g_e$ and a vacuum expectation value (VEV) for the Higgs, $v$.
Following the same procedure for neutrinos allows us to identify the
Dirac mass term with $m_\nu = g_\nu v/ \sqrt{2}$.  The VEV, $v$, has
to be the same as for all other leptons.  Therefore, the small mass
must come from a very small coupling, $g_\nu$.  This implies that $g_e
> 5 \times 10^4 g_\nu$.

There are several troublesome features to this procedure.  
The first issue which is often raised is: 
\begin{itemize}
\item Why would the Higgs
coupling vary across eleven orders of magnitude (the approximate
ratio of the neutrino mass to the top quark mass)?   
\end{itemize}
In fact, 
this question is rather odd.   Disregarding the neutrinos,
the masses of the charged fermions varies across six orders
of magnitude (from the electron mass to the top mass).   If
six orders of magnitude do not bother anyone, why should eleven?
Turning this around, if the Higgs couplings alreay seemed stretched
in the charged fermion case, the neutrinos stretch the argument 
much further.   This leads to the second troublesome issue,
\begin{itemize}
\item Physically, what is occurring?
\end{itemize}
The Higgs mechanism really gives little physical insight. While it
does introduce mass, it has simply shifted the arbitrariness of the 
magnitude of the mass into an arbitrary coupling to a new field.

These two questions have led theorists to look at other explanations
for small neutrino mass.  It has been noted that neutrinos have the
unique feature of carrying no electric or strong charge.  Thus,
neutrinos, alone among the Standard Model fermions, may be their own
antiparticle, {\it i.e.} they may be Majorana particles.  The nice
consequence of this is a somewhat more motivated theory of mass for
neutrinos.

To understand this, first consider what is meant to be a Dirac
versus a Majorana particle.
If neutrinos are Dirac particles, then 
the $\nu$ and the $\bar \nu$ are distinct particles,
just as the electron and positron are distinct.
The particle, $\nu$ has lepton number $+1$ and 
the antiparticle, $\bar \nu$ has lepton number $-1$.
Lepton number is conserved in an interaction.  Thus, using the muon
family as an example,
$\nu$s ($L=+1$) must produce $\mu^-$ ($L=+1$) and 
$\bar \nu$s ($L=-1$) must produce $\mu^+$ ($L=-1$).
The alternative viewpoint is that  
the $\nu$ and $\bar \nu$ are two helicity states of the same ``Majorana''
particle, which we can call ``$\nu^{maj}$.''   
The $\pi^+$ decay produces the left-handed 
$\nu^{maj}$ 
and the  $\pi^-$ decay produces 
the right-handed $\nu^{maj}$.
This model explains all of the data without invoking lepton number and
has the nice feature of economy of total particles and
quantum numbers, but it renders the neutrino different from all other
Standard Model fermions.

Saying that the neutrino is its own antiparticle is equivalent to saying that
the neutrino is its own charge conjugate, 
$\psi^c=\psi$.   
The operators which appear in the Lagrangian for the neutrino in this
case are the set 
$(\psi_L, \psi_R, \psi_L^c, \psi_R^c)$ and 
($\bar \psi_L, \bar \psi_R, \bar{\psi^c}_L, \bar{\psi^c}_R$).
Certain
bilinear combinations of these
in the Lagrangian can be identified as 
Dirac masses ({\it i.e.}
$m (\bar \psi_L \psi_R + ...) $).  However, we also get a set of terms
of the form:
\begin{equation}
(M_L/2) (\bar{\psi_L}^c \psi_L)+ (M_R/2) (\bar{\psi_R}^c \psi_R) +
...
\end{equation}
These are the 
``Majorana mass terms,'' which mix the pair of charge-conjugate states
of the fermion.  If the particle is not its own charge conjugate, then
these terms automatically vanish and we are left with only the Dirac 
terms.   Dirac particles have no Majorana mass terms, but Majorana 
particles will have Dirac mass terms.

The mass terms of the 
Lagrangian can be written in matrix form:
\begin{equation}
(1/2) (\bar\psi_L^c ~~\bar\psi_R)
\left( 
\begin{array}{ll}
M_L & m \\
m & M_R \\ 
\end{array}
\right) \left( 
\begin{array}{l}
\psi_L \\ 
\psi_R^c 
\end{array}
\right) + h.c., 
\end{equation}
The Dirac mass, $m$, is on the off-diagonal
elements, while the Majorana mass constants, $M_L, M_R$ are on the diagonal.
To obtain the physical masses, one diagonalizes the matrix.

One can now invoke ``see-saw models'' which motivate
small observable neutrino masses.  It turns out that 
GUT's motivate mass matrices that look like \cite{Boris}:
\begin{equation}
\left( 
\begin{array}{ll}
0 & m_\nu \\
m_\nu & M \\ 
\end{array}
\right),
\end{equation}
with $m_\nu \ll M$.
When you diagonalize this matrix to obtain the physical masses, 
this results in two states which can be measured experimentally:
\begin{eqnarray}
m_{light} & \approx & m_\nu^2/M,\\
m_{heavy} & \approx & M
\end{eqnarray}
Grand Unified Theories favor very large masses for the ``heavy
neutrino'' (often called a ``neutral heavy lepton'').  It is argued
that it is most ``natural'' to have $M$ be at the GUT scale.  If $M
\sim 10^{25} eV$, and $m_{light} < 1$ eV, as observed, then $m_\nu\sim
10^{12}$ eV, or is at the TeV scale.  This is rather high compared to
masses of other leptons, but not so far beyond the top quark mass to
regard the connection as crazy.  So while some arbitrariness remains in
this model, nevertheless there is a general feeling in the theory
community that this is an improvement.

In this theory neutrinos have only approximate handedness, where the
light neutrino is mostly LH with a very small admixture of RH and the
neutral heavy lepton is essentially RH.  Thus we have a LH neutrino
which is light, which matches observations, and a RH neutrino which is
not yet observed because it is far too massive.

\subsubsection{Majorana vs. Dirac?}

How can we experimentally tell the difference between the Dirac
($\nu$, $\bar \nu$) and Majorana ($\nu^{maj}$) scenarios?  One can
imagine a straight-forward thought experiment.  First, produce
left-handed neutrinos in $\pi^+$ decays.  These may be $\nu$s or they
may be $\nu^{maj}_{LH}$s.  Next, run the neutrino through a magic
helicity-flipping device.  If the neutrinos are Majorana, then what
comes out of the flipping-device will be $\nu^{maj}_{RH}$.  These
particles will behave like antineutrinos when they interact, showing
the expected RH $y$-dependence for the cross section.  But if the
initial neutrino beam is Dirac, then what comes out of the
flipping-device will be right-handed $\nu$s, which are sterile.  They
do not interact at all.  Such a helicity-flipping experiment is
presently essentially impossible to implement.  If neutrinos do have
mass, then they may have an extremely tiny magnetic moment and a very
intense magnetic field could flip their helicity.  But the design
requirements of such an experiment are far beyond our capability at
the moment.  Therefore, at the moment, we do not know if neutrinos are
Majorana or Dirac in nature.

Instead, experimentalists are pursing a different route.   The Majorana
nature of the neutrino can lead to an effect called 
neutrinoless double $\beta$ decay: $(Z,A) \rightarrow (Z+2,A) +
(e^- e^-)$.   
This is a beyond-the-Standard Model analogue to double $\beta$ decay:
$(Z,A) \rightarrow (Z+2,A) + (e^- e^- \bar \nu_e \bar \nu_e)$.  Double
$\beta$ decay is a standard nuclear decay process with a very low rate
because there is a suppression proportional to $(G_F \cos
\theta_C)^4$.  Therefore, in most cases, if the weak decay is
possible, single $\beta$ decay ($(Z,A) \rightarrow (Z+1,A) + e^- +
\bar \nu_e$) will dominate.  However, there are 13 nuclei, including
$^{136}{\rm Xe} \rightarrow ~^{136}{\rm Ba}$ and $^{76}{\rm Ge}
\rightarrow ^{76}{\rm Se}$, for which single $\beta$ decay is
energetically disallowed.  In these cases double $\beta$ decay with
two neutrinos has been observed \cite{2nudub}.  If the neutrino were
its own antiparticle, then the neutrinos produced in the double
$\beta$ decay process could annihilate, yielding neutrinoless double
$\beta$ decay.

\begin{figure}[t]
\centering
\scalebox{.7}{\includegraphics[clip=T, bb=0 500 450 760]{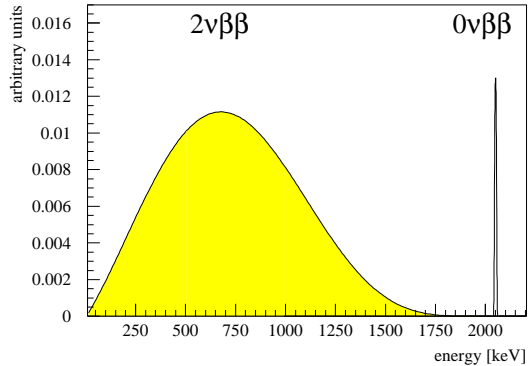}}
\caption{Spectrum for two-neutrino double $\beta$ decay and expected
peak for neutrinoless double $\beta$ decay.}
\label{fig:0bbspect}
\end{figure}

If there are Majorana neutrinos, then the amplitude for 
$0 \nu \beta \beta$ is proportional to the square of
\begin{equation}
 m_{0\nu\beta\beta} = \sum U_{ei}^2m_i.
\label{neutrinolessmass}
\end{equation}
This should be contrasted with eq.~\ref{betamass}.  The $0 \nu \beta
\beta$ searches are probing different effective masses than the direct
searches and the two results yield complementary information.  Like
the direct searches, the possibility of seeing $0 \nu \beta \beta$
depends on the amount of electron-flavor mixed in the most massive
neutrino state.  If this is small, then the rate of decay will be very
low.  Thus the hierarchy of the neutrino states affects our ability to
observe $0 \nu \beta \beta$.  To completely untangle the Dirac {\it
  vs.} Majorana question, three different experiments -- direct mass
measurement, hierarchy measurement and $0 \nu \beta \beta$ measurement
-- may be required\cite{APSwhitepaper}.

Extracting $m_{0\nu\beta\beta}$ from a measured half-life leads to a
theoretical error from the nuclear matrix element calculations.  A
favored style of calculation uses the ``QRPA'' (Quasiparticle Random
Phase Approximation) \cite{Rodin,Simkovic,Sunohen, Stoica} model.
Using $^{100}$Mo as an example, different matrix elements from QRPA
calculations cause $m_{0\nu\beta\beta}$ to vary by up to 2 eV for a
half-life of $4.5 \times10^{23}$ years \cite{NEMO3}.   So the error
is significant.

The $0 \nu \beta \beta$ events must be separated from the standard
two-neutrino double $\beta$ ($2 \nu \beta \beta$) decay background.
This can be done through simple kinematics cuts. The two-body nature
of $0 \nu \beta \beta$ decay will cause a peak at the endpoint of the
$2 \nu \beta \beta$ decay (4-body) spectrum, as shown in
Fig.~\ref{fig:0bbspect}.  
An advantage of observing $2 \nu \beta \beta$, however, is that
measurement of its half-life allows direct measurement of the matrix
element.  At this point the $2 \nu \beta \beta$ decay spectrum has
been observed in 10 elements.  In some cases, such as $^{100}$Mo, the
the $2 \nu \beta \beta$ half-life is well measured and can be used to
constrain nuclear matrix element calculation.  For this case, NEMO-3
reports a half life of $(7.68 \pm 0.02 {\rm (stat)} \pm 0.54 {\rm
  (sys)}) \times 10^{18}$ y \cite{NEMO32nu}.

At present, no signal for $0 \nu \beta \beta$ decay has been clearly
observed.  The present 90\% CL limit on the lifetime from CUORICINO on
$^{130}$Te is $1.8\times 10^{24}$ years, 
corresponding to limit of $m_{0\nu\beta\beta}<0.2-1.1$ eV
\cite{CUORICINO}.  The NEMO-3
experiment has set 90\% CL limits of $4.6\times 10^{23}$ and $1.0\times10^{23}$ on $^{100}$Mo and $^{82}$Se, respectively \cite{NEMO3}.  
The corresponding limits on $m_{0\nu\beta\beta}$ are 0.7-2.8 eV
and 1.7-4.9 eV \cite{NEMO3}. There is a
candidate signal observed at 4.2$\sigma$ from a Germanium detector
\cite{KK}, although the statistical significance is under debate
\cite{APSwhitepaper}.  The measured half-life was $1.19\times 10^{25}$
years.  Until this result is confirmed by further experiments, it is
best to reserve judgment.

Luckily, a range of future $0 \nu \beta \beta$ decay experiments are
on the horizon.  These are expected to probe an order of magnitude
further in lifetimes.  In particular, the germanium-based GERDA
experiment \cite{GERDA}, will turn on soon and will address the
existence of the possible signal.  CUORE \cite{CUORE}, SuperNEMO
\cite{SNEMO}, EXO \cite{EXO}, Majorana \cite{Majoranabb} and Moon
\cite{Moon} will extend the search even further using a wide range of
elements.  The reach of these near future $0 \nu \beta \beta$ covers
the prediction for the inverted mass hierarchy.

\subsubsection{CP Violation in the Neutrino Sector}

\begin{figure}[t]
\centering
\scalebox{.4}{\includegraphics[clip=T]{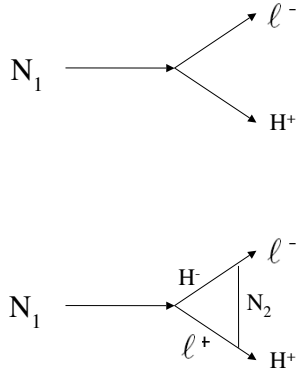}}
\caption{Example of two diagrams for Neutral Heavy Lepton decay which
  can interfere to produce CP violation.}
\label{NHLdecay}
\end{figure}

An intriguing aspect of the ``new Standard Model'' is the heavy
GUT-scale neutrinos which gain mass through the Majorana terms in the
Lagrangian.  There could be more than one, and likely, given the trend
in the Standard Model, there would be three, so we can label these
$N_1$, $N_2$ and $N_3$. These heavy neutrinos have mass prior to the
electroweak phase transition in which the Dirac terms appear.  As a
result, prior to the electroweak phase transition, decays shown in
Fig.~\ref{NHLdecay} are possible.  Both decays produce the same final
state, $N_1 \rightarrow \ell H$, where $\ell$ and $H$ are oppositely
charged.  These diagrams interfere, and can lead to a different decay
rate to $\ell$$^-$ and $\ell$$^+$, which is CP violation.

This form of CP violation would lead to a lepton asymmetry in the
early universe which could be transferred into a baryon asymmetry. A
mechanism for this already appears in the Standard Model, in which but
$B$, baryon number, and $L$, lepton number are not conserved, but the
difference, $B-L$, is exactly conserved.  $B$ and $L$ violation occurs
in transitions between vacuum states at high energies, called the
``sphaleron process.''  Variations on this mechanism,  called 
``leptogenesis,''
may explain the matter-antimatter asymmetry we see today.

$N_1$, $N_2$, and $N_3$ are far too massive to be produced at
accelerators in the near future. Thus observing CP violation in their
decays is out of the question.  However, observing CP violation in the
light neutrino sector would be a plausible hint that the theory is
correct.

To incorporate CP violation into the three-light-neutrino model, the
leptonic mixing matrix is expanded and written as: $U^{with~CP}= V K$.
In this case, $V$ is very similar to the $U$ of Eq. \ref{mns}, but
with a CP violating phase, $\delta$:
\begin{equation}
V =
\begin{pmatrix}
c_{12}c_{13} & s_{12}c_{13} & s_{13} e^{-i\delta} \cr
-s_{12}c_{23}-c_{12}s_{23}s_{13} e^{i\delta} &
c_{12}c_{23}-s_{12}s_{23}s_{13} e^{i\delta}
& s_{23}c_{13} \cr
s_{12}s_{23}-c_{12}c_{23}s_{13} e^{i\delta} &
-c_{12}s_{23}-s_{12}c_{23}s_{13} e^{i\delta} & c_{23}c_{13}
\end{pmatrix}.
\label{V}
\end{equation}
This is analogous to the CKM matrix of the quark sector.
The other term,
\begin{equation}
K~=~\mathrm{diag}\,(1, e^{i\phi_1},e^{i(\phi_2 + \delta)})
\end{equation}
has two further CP violating phases, $\phi_1$ and $\phi_2$.  

Now, we potentially have three non-zero CP violating parameters in the
light neutrino sector, $\delta$, $\phi_1$ and $\phi_2$, as well as one
or more CP violating parameters in the heavy neutrino sector, where
the number depends upon the total number of $N$.  In the Lagrangian,
these all come from a matrix of Yukawa coupling constants.  In
principle, all of these phases can take on the full range of values,
including exactly zero.  However, it is difficult to motivate a theory
in which some are nonzero and some are exactly zero.  It is expected
that these parameters will either all have non-zero values or all be
precisely zero.  If the latter case, then the difference between the
lepton sector, with no CP violation, and quark sector, with clear CP
violation, must be motivated.  As a result, observation of CP
violation in the light neutrino sector is regarded as the ``smoking
gun'' to CP violation in the heavy sector.

Returning to the light neutrino sector, how can the CP phases be
measured?  The $\phi$ phases arise as a direct consequence of the
Majorana nature of neutrinos.  Therefore, in principle, the the $\phi$
phase associated with the electron family is accessible in
neutrinoless double beta decay.  In practice, this will be extremely
hard to measure because this term manifests itself as a change in the
sum in eq.~\ref{neutrinolessmass}, which is proportional to the $0 \nu
\beta \beta$ decay amplitude.  Thus one seeks to measure a deviation
of the (as-yet-unmeasured) $0 \nu \beta \beta$ lifetime from the
prediction which depends upon the mixing angles (with relatively large
errors at present), the (unknown) neutrino masses, and the (poorly
known) nuclear matrix element.  Even if the effect is large,
observation of the effect is clearly hopeless in the near future.  On
the other hand, $\delta$, the ``Dirac'' CP violating term in $V$ may
be accessible though oscillation searches.

CP violation searches involve observing a difference in oscillation
probability for neutrinos and antineutrinos.  Only appearance
experiments can observe CP violation.  A difference between
oscillations of neutrinos and antineutrinos in disappearance searches
is CPT violating.  In oscillation appearance searches, the $K$ matrix does not
affect the oscillation probability because this diagonal matrix is
multiplied by its complex conjugate.  On the other hand, non-zero
$\delta$ can be observed.  To test for non-zero $\delta$, the
oscillation probability must depend upon the $U_{e3}$ component of
Eq.~\ref{V}.  In other words, the search needs to involve transitions
from or to electron flavor and involve the mass state $\nu_3$.  This
combination of requirements -- appearance signal, electron
flavor involvement, and $\nu_3$ mass state involvement -- leads to one
experimental option at present: comparison of $\nu_\mu \rightarrow
\nu_e$ to $\bar \nu_\mu \rightarrow \bar \nu_e$ at the atmospheric
$\Delta m^2$, which is $\Delta m^2_{13}$.  The oscillation probability
is given by:
\begin{eqnarray}
P_{long-baseline} & \simeq & \sin^2 2\theta_{13} \, \sin^2 \theta_{23}
\sin^2 {\Delta} \nonumber \\
& \mp &  \alpha\; \sin 2\theta_{13} \, \sin\delta_{CP}  \, \cos\theta_{13} \sin
2\theta_{12} \sin 2\theta_{23}
\sin^3{\Delta} \nonumber \\
&+&  \alpha\; \sin 2\theta_{13}  \, \cos\delta_{CP} \, \cos\theta_{13} \sin
2\theta_{12} \sin 2\theta_{23}
 \cos {\Delta} \sin^2 {\Delta} \nonumber  \\
&+& \alpha^2 \, \cos^2 \theta_{23} \sin^2 2\theta_{12} \sin^2 {\Delta},
\label{equ:beam}
\end{eqnarray}
where $\alpha$ and $\Delta$ are defined in eq.~\ref{alfdelt}.  The
second term is negative for neutrino scattering and positive for
antineutrino scattering.

Unfortunately, eq.~\ref{equ:beam} convolutes two unknown parameters,
the sign of $\Delta m^2_{13}$ (the mass hierarchy) and the value of
$\theta_{13}$, with the parameter, $\delta$, that we want to measure.
The problem of the mass hierarchy can be mitigated by the experimental
design.  The sign of $\Delta m^2_{13} = m_3^2-m_1^2$ affects the terms
where $\Delta m^2$ is not squared. These terms arise from matter
effects and can so be reduced if the pathlength in matter is
relatively short.  For long baseline experiments, which must shoot the
beam through the Earth, this means that $L$ must be relatively short.
In order to retain the same $L/E$ and, hence, the same sensitivity to
$\Delta m^2_{13}$, $E$ must be comparably reduced.  On the other hand,
the problem of $\theta_{13}$ cannot be mitigated.  From eq.~\ref{V},
one sees that we are in the unfortunate situation of having the CP
violating term multiplied by $\sin \theta_{13}$.  The smaller this
factor, the harder it will be to extract $\delta$.  If $\sin^2
2\theta_{13}$ is smaller than $\sim 0.01$ at 90\% CL, then substantial
improvements in beams and detectors will be required.

Eq.~\ref{equ:beam} also depends on two other as-yet-poorly understood
parameters, $\theta_{23}$ and the magnitude of $\Delta m^2_{23}$.
Disappearance experiments measure $\sin^2
2\theta_{23}=1.00^{+0.16}_{-0.14}$ \cite{Schwetzglob}, thus there is
an ambiguity as to whether $\theta_{23}$, is larger or smaller than
$45^\circ$.  $\Delta m^2_{23}$ is only known to about 10\%
\cite{Schwetzglob, Nikkitalk}.  This measurement is extracted from the
location of the ``dip'' in the rate versus $L/E$ distribution of
disappearance experiments, and is already systematics-dominated.
Improvement requires experiments with better energy resolution
\cite{Nikkitalk} and better understanding of the CCQE and background
cross sections \cite{WaltersNuInt}.  These errors lend a significant
error to the analysis.

\begin{figure} [t]
\centering
\scalebox{.5}{\includegraphics[clip=T]{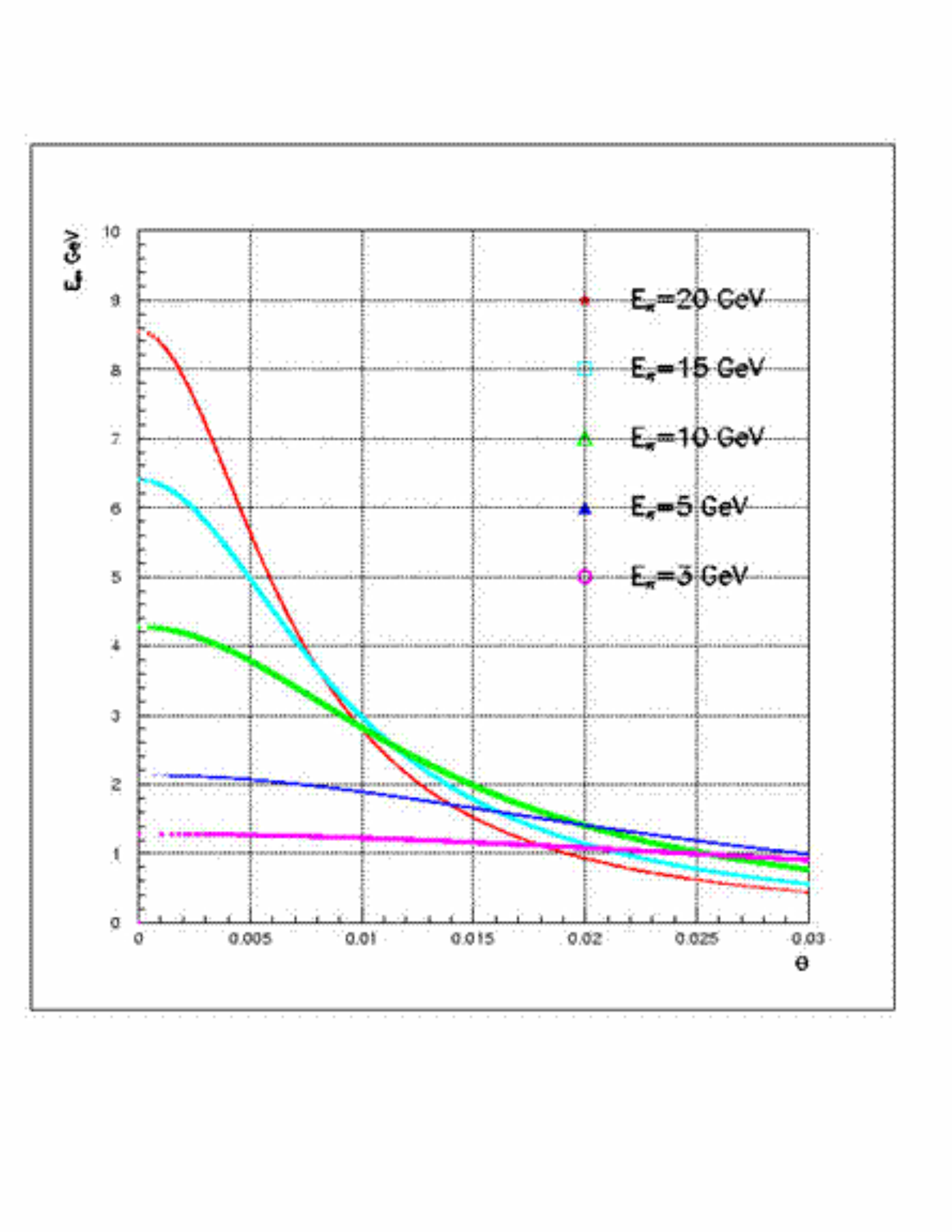}}
\caption{Neutrino energy versus angle off-axis for various values of
pion energy. In this example one can see that 
for moderate off-axis angles, between 15-30 mrad, all pion energies
between 3 and 20 GeV map to approximately 1 GeV neutrino energy.}
\label{offaxis}
\end{figure}

Lastly, it is difficult to measure a $\nu_e$ 
signal which is at the $\sim 1\%$ level.  
Most $\nu_\mu$ beams have a substantial $\nu_e$
intrinsic contamination from $\mu$ and $K$ decays.  Given
that $\theta_{13}$ is small, this contamination is a serious issue.
One solution to this problem is to go to an off-axis beam design.
This relies on the tight correlation between energy and off-axis
angle, $\theta$, 
in two-body decays.  For pion decay, which dominates most beams,
\begin{equation}
E_\nu = {{0.43 E_\pi}\over{1+\gamma^2 \theta^2}},
\end{equation}
where $\gamma = E_\pi/m_\pi$ is the Lorentz boost factor.  The
solution for two-body $K$ decay replaces 0.43 with 0.96. Thus at
$\theta=0$, the relationship between $E_\nu$ and $E_\pi$ is linear.
However, for larger $\theta$, above a moderate energy threshold,
all values of $E_\pi$ map to the same $E_\nu$.  This is illustrated in
Fig.~\ref{offaxis}.  The result is that an off axis $\nu_\mu$ beam
which comes largely from pion decay is tightly peaked in energy while
the $\nu_e$ intrinsic background is spread across a range of energies.
This also helps to reduce $\nu_\mu$ events which are mis-reconstructed
as $\nu_e$ scatters, such as NC $\pi^0$ production where a photon is
lost.  These ``mis-ids'' tend to be spread across a range of energies,
since the true energy is misreconstructed.  Thus a peaked signal, as
one expects from an off-axis beam, is helpful in separating signal and
background.

The major problem with this design is that off axis beams have
substantially lower flux.  The flux scales \cite{NOvAProp} as:
\begin{equation}
F= ( {{2\gamma}\over{1+\gamma^2\theta^2}})^2 {{A}/{4\pi z^2}}.
\end{equation}
In this equation, $A$ is the area of the detector and $z$ is the
distance to the detector.  Two future long baseline experiments, NOvA
\cite{NOvA} and T2K \cite{T2K} are proposing off-axis beams for a
$\nu_\mu \rightarrow \nu_e$ search.  Because of
the low flux, very large detectors are required.

In summary, the path to a test for non-zero $\delta$ is clear but
will take several steps and requires some luck.  First, a clean measure
$\theta_{13}$ from Double Chooz and Daya Bay is needed.  If $\sin^2
2\theta_{13}<0.005$ at 90\% CL, then a significant measurement of CP
violation is unlikely to be possible in the near future.  At the same
time, improvements in the $\theta_{23}$ and $\Delta m^2_{23}$ from
disappearance ($\nu_\mu \rightarrow \nu_\mu$) measurements at MINOS
and T2K will improve the situation.  T2K and NOvA \cite{NOvA} may be
able to make a first exploration of CP parameter space, from $\nu_e$
appearance measurements, depending on statistics.  NOvA may also be
able to address the mass hierarchy question. This will open up the
possibility of measuring CP violation to the next generation of very
long baseline experiments \cite{VLBLstudy}.  The most sensitive of
these use a beam originating at Fermilab and a LAr detector located at
Ash River, Minnesota \cite{Ashletter} or a Cerenkov or LAr detector
located at the Deep Underground Science Laboratory at Homestake
\cite{Patrick}.

At some point in the future, a beta beam or a neutrino factory beam
could provide an intense source of $\nu_e$ and $\bar \nu_e$ fluxes,
allowing comparison of $\nu_e \rightarrow \nu_\mu$ to $\bar \nu_e
\rightarrow \bar \nu_\mu$.  In this case one would search for events
with wrong-sign muons in a calorimeter-style detector.  This would be
a striking signature with low background, especially in the case of a
beta beam.  This could allow a very precise measurement of $\delta$
\cite{nufact}.

\subsection{Neutrinos and the Unexpected}
\label{sub:unexpected}

While it is nice to have a tidy, well-motivated theory of neutrino
masses, it is disconcerting to have essentially no experimental
evidence for this theory.  Moreover, neutrino theories have a history
of being incorrect. Only a decade ago, most theorists would have told
you that neutrinos have no mass.  Those who thought neutrinos might
have mass believed it would be relatively large ($>5$ eV), explaining
dark matter.  Most theorists also believed that if the solar neutrinos
were experiencing oscillations, the correct solution would be the
small mixing angle MSW solution, because the mixing matrix should look
like the quark matrix.  Using the same logic, the atmospheric neutrino
deficit, which could only correspond to large mixing angle, was
routinely dismissed as an experimental effect.

On the basis of this, it is wise not to constrain ourselves to the
``New Paradigm.'' The reason the APS neutrino study chose to devote a
chapter to ``the Unexpected'' was to emphasize the importance of being
open to what nature is telling us about neutrinos.  There are two ways
to approach this idea: (1) the theory-driven approach: explore for
properties which could, theoretically, exist and (2) the
experiment-driven approach: follow up on anomalous results which have
been observed.

For lack of time, I will only briefly consider two examples of the
first case: searching for a neutrino magnetic moment and searching for
CPT violation.  In the Standard Model, the neutrino magnetic moment is
expected to be $\sim 10^{-19} \mu_B$.  Laboratory experiments and
astrophysical limits are many orders of magnitude away from this level
\cite{pdgmagmom}.  Nevertheless, if a new experiment could advance
this measurement by an order of magnitude, that would be worth
pursuing.  A more startling discovery would be a difference in the
oscillation disappearance probability of neutrinos versus
antineutrinos.  In a three-neutrino model, a difference in the rate of
disappearance of neutrinos and antineutrinos would imply CPT
violation.  MINOS will be the next experiment to pursue such a search
\cite{MINOS}.  If CPT violation were discovered we would need to
rethink the very basis of our theory.  However, there are theorists
exploring these ideas.

The remainder of this section will focus on the second approach,
exploring ``anomalies'' which have appeared in various experiments.
Physicists today are always cautious about pursuing deviations from
the Standard Model.  Most do not, in the end, point to new physics.
The Standard Model has been very resilient Most anomalies are
arguably more likely due to systematic effects or statistical
fluctuations, than to new physics.  However, those which do ``pan
out''completely change the way we think.  The solar neutrino deficit
is a perfect example.  So, if a new, unexpected result withstands
questions by the community on the systematics of the experiment, then
the anomaly becomes worth pursuing further.  

There are several examples of $> 3 \sigma$ unexpected results in the
neutrino sector which are worth pursuing and two cases are covered
here. The first, the LSND anomaly, is being actively pursued.  The
second, the NuTeV anomaly, will require a new experiment.  Unlike most
of the topics in these lectures, the NuTeV anomaly is not directly
related to neutrino oscillations and neutrino mass, and so expands the
discussion, which has so far been rather narrowly focussed.

Along with the known discrepancies which have reached the level of
full-fledged anomalies, there are also examples of ``unexpected
results to watch.''  These results which have not yet reached the $3
\sigma$ level, but are showing interesting trends.  For example,
unconstrained fits to atmospheric oscillation data from a wide range
of experiments consistently result in $\sin^2 2\theta_{23}$ best fit
value greater than unity.  While in each case, the best fit is
$\sim 1 \sigma$ from unity, it is the trend which is interesting,
since the experiments involved are all very different.  There is
simply not enough space to cover this and other examples of ``results
to watch.''

The take-away message of this section is: the neutrino sector is a
rich place for new physics to appear, and physicists need to be alert
and open-minded to what nature is saying.

\subsubsection{The LSND Anomaly}

The LSND experiment ran at the LAMPF accelerator at Los Alamos
National Laboratory between 1993 and 1998.  The decay-at-rest (DAR)
beam was produced by impinging 800 MeV protons on a beam dump. These
produced $\pi^+$s which stop and decay to produce $\mu^+$s, which
also stop and decay to produce $\bar \nu_\mu$ and $\nu_e$.  These were
studied in the range of 20 to 55 MeV.  The $\pi^-$s capture, so the
beam has a $<8\times10^{-4}$ contamination of $\bar \nu_e$.  The
neutrino events were observed in a detector located 30 m downstream of
the beam dump.  1220 phototubes surrounded the periphery of a
cylindrical detector filled with 167 tons of mineral oil, lightly
doped with scintillator.  The signature of a $\bar \nu_e$ appearance
was $\bar \nu_e + p \rightarrow e^+ n$.  This resulted in a
two-component signature: the initial Cerenkov and scintillation light
associated with the $e^+$, followed later by the scintillation light
from the $n$ capture on hydrogen, producing a 2.2 MeV $\gamma$.  The
experiment observed $87.9 \pm 22.4 \pm 6.0$ events, a 4$\sigma$ excess
\cite{LSNDfinal} above expectation.

\begin{figure}[t]
\vspace{5mm}
\centering
\scalebox{0.3}{\includegraphics{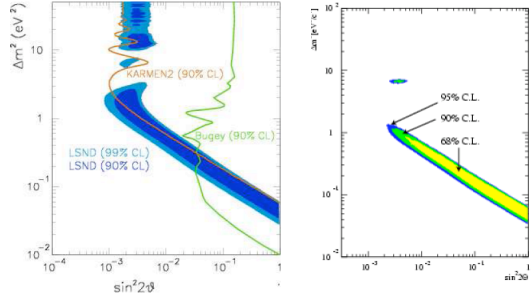}}\\
\vspace{1mm}
\caption{
Left:
LSND allowed range compared to short baseline experiment limits.
Right: Allowed range from the Karmen-LSND joint analysis\cite{joint}.  }
\label{fig:joint}
\end{figure}

LSND is a short baseline experiment, with an $L/E \sim1$ m/MeV.  Thus,
from the two-generation oscillation formula, eq.~\ref{prob}, one can
see that this experiment is sensitive to $\Delta m^2 \ge 0.1$ eV$^2$.
Other experiments have searched for oscillations at high $\Delta m^2$,
and the two most relevant to LSND are Karmen\cite{Karmen} and Bugey
\cite{Bugey}.  The KARMEN experiment, which also used a DAR muon beam
and was located 17.7 m from the beam dump, had sensitivity to address
only a portion of the LSND region, and did not see a signal there.
Since the design of the experiments are very similar,
one can think of the Karmen experiment as a ``near detector,'' which
measures the flux before oscillation.  The results were combined in a
joint analysis performed by collaborators from both experiments; the
allowed range for oscillations is shown in Figure~\ref{fig:joint}
\cite{joint}.  The Bugey experiment was a reactor-based $\bar
\nu_e$ disappearance search which set a limit on oscillations.
Because this is disappearance and not explicitly $\bar \nu_e
\rightarrow \bar \nu_\mu$, its limit is applicable to LSND in many,
but not all, oscillation models.  This limit is shown in
Figure~\ref{fig:joint}.

\paragraph{{ Why can't we fit LSND into the three-neutrino theory?}}

The LSND signal cannot be accommodated within the standard
three-neutrino picture, given the solar and atmospheric oscillations.
To see the
incompatibility, first consider the case where the oscillation signals
de-couple into, effectively, two-generation oscillations
(eq.~\ref{prob}).  For three generations, then 
$\Delta m_{31}^2 = \Delta m_{32}^2 + \Delta m_{21}^2$,
which is clearly not the 
the case for these three signals.   
The more general case allows the atmospheric result to be due to 
a mixture of high (LSND-range)  and low  (solar-range) $\Delta m^2$ values. 
In order for this model to
succeed, the atmospheric $\Delta m^2$ from a shape analysis must shift
up from its present value of $\sim 2 \times 10^{-3}$ eV$^2$ and the chlorine
experiment must have overestimated the deficit of $^7$Be solar
neutrinos.  However, the largest clash between data and
this model arises from the Super-K $\nu_e$ events.  This model
requires that Super-K has missed a $\nu_e$ appearance signal of
approximately the same size and shape as the $\nu_\mu$ deficit before
detector smearing and cuts.  Neutrino measurements are experimentally
difficult and parameters do sometimes shift with time as systematics
are better understood, but it seems unlikely that all of the above
results could change sufficiently to accommodate LSND.  

\paragraph{{ Sterile Neutrinos as a Solution}}

Additional neutrinos which do not interact via exchange of $W$ or $Z$
are called ``sterile;'' they may mix with active neutrinos, and
thereby can be produced in neutrino oscillations.  Experimental
evidence of this would be the disappearance of the active flavor from
the beam.  In contrast to the GUT-scale sterile neutrinos we have
already discussed, the sterile neutrinos which could explain LSND must be
light (in the eV range), and this narrows the class of acceptable
theories.  Nevertheless, a number of possible explanations remain.\cite{plausibletheories}

Sterile neutrinos solve the LSND problem by adding extra mass
splittings.  The additional mass states must be mostly sterile, with
only a small admixture of the active flavors in order to accommodate
the limits on sterile neutrinos from the atmospheric and solar
experiments.    In principle, 
one might expect three sterile neutrinos.   In practice, 
the data cannot constrain information on more than two sterile neutrinos.
Therefore these are called ``3+2'' models.
The method for fitting the data is described in
reference~\cite{sorel}.  One is fitting for two
additional mass splittings, $\Delta m^2_{14}$ and $\Delta m^2_{15}$.
In the fit, the three mostly active neutrinos
are approximated as degenerate.  
The mixing matrix is also expanded by two rows and two columns. 

\begin{table}[tbp] \centering%
{
\begin{tabular}{c|c|c|c|c}
\hline
Channel & Experiment & Lowest       & $\sin^2 2 \theta$   &  Best reach in  \\
        &            & $\Delta m^2$ & at high $\Delta m^2$ & $\sin^2 2 \theta$ \\ 
\hline
$\nu_\mu \rightarrow \nu_e$
       & LSND      & 0.03 eV$^2$  & $> 2.5\times10^{-3}$ & $> 1.2\times10^{-3}$ \\
       & KARMEN    & 0.06 eV$^2$  & $< 1.7\times10^{-3}$ & $< 1.0\times10^{-3}$ \\
       & NOMAD     & 0.4  eV$^2$ & $< 1.4\times10^{-3}$  & $< 1.0\times10^{-3}$ \\ 
\hline
$\nu_e$ disappearance
       & Bugey     & 0.01 eV$^2$  & $< 1.4\times10^{-1}$   & $< 1.3\times10^{-2}$ \\
       & Chooz     & 0.0001 eV$^2$  & $< 1.0\times10^{-1}$ & $< 5\times10^{-2}$ \\
\hline
$\nu_\mu$ disappearance
       & CCFR84      & 6 eV$^2$     &   NA                & $< 2\times10^{-1}$   \\
       & CDHS      & 0.3 eV$^2$  &   NA                   & $< 5.3\times10^{-1}$ \\
\hline
\end{tabular}}
\caption{Results used in 3+2 fit. $\sin^2 2\theta$ limit is 90\% CL
}\label{sbresults}
\end{table}%

The data that drive the fits are the ``short baseline'' experiments
that provide information on high $\Delta m^2$ oscillations, 
summarized in Tab.~\ref{sbresults}.  The
combination of $\bar \nu_\mu \rightarrow \bar \nu_e$
(LSND\cite{LSNDfinal}, Karmen~II\cite{Karmen}), $\nu_\mu \rightarrow
\nu_e$ (NOMAD \cite{NOMAD}), $\nu_\mu$ disappearance (CDHS\cite{cdhs},
CCFR84\cite{ccfr84}), and $\nu_e$ disappearance (Bugey\cite{Bugey},
CHOOZ\cite{chooz}) must all be accommodated within the model.  A
constraint for Super K $\nu_\mu$ disappearance is also included.  None
of the short baseline experiments except for LSND provide evidence for
oscillations beyond 3$\sigma$.  However, it should be noted that CDHS
has a $2\sigma$ (statistical and systematic, combined) effect
consistent with a high $\Delta m^2$ sterile neutrino when the data are
fit for a shape dependence, and Bugey has a 1$\sigma$ pull at $\Delta
m^2 \sim 1$eV$^2$.  As a result, these two experiments define the best
fit combination of high and low $\Delta m^2$ for the 3+2 model.
However, there are acceptable solutions with a combination of low
$\Delta m^2$ values.  The best fit \cite{sorel}, has $\Delta
m^2_{14}=0.92$ eV$^2$, $\Delta m^2_{15}$=22 eV$^2$, although there are
combinations which work with two relatively low $\Delta m^2$ values.
A wide range of mixing angles can be accommodated, and the best fit
has $U_{e4}=0.121$, $U_{\mu4}=0.204$, $U_{e5}=0.036$ and
$U_{\mu4}=0.224$.  The other mixing angles involving the sterile
states are not probed by the $\nu_\mu$ disappearance, $\nu_e$
disappearance and $\nu_\mu \rightarrow \nu_e$ appearance experiments
listed above.  There is 30\% compatibility for all other experiments
and LSND.

Introducing extra neutrinos, including sterile ones, would have
cosmological implications, compounded if the extra neutrinos have
significant mass ($>$1 eV).  However, there are several ways around
the problem.  The first is to note that while the best fit requires a
high mass sterile neutrino, there are low-mass fits which work within
the 3+2 model.  The second is to observe that there are a variety
of classes of theories where the neutrinos do not thermalize in the early
universe \cite{cosmoalternatives}.  In this case, there is no conflict
with the cosmological data, since the cosmological neutrino abundance
is substantially reduced. 

If more than one $\Delta m^2$ contributes to an oscillation appearance
signal, then the data can be sensitive to a CP-violating phase in the
mixing matrix.  Experimentally, for this to occur, the $\Delta m^2$
values must be within less than about two orders of magnitude of one
another.

In 3+2 CP-violating models\cite{cpv3}:
\begin{eqnarray}
P(\stackrel{(-)}{\nu_{\mu}}\to\stackrel{(-)}{\nu_e}) & = & 
4|U_{e4}|^2|U_{\mu 4}|^2\sin^2 x_{41}\\ \nonumber
~~ & & +4|U_{e5}|^2|U_{\mu 5}|^2\sin^2 x_{51} \\ \nonumber
~~ & & +8|U_{e4}||U_{\mu 4}||U_{e5}||U_{\mu 5}|\\ \nonumber
~~&  & ~~~~~~\sin x_{41}\sin x_{51}\cos(x_{54}\mp\phi_{54}),\\ \nonumber
\end{eqnarray}
where in the last line, the negative sign is for neutrino oscillations
and the positive sign is for antineutrino oscillations, and
defined:
\[ { x_{ji}\equiv 1.27\Delta m_{ji}^2 L/E, \hspace{0.25cm}
  \phi_{54}\equiv arg(U_{e4}^*U_{\mu 4}U_{e5}U_{\mu 5}^*).  }
\] 
Thus the oscillation probability is affected by CP violation
through the term $\phi_{54}$.   The CP conserving cases are 
$\phi_{54}=0$ and 180 degrees.  

\paragraph{{MiniBooNE First Results}}

The main purpose of the MiniBooNE experiment was to resolve the
question of the LSND signal.  First results of this experiment,
presented in April, 2007, considered those explanations with a high
expectation for $\nu_\mu \rightarrow \nu_e$ oscillations.  This
includes the CP conserving 3+2 model and many cases of CP violating
3+2 models described above.  As will be described below, the first
results are incompatible with $\nu_\mu \rightarrow \nu_e$
oscillations, but show an unexpected low energy excess, very much in
keeping with the subsection title of ``Neutrinos and the Unexpected.''

The MiniBooNE experiment uses the Fermilab Booster Neutrino Beam,
which is produced from 8 GeV protons incident on a beryllium target
located within a magnetic focusing horn.  The current of the horn can
be reversed such that the beam is dominantly neutrinos or
antineutrinos.  The first results are from neutrino running.  The
MiniBooNE detector is located $L=541$ m from the primary target, and
the neutrino flux has average energy of $\sim 0.75$ GeV.  The detector
is located 541~m from the front of the beryllium target and consists
of a spherical tank of radius 610~cm that is covered on the inside by
1520 8-inch photomultiplier tubes and filled with 800 tons of pure
mineral oil (CH$_2$).  Neutrino events in the detector produce both
Cerenkov and scintillation light.

In order to test the LSND result, the MiniBooNE design maintains
$L/E\sim 1$~m/MeV while substantially changing the systematic errors
associated with the experiment.  This is accomplished by increasing
both $L$ and $E$ by an order of magnitude from the LSND design.  This
changes the source of the neutrinos ($\nu_\mu$ from energetic pions
rather than $\bar\nu_\mu$ from stopped muons), the signature for the
signal, and the major backgrounds in the detector.  In its first run,
in neutrino mode, MiniBooNE collected over a million clean, neutrino
events.  About 99.5\% of the MiniBooNE neutrino events are estimated
to be $\nu_\mu$-induced, while 0.5\% are estimated to be due to
``intrinsic'' $\nu_e$ background in the beam.

The initial MiniBooNE results were analyzed within an appearance-only,
two neutrino oscillation context.  While the LSND signal must be a
result of a more complex oscillation model, in most cases a $\nu_\mu
\rightarrow \nu_e$-like oscillation signal is predicted.  After the
complete $\nu_e$ event selection is applied, the total background was
estimated to be $358 \pm 35$ events, while $163 \pm 21$ signal events
were expected for the LSND central expectation of 0.26\% $\nu_\mu
\rightarrow \nu_e$ transmutation.

\begin{figure}[t]
\centerline{\includegraphics[height=3.4in]{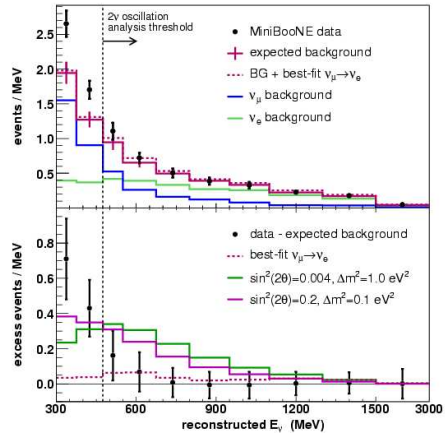}}
\caption{The top plot shows the number of candidate $\nu_e$ events
as a function of $E^{QE}_\nu$.
Also shown are the best-fit oscillation
spectrum (dashed histogram) and the background contributions from
$\nu_\mu$ and $\nu_e$ events. The bottom plot shows the number of 
events with the predicted background subtracted
as a function of $E^{QE}_\nu$.  The two histograms correspond to
LSND solutions at high and low $\Delta m^2$.} 
\label{fig:excess}
\end{figure}

The top plot of Fig.~\ref{fig:excess} shows candidate $\nu_e$ events
as a function of reconstructed neutrino energy ($E^{QE}_\nu$).  The
vertical dashed line indicates the minimum $E^{QE}_\nu$ used in the
two-neutrino oscillation analysis.  There is no significant excess of
events ($22 \pm 19 \pm 35$ events) for $475<E^{QE}_\nu<1250$ MeV;
however, an excess of events ($96 \pm 17 \pm 20$ events) is observed
below 475 MeV.  In the top plot, the points show the statistical
error, while the histogram is the expected background with systematic
errors from all sources.  The background subtracted excess as a
function of $E^{QE}_\nu$ is shown in the bottom plot, where the points
represent the data with total errors.  Oscillation scenarios are
indicated by the histograms.

The low-energy excess cannot be explained by a two-neutrino
oscillation model, and its source is under investigation.  The low
energy events isolated by the cuts, including the excess events, are
single-ring and electromagnetic-like, with no unusual detection
issues.  The low energy excess events are neither consistent with the
spatial nor energy distributions of photons coming from interactions
outside of the tank.  Nor are they consistent with the energy
distribution from single photons from radiative $\Delta$ decays
($\Delta \rightarrow N + \gamma$).  Mis-identification of $\pi^0$ events
is well constrained in MiniBooNE by the rate of reconstructed $\pi^0$
events, studied as a function of $\pi^0$ momentum \cite{pi0confproc}.
This rate would need to be mis-measured by well over a factor of three
to explain the excess, far outside of the error of the analysis.

With that said, as shown in Fig.~\ref{fig:excess}, the excess is not
in agreement with a simple two-neutrino $\nu_\mu \rightarrow \nu_e$
oscillation signal.  This figure shows the predicted spectrum when the
best-fit two-neutrino oscillation signal is added to the predicted
background.  The bottom panel of the figure shows
background-subtracted data with the best-fit two-neutrino oscillation
and two oscillation points from the favored LSND region.

\begin{figure}[tp]
\centerline{\includegraphics[height=4.0in]{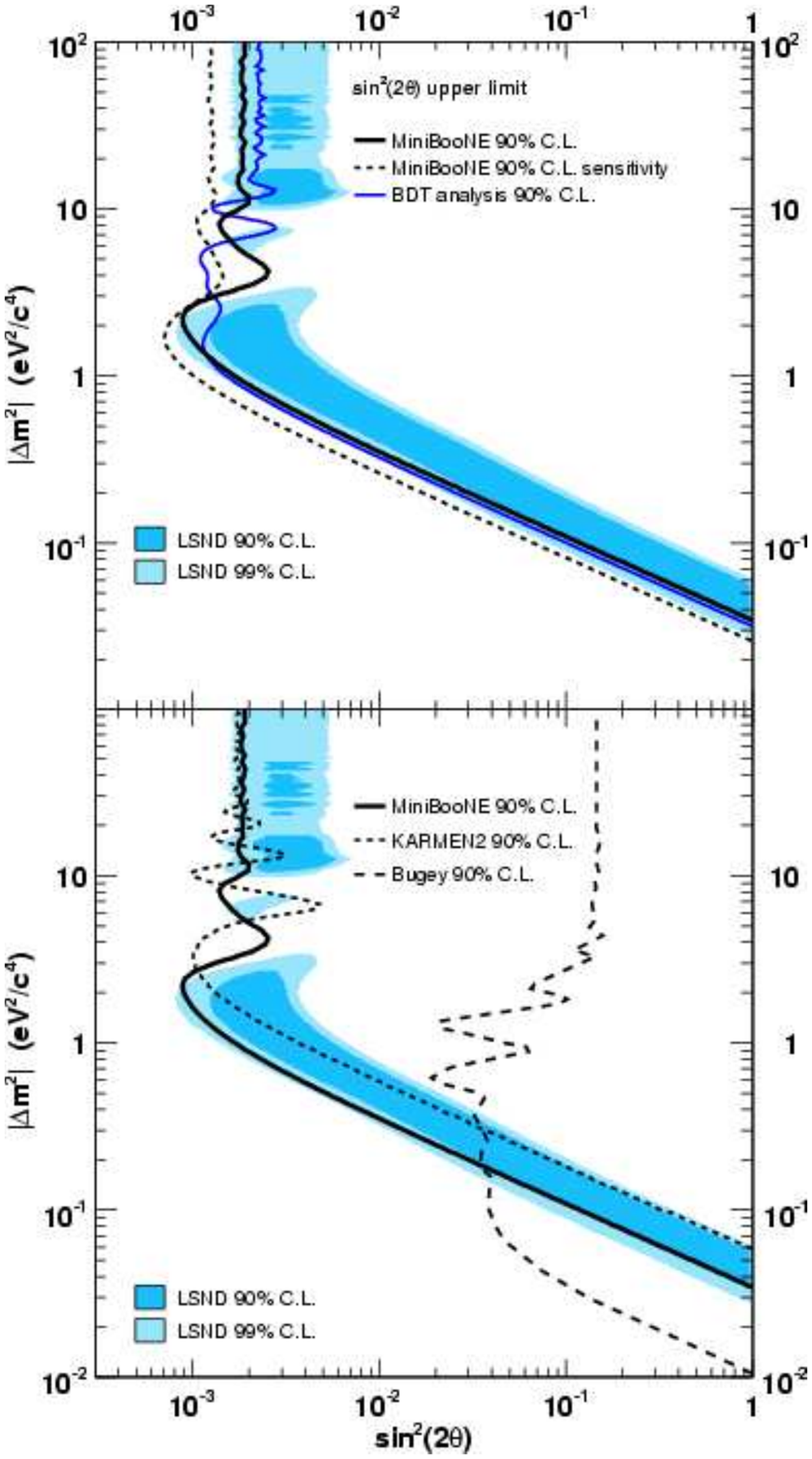}}
\caption{The top plot shows the $\nu_\mu \rightarrow \nu_e$ 
MiniBooNE 90\% CL limit (thick solid curve) 
and sensitivity (dashed curve) for events with
$475<E_\nu^{QE}<3000$ MeV. Also shown 
is the limit from a second cross-check analysis (thin solid curve).
The bottom plot shows the
limits from the KARMEN \cite{Karmen} and Bugey \cite{Bugey} experiments.
The
shaded areas show the 90\% and 99\% CL allowed regions from the LSND
experiment.}
\label{combined_fit}
\end{figure}

A single-sided raster scan to a two neutrino appearance-only
oscillation model is used in the energy range $475<E_\nu^{QE}<3000$
MeV to find the 90\% CL limit corresponding to $\Delta\chi^2 =
\chi^2_{limit} - \chi^2_{best fit}= 1.64$.  As shown by the top plot
in Fig. \ref{combined_fit}, the LSND 90\% CL allowed region is
excluded at the 90\% CL.  A joint analysis as a function of $\Delta
m^2$, using a combined $\chi^2$ of the best fit values and errors for
LSND and MiniBooNE, excludes at 98\% CL two-neutrino appearance-only
oscillations as an explanation of the LSND anomaly. The bottom plot of
Fig. \ref{combined_fit} shows limits from the KARMEN \cite{Karmen} and
Bugey \cite{Bugey} experiments.  This is plot represents an example of
the problem of apples-to-apples comparisons raised in
sec.~\label{subsub:design}.  The MiniBooNE and Bugey curves are
1-sided upper limits on $\sin^22\theta$ corresponding to $\Delta
\chi^2 = 1.64$ -- hence directly comparable -- while the published
KARMEN curve is a ``Feldman Cousins'' contour.

\paragraph{{Next Steps}}

From the initial MiniBooNE result, one can draw two conclusions: (1)
there is excellent agreement between data and prediction in the
analysis region originally defined for the two-neutrino oscillation
search and (2) there is a presently unexplained discrepancy with data
lying above background at low energy.   This combination of information
severely limits models seeking to explain the LSND anomaly.

Interpreting the MiniBooNE data as appearance-only and combining this
result with other data in a 3+2 fit does not give a satisfactory
result \cite{Maltoni}.  While MiniBooNE and LSND are compatible if CP
violation is allowed in the 3+2 model, there is tension between these
results and the $\nu_\mu$ disappearance experiments.  This might be
addressed if a 3+2 interpretation of the MiniBooNE result were
expanded to include the possibility of $\nu_\mu$ disappearance and
intrinsic $\nu_e$ disappearance.  This analysis is underway by the
MiniBooNE collaboration \cite{Georgia}.  Most likely, if a good fit is
obtained in a 3+2 scenario, it will require some level of CP
violation.  MiniBooNE is presently collecting data in antineutrino
mode.  However, this is a small data set ($\sim 2 \times 10^{20}$
protons on target producing the beam) and future running to reach
roughly three times the statistics will be required to make a decisive
statement.  Other, alternative explanations are also being explored
\cite{LorentzViolation, Shortcuts, Yong} .

An upcoming result which will shed light on the question is the
analysis of the MiniBooNE data from the NuMI beam.  This beam is 110
mrad off-axis, with a $\pi$ peak of average $\nu_\mu$ energy of
about 200 MeV and a $K$ peak of about 2 GeV,  and a length of
750 m.  If an excess of events is observed in this analysis, this rules
out mis-estimate of intrinsic $\nu_e$ in the Booster Neutrino Beam as
the source of the MiniBooNE excess.  Results from this study are
expected in autumn, 2007.

If the unexplained excess persists after the above studies, then it
will be valuable to introduce a detector which can differentiate
between electrons and photons.  That is the goal of MicroBooNE
\cite{Bonnieletter}, which uses a Liquid argon TPC (LArTPC) detector.
This is particularly sensitive at low energies and nearly
background-free.  Specifically, this detector has a $\nu_e$ efficiency
$>80\%$ and rejects photons efficiently through $dE/dx$ deposition in
the first $\sim$2 cm of the shower.  With these qualities the detector
can be an order of magnitude smaller in size than MiniBooNE, making
quick construction feasible.  A proposal for this experiment will be
submitted to the Fermilab PAC in autumn 2007.

\subsubsection{The NuTeV Anomaly}

Neutrino scattering measurements offer a unique tool to probe the
electroweak interactions of the Standard Model (SM).  The NuTeV
anomaly is a $3 \sigma$ deviation of $\sin^2 \theta_W$ 
from the Standard Model prediction 
\cite{NuTeVanomaly}.  $\sin^2 \theta_W$ parameterizes the mixing
between the weak interaction $Z$ boson and the photon in electroweak
theory.  Deviations of measurements of this parameter, and its partner
parameter, $\rho$, the relative coupling strength of the
neutral-to-charged-current interactions, may indicate
Beyond-Standard-Model physics. This section also highlights that fact
that new neutrino properties may be revealed in TeV-scale interactions
at LHC, which has not been addressed previously.

The NuTeV experiment represents a departure from the previous train of
thought in several ways.  NuTeV was a deep inelastic neutrino
scattering experiment, and thus is performed at significantly higher
energy than the experiments previously discussed.  Also, while NuTeV
did an oscillation search, it was mainly designed for another purpose:
precision measurement of electroweak parameters.  We will focus on
that purpose here.  As a result, this analysis allows new issues
related to neutrino physics to be brought into the discussion.

\vspace{0.1in}
{\it $\sin^2 \theta_W$ in Neutrino Scattering and Other Experiments}

In neutrino scattering, the neutral current cross section depends upon
$\sin^2 \theta_W$.  The dependence is a function of the neutrino flavor
and the target.   NuTeV was a muon-neutrino-flavor scattering experiment.
In this case, the NC cross sections for scattering from a light
fermion target are:
\begin{eqnarray}
\frac{d\sigma (\nu _\mu f\rightarrow \nu _\mu f)}{dy} &=&\frac{G_F^2s}\pi
\left( \ell _f^2+r_f^2(1-y)^2\right) \left( 1+\frac{sy}{M_Z^2}\right) ^{-2},
\\
\frac{d\sigma (\bar{\nu}_\mu f\rightarrow \bar{\nu}_\mu f)}{dy} &=&\frac{%
G_F^2s}\pi \left( \ell _f^2(1-y)^2+r_f^2\right) \left( 1+\frac{sy}{M_Z^2}%
\right) ^{-2}.
\end{eqnarray}
In this equation, $f$ is the type of light fermion: $f=e^{-},u,d,s,c$.
Several constants appear: $G_F$ is the Fermi constant, $M_Z$ is the
mass of the $Z$.  The two kinematic variables are: $s$, the effective
center of mass energy, which depends on the mass of $f$ and $y$, the
inelasticity (see definitions in Sec.~\ref{sub:interacts}).  $\ell
_f,r_f$ are left and right handed coupling constants which are given
in Table~\ref{couplings}.

\begin{table}[tbp] \centering%
{
\begin{tabular}{c|c|c}
\hline
~~~~~~~~$f~~~~~~~~$ & ~~~~~~~~$\ell_f$~~~~~~~~ & ~~~~~~~~$r_f$~~~~~~~~ \\ \hline
$e^{-}$ & $-\frac 12+\sin ^2\theta _W$ & $\sin ^2\theta _W$ \\ 
$u,c$ & $\frac 12-\frac 23\sin ^2\theta _W$ & $-\frac 23\sin ^2\theta _W$ \\ 
$d,s$ & $-\frac 12+\frac 13\sin ^2\theta _W$ & $\frac 13\sin ^2\theta _W$ \\ 
\hline
\end{tabular}}
\caption{left and right handed coupling constants.
}%
\label{couplings}
\end{table}%

While neutrino scattering has traditionally been a method for
measuring $\sin^2 \theta_W$, the ``Standard Model Prediction'' quoted
in literature comes from the very precise measurements made by the LEP
and SLD experiments, which have been summarized by the Electroweak
Working Group \cite{EWWG}.   $\sin^2 \theta_W$ appears in various
measurements from $e^+e^-$ scattering at the $Z$ pole.  An example which
leads to a highly precise measurement is the ``left-right asymmetry''
measured from polarized scattering at SLD: 
$A_{LR} = (\sigma_L-\sigma_R)/(\sigma_L+\sigma_R)$
where $\sigma_L$ and $\sigma_R$ refer to the scattering cross sections
for left- and right- polarized electrons, respectively.  In this 
case the asymmetry is given by:
\begin{equation}
A_{LR}(Z^0)\equiv \frac{\left( \frac 12-\sin ^2\theta _W^{({\rm (eff)}}%
\right) ^2-\sin ^4\theta _W^{({\rm {eff})}}}{\left( \frac 12-\sin ^2\theta _W(%
^{{\rm {eff})}}\right) ^2+\sin ^4\theta _W^{({\rm {eff})}}},
\label{ALR}
\end{equation}

When comparing quoted values of the weak mixing angle $\sin ^2\theta
_W$, care must be taken because this parameter is defined in various
ways.  The simplest definition is the ``on shell'' description:
\begin{equation}
1-M_W^2/M_Z^2\equiv \sin ^2\theta _W^{({\rm {on-shell})}}.  \label{on-shell}
\end{equation}
This is the definition commonly used in neutrino physics.  In the
discussion which follows, if not explicitly labeled, the on-shell
definition for $\sin^2 \theta_W$ is used.  A variation on this
definition uses the renormalized masses at some arbitrary scale $\mu$
which is usually taken to be $M_Z$:
\begin{equation}
1-M_W(\mu)^2/M_Z(\mu)^2\equiv \sin ^2\theta _W^{({\rm {\bar{MS}})}}.  \label{msbar}
\end{equation}
However, the LEP experiments used the ``effective'' weak mixing angle
which is related to the vector and axial vector couplings:
\begin{equation}
{{1}\over{4}}
(1-g^l_V/g_A^l)\equiv \sin ^2\theta _W^{({\rm {eff})}}.  \label{effective}
\end{equation}
This is what appears in eq.~\ref{ALR} above.
One must convert between definitions, which have different radiative
corrections and renormalization prescriptions, in order to make
comparisons.

\begin{figure}[t]
\vspace{5mm}
\centering
\scalebox{0.5}{\includegraphics[clip=true]{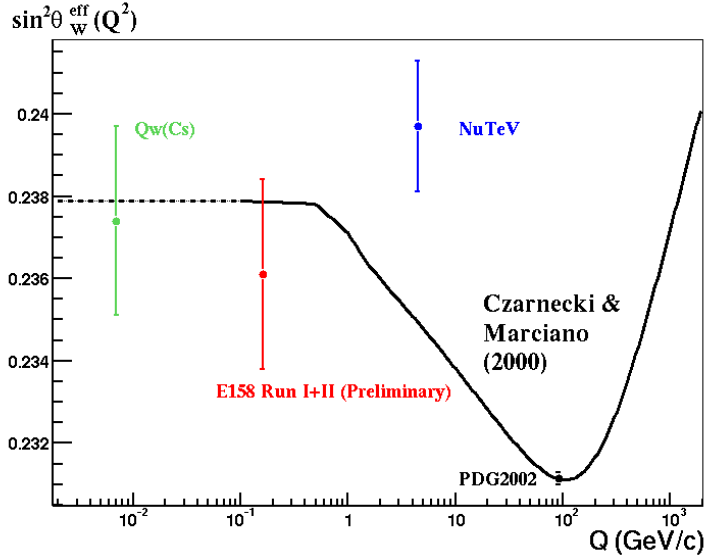}}
\vspace{-2.cm}
\caption{Measurements of $\sin^2 \theta_W$ as a function of $Q$.\cite{E158web}   The curve shows 
the Standard Model expectation. }
\label{Marciano}
\end{figure}

The parameter $\sin^2 \theta_W$ evolves with $Q^2$, the squared
4-momentum transfer of the interaction.  Fig.~\ref{Marciano}
illustrates this evolution.  The highest $Q^2$ measurements are from
LEP and SLD, with $Q^2 = M_Z^2$.  There are several types of
experiments, including neutrino experiments, which measure $\sin^2
\theta_W$ with $Q^2 \ll m_Z^2$.  NuTeV was performed at $Q^2 =$ 1 to
140 GeV$^2$, $\langle Q^2_\nu \rangle=26$ GeV$^2$, $\langle Q^2_{\bar
  \nu} \rangle =15$ GeV$^2$. The lowest $Q^2$ measurements are from
studies of atomic parity violation in the nucleus\cite{APV} (APV),
which arises due to the electroweak interference of the photon and the
$Z$ in the boson exchange between the electrons and the nucleus.  This
samples $Q^2 \sim 0$.  At higher $Q^2$, there is the result from SLAC
E158, a M{\o}ller scattering experiment at average $Q^2=0.026$ GeV$^2$
\cite{E158}.  Using the measurements at the $Z$-pole with $Q^2=M_z^2$
to fix the value of $\sin^2 \theta_W$, and evolving to low $Q^2$,
Fig.~\ref{Marciano}\cite{E158web} shows that APV and SLAC E158 are in
agreement with the Standard Model.

NuTeV is strikingly off the prediction of Fig.~\ref{Marciano}.
Neutrino scattering may measure a different result because new physics
enters the neutrino process differently than the other
experiments. Compared to the colliders, neutrino physics measures
different combinations of couplings.  Also neutrino scattering
explores new physics through moderate space-like momentum transfer, as
opposed to the time-like scattering at the colliders.  With respect to
the lower energy experiments, the radiative corrections to neutrino
interactions allow sensitivity to high-mass particles which are
complementary to the APV and M{\o}ller-scattering corrections.

\paragraph{{The NuTeV Result}} 

The NuTeV experiment provides the most precise measurement of 
$\sin^2 \theta_W$ from neutrino experiments.   The measurement
relied upon deep inelastic scatter (DIS).   It was performed
using a ``Paschos-Wolfenstein
style'' \cite{PW} analysis which is designed to minimize the 
systematic errors which come from our understanding of parton distributions
and masses.

This method requires separated $\nu$ and $\bar
\nu$ beams.   In this case, the following ratios could be formed:
\begin{eqnarray}
R^\nu &=& \frac{\sigma_{NC}^\nu}{\sigma_{CC}^\nu} \\
R^{\bar \nu} &=& \frac{\sigma_{NC}^{\bar \nu}}{\sigma_{CC}^{\bar \nu}} \\
\end{eqnarray}
Paschos and Wolfenstein \cite{PW} recast these as:
\begin{equation}
R^- = \frac{\sigma_{NC}^\nu - \sigma_{NC}^{\bar \nu}}{\sigma_{CC}^\nu - \sigma_{CC}^{\bar \nu}} = \frac{R^\nu - r R^{\bar \nu}}{1-r},
\end{equation}
where $r=\sigma_{CC}^{\bar \nu}/\sigma_{CC}^{\nu}$.  In the case of
$R^-$, many systematics cancel to first order.  In particular, the
quark and antiquark seas for $u, d, s,$ and $c$, which are less
precisely known than the valence quark distributions, will
cancel. Charm production only enters through $d_{valence}$ which is
Cabbibo suppressed and at high $x$, thus the error from the charm mass
is greatly reduced.  One can also form $R^+$, but this will have much
larger systematic errors, and so the strength of the 
NuTeV analysis lies in the measurement of $R^-$.

According to the ``Paschos-Wolfenstein'' method, an experiment should
run in neutrino and antineutrino mode, categorize the events as CC or
NC DIS, and then form $R^-$ to extract $\sin^2 \theta_W$.  This
requires identifying the CC or NC events properly in NuTeV's
iron-scintillator/drift-chamber calorimeter.  Most CC DIS events have
an exiting muon, which causes a long string of hits in the
scintillator and are therefore called ``long.''  Most NC DIS events
are relatively ``short'' hadronic showers.  However, there are
exceptions to these rules.  A CC event caused by interaction of an
intrinsic $\nu_e$ in the beam will appear short.  An NC shower which
contains a pion-decay-in-flight, producing a muon, may appear long.
The connection between long vs. short and CC vs. NC must be made via
Monte Carlo.

NuTeV measurement is in agreement with past neutrino scattering
results, although these have much larger errors.  However, the NuTeV
result is in disagreement with the global fits to the electroweak data
which give a Standard Model value of $sin^2\theta_W =0.2227$
\cite{NuTeVanomaly}.

\paragraph{{Explanations}} 

In the case of any anomaly, it is best to start with the commonplace
explanations.  Three explanations for the NuTeV anomaly that are
``within the Standard Model'' have been proposed: the QCD-order of the
analysis, isospin violation, and the strange sea asymmetry.  The
NuTeV analysis was not performed at a full NLO level.  However, the
effect of going to NLO on NuTeV can be estimated \cite{nlomodels}, and
the expected pull is away from the Standard Model.  The NuTeV
analysis assumed isospin symmetry, that is, $u(x)^p = d(x)^n$ and
$d(x)^p = u(x)^n$.  Various models for isospin violation have been
studied and their pulls range from less than $1\sigma$ away from the
Standard Model to $\sim 1 \sigma$ toward the Standard
Model. \cite{isomodels}.  Variations in the
strange sea can either pull the result toward or away from the
Standard Model expectation \cite{isomodels}, but not by more than one
sigma.   

With respect to Beyond-Standard-Model explanations, Chapter 14 of the
APS Neutrino Study White Paper on Neutrino Theory \cite{APSwhitepaper}
is dedicated to ``The Physics of NuTeV'' and provides an excellent
summary.  The discussion presented here is drawn from this source.

The NuTeV measurements of $R^\nu$ and $R^{\bar \nu}$, the NC-to-CC
cross sections, are low compared to expectation. For this to be a 
Beyond-Standard-Model effect, it therefore requires introduction
of new physics that suppresses the
NC rate with respect to the CC rate.  Two types of models produce this
effect and remain consistent with the other electroweak measurements:
(1) models which affect only the $Z$ couplings, {\it e.g.},  the
introduction of a heavy $Z^\prime$ boson which interferes with the
Standard Model $Z$; or (2) models which affect only the neutrino
couplings, {\it e.g.},  the introduction of moderate mass neutral heavy
leptons which mix with the neutrino.

Any $Z^\prime$ model invoked to explain NuTeV must selectively
suppress NC neutrino scattering, without significantly affecting the
other electroweak measurements.  This rules out most models,
which tend to increase the NC scattering rate. 
Examples of successful models are those where the $Z^\prime$
couples to $B-3L_\mu$ \cite{hep-ph/209316} or to $L_\mu -L_\tau$
\cite{hep-ph/0110146}. 

Moderate-mass neutral heavy leptons, {\it a.k.a.} `` neutrissimos,'' can
also produce the desired effect.  Suppression of the coupling comes
from intergenerational mixing of heavy states, so that the $\nu_\mu$
is a mixture:
\begin{equation}
\label{mixeq}
\nu_\mu = (\cos \alpha) \nu_{{\rm light}} + (\sin \alpha) \nu_{{\rm heavy}}.
\end{equation} 
The $Z\nu_\mu\nu_\mu$ coupling is modified by $\cos^2 \alpha$ and
the $W\mu\nu_\mu$ coupling is modified by $\cos \alpha$.  
Neutrissimos may have masses as light as $\sim 100$ GeV
\cite{0706.1732v1.pdf}.  These new particles can play the role of the
seesaw right-handed neutrinos, as long as one is willing to admit
large tuning among the neutrino Yukawa couplings
\cite{0706.1732v1.pdf}.  So this offers an alternative to the GUT-mass
heavy neutrino model discussed in sec.~\ref{sub:paradigm}.

If neutrissimos exist, they would be expected to show up in other
precision experiments.  One must avoid the constraints on mixing from
$0 \nu \beta \beta$ (recall eq.~\ref{neutrinolessmass} to see why
these experiments have sensitivity to the mixing).  These experiment
place a limit of $|U_{e4}|^2$ at less than a few $\times 10^{-5}$ for
a 100 GeV right-handed neutrino.  Rare pion and tau decays constrain
$|U_{\mu 4}|^2$ to be less than 0.004 and $|U_{\tau 4}|^2$ to be less
than 0.006, respectively.

Neutrissimos would be produced at LHC, thus neutrino
physics can be done at the highest energy scales!  However, they may
be difficult to observe.  One would naturally look for a signal of
missing energy.  However, neutrissiomos will not necessarily decay
invisibly; for example one can have $N \rightarrow \ell + W$ and the
$W$ may decay to either two jets or a neutrino--charged-lepton pair.
Only the latter case has missing energy.  This may make them difficult 
to identify.

If the neutrissimo is a Majorana particle, then these could provide a
clue to the mechanism for leptogenesis.  The present models of
leptogenesis require very high mass scales for the neutral lepton.
However, theorists are identifying ways to modify the model to
accommodate lower masses \cite{hep-ph/0410075v2}.  There also may be a
wide mass spectrum for these particles, with one very heavy case that
accommodates standard leptogenesis models, while the others have masses
in the range observable at LHC \cite{hep-ph/0608147}.

\paragraph{{NuSOnG and Other Possibilities}} 

A new round of precision electroweak measurements can be motivated by
the NuTeV anomaly as well as the imminent turn-on of LHC.  These
measurements are best done using neutrino-electron scattering, because
this removes the quark-model related questions discussed in the
previous section.  Two possible methods for such a measurement are
$\nu_\mu$ scattering with higher statistics, using a NuTeV-style beam,
or a $\bar \nu_e$s scattering measurement from a reactor.  In either
case, to provide a competitive measurement, the error from the best
present neutrino-electron scattering measurement, from CHARM II, must
be reduced by a factor of five.

The NuSOnG (Neutrino Scattering On Glass) Experiment \cite{NuSOnG}is
proposed to run using a $\nu_\mu$ beam produced by 800 GeV protons on
target from the TeVatron.  The plan is to use a design which is
inspired by the CHARM II experiment: a target of SIO$_2$ in one
quarter radiation length panels, with proportional tubes or
scintillator to allow event reconstruction.  The detector will have a
2.6 kton fiducial volume.  The major technical challenge of such an
experiment is in achieving the required rates from the TeVatron, as
$\times 20$ the rate of NuTeV proton delivery is required.

Alternatively, a measurement of the weak mixing angle using
anti-neutrinos from reactors may be possible \cite{braidwoodsin2thw}.
The weak mixing angle can be extracted from the purely leptonic $\bar
\nu_e e$ ``elastic scatter'' (ES) rate, which is normalized using the
$\bar \nu_e p$ ``inverse beta decay'' (IBD) events, to reduce the
error on the flux.  Thus, a hydrocarbon (scintillator oil) based
detector, which has free proton targets for the IBD events, is ideal.
Gadolinium (Gd) doping is necessary for a high rate of neutron
capture, which constitutes the signal for the IBD events.  A window
in visible energy of 3 to 5 MeV is selected to reduce backgrounds from
contaminants in the oil and cosmic-muon-induced isotopes.  In this
energy range, the dominant contamination comes from the progeny of the
uranium and thorium chain.  This would clearly be an ambitious,
state-of-the-art measurement, but could be done at a new reactor
experiment where the detector is in close proximity to the source and
had high shielding from cosmic rays.

\subsection{Neutrinos and the Cosmos}
\label{sub:cosmos}

Neutrinos are ubiquitous in the universe, and their presence and
interactions must be incorporated into astrophysical and cosmological
models.  Nearly any new neutrino property will have direct
consequences in these fields, which must be examined.  As an
illustrative example, the first discussion considers the impact
of introducing of relatively light sterile neutrinos to the theory.
Sterile neutrinos with keV-scale masses can explain dark matter as
well other astrophysical questions.

As we improve our capability for detecting astrophysical neutrinos,
these become a new source of neutrinos for study.  The second example
is a case in point: the search for ultra-high energy
sources of neutrinos.  The discovery of such sources would be of great
interest to astrophysics, and the particle physics we can do with such
a ``beam'' is remarkable.

While this section concentrates on the unknown, it is interesting to
note that the known astrophysical sources of neutrinos are
sufficiently intense that these neutrinos are already a possible
background to other physics measurements.  An example is the case of
dark matter searches, which aim to measure cross sections as small as
$10^{-46}$ cm$^2$.  These experiments will have to contend with the
background from coherent scattering ($\nu + N \rightarrow \nu + N$) of
solar neutrinos, which has a cross section of $10^{-39}$ cm$^2$ and
which produces a recoil nucleon that is very much like the expected
dark matter signal \cite{Joc}.

\subsubsection{Neutrinos as Dark Matter}

From the mid-1980's through mid-90's a 5 eV $\nu_\tau$ was considered
a likely candidate for dark matter.  This was the motivation for the
Chorus and NOMAD search for $\nu_\mu \rightarrow \nu_\tau$ oscillations
in the $> 10$ eV$^2$ range \cite{Chorusnutau, NOMADnutau} as well as the
proposed COSMOS experiment \cite{COSMOS}.

In the late 90's and early 2000's, two measurements led to a shift in
opinion about neutrinos as candidates for dark matter.  The first was
the Super-K confirmation of $\nu_\mu$ oscillations.  As discussed
in sec.~\ref{subsub:evidence}, the cleanest explanation, which fits within a three-neutrino
model, is that this effect is $\nu_\mu \rightarrow \nu_\tau$.
Combining this information with the direct limit on the $\nu_e$
implied that neutrinos were unlikely to have masses in the 5 to 10 eV
range, as required for dark matter.  Also, at the same time, studies
of the large scale structure of the universe indicated that dark
matter must be non-relativistic, or ``cold.''  Relativistic, or ``hot
dark matter,'' like neutrinos, would smooth the large scale structure
far beyond observations \cite{clumpy}.   

The idea of neutrinos as dark matter candidates fell into disfavor.
For some time, the more likely solution was assumed to be WIMPs,
Weakly Interacting Massive Particles.  The ``weak'' in this name is
somewhat confusing, since it does not refer strictly to the weak 
interaction -- other Beyond-Standard-Model interactions are involved.
It is simply meant to say that the interaction rate is very low.  

For some time, the lightest supersymmetric particle has been the most
favored candidate for the WIMP.  However, this is now starting to be
questioned, as no evidence for supersymmetry has been observed at
colliders \cite{susylimit}.  This makes the formulation of the theory
more awkward, and SUSY explanations for dark matter have been pushed
from the ``Minimal Super Symmetric Model'' to the ``Next-to-Minimal
Super Symmetric Model,'' and even this is challenged \cite{NMSSM}.
If supersymmetry does not show up at LHC, then a new explanation for
dark matter must be found.

\begin{figure}[t]
\vspace{5mm}
\centering
\scalebox{0.5}{\includegraphics[clip=true]{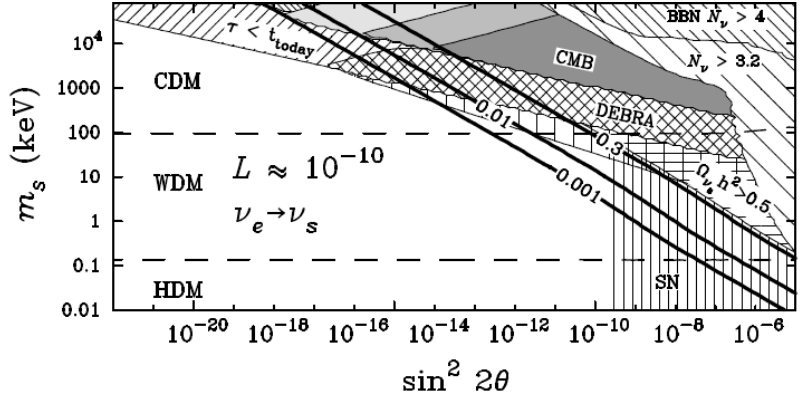}}
\vspace{-2.cm}
\caption{Bounds for $\nu_e \rightarrow \nu_s$ oscillations from
astrophysics and cosmology.  Allowed regions for
neutrino cold, warm and hot dark matter are shown.\cite{AFP}.
}
\label{AFP}
\end{figure}

As a result, neutrino models are being reconsidered
\cite{DM1,DM2,XSF,AFP,DH,AF,Kev,BiermannKusenko,AbazajianKoushiappas,kevlimits}.
The new models involve neutrinos which are mostly sterile, with a very
tiny mixing with the light flavors and with keV masses ($0.5\,{\rm
  keV} < m_\nu < 15\,{\rm keV}$ and $\sin^22\theta \ge 10^{-12}$)
Because of the high mass, they are not relativistic and are regarded
as ``warm'' or ``cold.''  At high mass, the large scale structure
limits are much less stringent.  A recent analysis from the Sloan
Digital Sky Survey (SDSS) finds that a sterile neutrino mass above 9
keV can reproduce the power spectrum\cite{SDSS}.  Only tiny mixing
with the active neutrinos is required in order to produce the dark
matter.  With such small mixings, this model easily evades all
accelerator-based bounds on $\nu_e \rightarrow \nu_s$ oscillations.
As shown in Fig.~\ref{AFP}, this model also escapes the cosmological limits
on $\nu_e \rightarrow \nu_s$ from the CMB measurements, from Big Bang
Nucleosynthesis (BBN), and supernova limits (SN), assuming negligible
lepton number asymmetry, $L$, in the early universe.

Currently the only constraints on keV sterile neutrinos come from
X-ray astronomy\cite{kevlimits} which are searching for evidence of
radiative decay of the massive neutrino into a lighter state, $\nu_2
\rightarrow \nu_1 + \gamma$.  This proceeds through loop diagrams where
the photon is coupling to a $W$ or a charged lepton in the loop
\cite{PalWolf}.  Because this is a 2-body decay, one is searching for
a spectral line in the x-ray region.  For a Dirac-type sterile
neutrinos of mass $m_s$ the decay rate is given by:
\begin{equation}
\Gamma_\gamma(m_s) = 1.36\times 10^{-29} {\rm s}^{-1} ({{\sin^2 2\theta}\over{10^{-7}}})
({{m_s}\over{1 {\rm keV}}})^5,
\end{equation}
which is clearly tiny, even for a keV scale neutrino.  This is
important as the dark matter neutrinos must be stable on the scale of
the lifetime of the universe.  No signal has been observed and the
current mass limit, from the Chandra X-ray telescope observations,
ranges from $>$3 to $>$6 keV depending on model assumptions.

Having motivated a keV-mass sterile state using dark matter, 
one can explore the consequences in other areas of cosmology and
astrophysics.   The small mixing allows these neutrinos to evade bounds from 
big bang nucleosynthesis \cite{fullpulse}.   Their presence may
be beneficial to models of supernova explosions and pulsar kicks,
as discussed below.

The existence of these neutrinos may improve the supernova models
substantially.  The problem faced by most models is that the supernova
stalls and fails to explode.  As modest increase in neutrino luminosity
during the epoch when the stalled bounce shock is being reheated
($t_{\rm pb} < 1\,{\rm s}$) will incite the explosion.\cite{hep-ph/0609425}  
As the
supernova occurs, neutrinos will oscillate and even be affected by MSW
resonances.  If neutrinos oscillate to a sterile state, then their
transport-mean-free-paths become larger.  This increases the neutrino
luminosity at the neutrino sphere and
makes ``the difference between a dud and an
explosion.'' \cite{hep-ph/0609425}

Also, sterile neutrinos in the 1 to 20 keV mass range can also be used to
explain the origin of pulsar motion.  Pulsars are known to have large
velocities, from 100 to 1600 km/s.  This is a much higher velocity
than an ordinary star which typically has 30 km/s.  Pulsars also have
very high angular velocities.  Apparently, there is some mechanism to
give pulsars a substantial ``kick'' at birth, which sends them off
with high translational and rotational velocities.  One explanation
for the kick is an asymmetric neutrino emission of sterile neutrinos
during or moments after the explosion which forms the pulsar
\cite{fullpulse}.  An asymmetry in neutrino emission occurs because
the ``urca reactions''
\begin{eqnarray}
\nu_e + n &\leftrightarrow& p+e^-, \\
\bar \nu_e + p &\leftrightarrow& n+e^+, 
\end{eqnarray}
are affected by magnetic fields which trap electrons and positrons.
If the neutrinos oscillate to a sterile state, they stream out of the
pulsar.  If the sterile neutrinos have high mass, they can provide a
significant kick.  There are a number of solutions, either with
standard neutrino oscillations, as described by eq.~\ref{prob} or with
an MSW-type resonance.  All require a sterile state in the 1 to 20 keV
range with very small mixing, compatible with the dark matter scenario
described above.

In summary, this is an example of how introducing a new neutrino
property, {\it i.e.} sterile companions to the known neutrinos, can
have a major impact on astrophysical models.  This is an interesting
case in point, because it is unlikely that these neutrinos will be
observable in particle physics experiments in the near future.  At
present, the only detection method is through X-ray emission due to
the radiative decay.  Thus this is, at the moment, an example of a
neutrino property which is entirely motivated and explored in the
context of astrophysics.

\subsubsection{Ultra High Energy Neutrinos}
\label{subsub:ultra}

All of the experiments so-far discussed have
used neutrino sources in the energy range of a few MeV to many GeV.
We do not know how to produce neutrino beams at higher energies.
However, nature clearly has high energy acceleration mechanisms,
because cosmic rays with energies of  $10^8$ GeV have been measured.
A new generation of neutrino experiments is now looking for neutrinos
at these energies and beyond.  These include  AMANDA \cite{AMANDA},
ICEcube \cite{ICE}, Antares \cite{Antares}, and Anita \cite{Anita}.

These experiments make use of the fact that the Earth is opaque to
ultra-high energy neutrinos.  The apparent weakness of the weak
interaction, which is due to the suppression by the mass of the $W$ in
the propagator term, is reduced as the neutrino energy increases.
Amazingly, when you reach neutrino energies of $10^{17}$ eV, the Earth
becomes opaque to neutrinos.  To see this, recalling the kinematic
variables defined in sec.~\ref{sub:interacts}, consider the following
back-of-the-envelope calculation.  For a $10^8$ GeV $\nu$,
$s=2ME_\nu=2\times10^{8}$ GeV$^2$. Most interactions occur at low $x$;
and at these energies $x_{typical}\sim 0.001$.  For neutrino
interactions, the average $y$ is 0.5.  Therefore, using $Q^2 = sxy$,
we find $Q^2_{typical}= (2\times 10^8) (1\times 10^{-3}) (0.5) = 1
\times 10^5$ Gev$^2$.  The propagator term goes as
\begin{equation}
\Bigg({{M_W^2}\over{M_W^2+Q^2}}\Bigg)^2 =
\Bigg({{1}\over{{1+Q^2/M_W^2}}} \Bigg)^2 \approx {{M_W^4}\over{Q^4}},
\end{equation}
which is approximately $10^{-3}$ for our ``typical'' case.
The typical cross section $\sigma_{typical}$ is:\\  
$E \times (\sigma_{tot}/E) \times  ({\rm prop~term}) =
10^8 {\rm (GeV)} (0.6\times 10^{-38} {\rm cm}^2/{\rm GeV}) 10^{-3}\\ =
0.6 \times 10^{-33} {\rm cm}^2$.
From this we can extract the interaction length, $\lambda_0$, on iron,
by scaling from hadronic interactions, which tells us that 
at 30 mb ($=0.3 \times 10^{-25}$ cm$^2$), $1 \lambda_0 \sim 10$ cm.   
This implies that 
$\lambda_0$ for our very high energy $\nu$s is $\approx 5 \times 10^3$ km.
However, the Earth is a few $\times 10^4$ km.  Thus all of the neutrinos 
interact; the Earth is opaque to them.    

This opens up the opportunity to instrument the Earth and use it as a
neutrino target.  One option is to choose a transparent region of the
Earth, ice or water, and instrument it like a traditional neutrino
detector.  This has been the design chosen by AMANDA, ICECube and
Antares, which use phototubes to sense the Cerenkov light produced
when charged particles from neutrino interactions traverse the
material.  The largest of these detectors are on the order of (1
km)$^3$ of instrumented area.  The second method exploits the Askaryan
effect \cite{Askaryan} in electromagnetic showers.  Electron and
positron scattering in matter have different cross sections.  As the
electromangetic shower develops, this difference leads to a negative
charge asymmetry, inducing strong strong coherent Cerenkov radiation
in the radio range.  The pulse has unique and easy-to-distinguish
broadband (0.2 - 1.1 GHz) spectral and polarization properties which
can be received by detecting antennas launched above the target area.
In ice, the radio attenuation length is 1 km.  The Anita Experiment,
which uses such a detector, can view $2\times 10^6$ km$^3$ of volume.
This makes it by far the world's largest tonnage experiment.

Neutrinos with energies above 10$^4$ GeV have yet to be observed.
However, they are expected to accompany ultra high energy cosmic rays,
which have been observed.  Nearly all potential sources of ultra-high
energy cosmic rays are predicted to produce protons, neutrinos, and
gamma rays at roughly comparable levels.  Ultra high
energy protons, which have been observed by the HiRes \cite{HiRes} and
Auger \cite{Auger}  experiments are guaranteed sources of neutrinos
through the Gresein-Zatsepi-Kuzmin (GZK) interaction.  In this effect,
protons above $E_{GZK}=6\times 10^{10}$ GeV scatter from the
cosmic microwave background: $p \gamma \rightarrow \Delta^+
\rightarrow n \pi^+$.  This degrades the energy of protons above
$E_{GZK}$, leading to an apparent cutoff in the flux called the ``GZK
cutoff.''  There are several $\Delta$ resonances and the CMB photons
have an energy distribution, so the cutoff is not sharp.  But it has
been clearly observed by both HiRes and Auger \cite{GZKcutoff}.  As a
result, ultra-high energy pions are produced and these must decay to
ultra-high-energy neutrinos.  There may be other, more exotic
mechanisms for producing an ultra-high energy flux, possibly with
energies beyond the GZK cutoff.  Since, unlike the protons, these
neutrinos do not interact with the cosmic microwave background, they
can traverse long distances and can be messengers of distant point
sources.

There many reviews of the exotic physics one can do with ultra high
energy neutrino interactions.  The opportunities include\cite{Quigg}
gravitational lensing of neutrinos, the search for bumps or steps in
the NC/CC ratio, the influence of new physics on neutrino oscillations
at high energies, the search for neutrino decays, neutrino interaction
with dark matter WIMPs, and the annihilation of the ultra high energy
neutrinos by the cosmic neutrino background.  This final example is
interesting because significant limits have been set by Anita-lite, a
small prototype for Anita that flew only 18.4 days.  This illustrates
the power of even a small experiment entering an unexplored frontier
of particle physics.

The 1 eV mass
neutrino implied by the LSND anomaly could be a candidate for the
source of the ultra-high energy cosmic rays observed on Earth.\cite{Weiler}  
This neutrino, if produced at ultra-high energies,
could annihilate on the cosmic neutrino
background producing a ``Z-burst'' of ultra high energy hadrons.  This
was offered as an explanation for ultra-high energy cosmic which were
observed by the AGASA experiment \cite{AGASA}.  Scaling from the AGASA
rate, a prediction for the flux of ultra-high neutrinos in the energy
range of $10^{18.5}<E_\nu<10^{23.5}$ eV for Z-burst models was made
\cite{Fodor, Kalashev}.  Based on this flux, Anita-lite was predicted
to see between 5 and 50 events at $>$99\% CL.  During its short run,
this prototype detector observed no events in the energy range and
therefore could definitively rule out this model \cite{AnitaLite}.
Shortly thereafter, the AGASA events were shown to be due to energy
miscalibration \cite{Agasaexplained}.

\section{Conclusions}

The goal of this review was to sketch out the present questions in
neutrino physics, and discuss the experiments that can address them.
Along the way, I have highlighted the experimental techniques and
challenges.  I have also tried to briefly touch on technological
advances expected in the near future.  This text followed the
structure of the set of lectures entitled ``Neutrino Experiments,''
given at the 2006 TASI Summer School.

Neutrino physics is an amalgam of astrophysics, cosmology, nuclear
physics, and particle physics, making the field diverse and exciting,
but hard to review comprehensively.  In this paper, I have tried to
touch on examples which are particularly instructive and have been
forced to leave out a wide range of other interesting points.  What
should be clear, however, is that the recent discoveries by neutrino
experiments have opened up a wide range of interesting questions and
opportunities.  This promises to be a rich field of research for both
theorists and experimentalists for years to come.

~~~~\\
~~~~\\
~~~~\\
I wish to thank A. Aguilar-Arevalo,  G. Karagiorgi, B. Kayser, P. Nienaber,
J. Spitz, and E. Zimmerman for their suggestions concerning this
text.

~~~~\\

\end{document}